\documentclass[11pt]{article}
\usepackage[utf8]{inputenc}
\usepackage[margin=1in]{geometry}
\usepackage{graphicx} % Required for inserting images
\usepackage{placeins}
\usepackage{amsfonts}
\usepackage{amsmath}
\usepackage{dirtytalk}
\usepackage{algorithm}
\usepackage{mathtools}
\usepackage{xcolor}
\usepackage{amssymb}
\usepackage[title]{appendix}
\usepackage[noend]{algpseudocode}
\usepackage{amsthm}
\usepackage{bigints}
\usepackage{thm-restate}
\usepackage{hyperref}
\hypersetup{
    colorlinks = true,
    linkcolor = blue,
    citecolor = blue
    }
    
\usepackage[
        noabbrev,
        capitalise,
        nameinlink,
    ]
    {cleveref}
\usepackage{appendix} 
\usepackage{makecell}
\usepackage{bigints}
\usepackage{mathtools}
\usepackage{subfigure}
\usepackage{booktabs}
\usepackage{bbding}
\usepackage{wrapfig}
\usepackage{bbm}
\usepackage{cleveref}
\usepackage{romannum}
\usepackage{siunitx}
\newtheorem{thm}{Theorem}[section]

\newtheorem{lemma}[thm]{Lemma}
\newtheorem{prop}[thm]{Proposition}

\theoremstyle{definition}

\newtheorem{rmk}[thm]{Remark}

\usepackage[round]{natbib}
\title{Locally Linear Generalized Single Index Regression}

\author{
\begin{tabular}{c c c}
\makecell[c]{Tristan Luca Saidi\textsuperscript{\(\dagger\)}\\
\href{mailto:tsaidi@andrew.cmu.edu}{\nolinkurl{tsaidi@andrew.cmu.edu}}}
&
\makecell[c]{Arun Kuchibhotla\textsuperscript{\(\dagger\)}\\
\href{mailto:karun3kumar@gmail.com}
{\nolinkurl{arunku@cmu.edu}}}
\end{tabular}\\[0.75em]
\textsuperscript{\(\dagger\)}Department of Statistics and Data Science, Carnegie Mellon University
}

\begin{document}
\pagenumbering{arabic}

\maketitle

\begin{abstract}
    In this work we propose and analyze a procedure for estimating a generalized single-index regression model. This generalized single-index regression framework models the conditional mean of the responses $Y$ as a function of the covariates $X$ where the mapping is assumed to have the form $f \circ g$, where $g$ is $\beta$-H\"older with $\beta \geq 2$ and $f$ can be a discontinuous function of bounded variation. We propose a general procedure that operates under minimal assumptions about the distribution of covariates $X$, requiring only boundedness, sufficient smoothness and absolute continuity with respect to Lebesgue. Our procedure jointly estimates the gradient vector $\hat v(x) \approx \nabla g(x)$ and the conditional mean $\mu(x) = f(g(x))$ by solving a localized weighted linear least squares problem. We derive central limit theorems for gradient vector estimates $\hat{v}_x$ and regression estimates $\hat{\mu}(x)$, and we derive rates of convergence for our approach, illustrating that it indeed achieves a stochastic convergence rate matching the minimax mean squared error rate when $\beta = 2$ and $f$ has a locally Lipschitz derivative. Unlike prior approaches our velocity estimation results accommodate link functions in the local bounded-variation class, allowing for nonsmooth and even discontinuous links while retaining consistency and asymptotic normality at almost every inference point. Finally, we instantiate this model in the situation where $g = \Pi$ is a closest point projection onto a compact and regular embedded $C^3$ curve $\gamma$, a setting recently studied in the literature. We also supplement our theoretical results with experiments on real and synthetic data to demonstrate the utility of our procedure.
\end{abstract}

% \begin{keywords}
%   single-index model, nonparametric regression, local linear estimation, density weighting, sufficient dimension reduction
% \end{keywords}

% \tableofcontents

\section{Introduction}
Single-index regression is a classical tool in semi-parametric statistics for regression analysis. In a regression setting, we assume access to data $(X, Y)$ consisting of a covariate $X$ and a  (univariate) response variable $Y$ where the goal is to estimate $\mu(x) = \mathbb{E}[Y|X = x]$. The classical single-index model stipulates that $\mu(x) = f(x^{\top}\theta)$ for some unknown $f:\mathbb{R}\to\mathbb{R}$ and $\theta \in\mathbb{S}^{d-1}$. As a generalization, we consider the assumption
\begin{equation}\label{eq:non-linear-index-model}
    \mu(x) = f(g(x))\quad P_X-\text{a.s.}
\end{equation}
where $f:\mathbb{R}\to\mathbb{R}$ and $g:\mathbb{R}^d\to\mathbb{R}$ are some unknown functions. Alternatively, this can be expressed as 
\[
Y = f(g(X)) + \varepsilon\quad\mbox{where $\mathbb{E}[\varepsilon|X] = 0$ almost surely.}
\]
No substantive additional assumptions on $\varepsilon$ are needed in this formulation. In particular, we do not need independence between $\varepsilon$ and $X$, nor do we need conditional Gaussianity of $\varepsilon$ given $X$.
Many techniques have been developed for estimation and inference in this model under a variety of assumptions \citep{li1991sliced,dennis2000save,li2005contour,li2007directional, li1989regression}, as the model simultaneously offers flexibility and interpretability. Most of these methods rely heavily on the linear structure of $g$ for computational tractability, thereby avoiding the need to jointly optimize over $f$ and $g$. 

In this work we consider a modified version of the classical single-index regression model which ablates the linear structure of $g$, that we call the \emph{generalized single-index regression model}. In particular, the model instead assumes that $g: \mathbb{R}^d \to \mathbb{R}$ is an arbitrary $\beta$-H\"older function, for some $\beta \geq 2$. Such a model has been studied in the literature on a few occasions and in various forms \citep{juditsky2009nonparametric, horowitz2007rate, wu2026conditional, kereta2021nonlinear, horowitz2001nonparametric}. Na\"ively, estimation in such a model would require a bi-level optimization procedure of the form
\begin{equation}\label{bilevel optimization}
    \hat f_n, \hat g_n \in \arg\min_{ f \in \mathcal{F},  g \in \mathcal{G}}\frac{1}{n}\sum_{i = 1}^n\left|Y_i -  f( g(X_i))\right|^2.
\end{equation}
In this work we instead propose and analyze a conceptually simple and computationally tractable estimation procedure for the generalized single-index model that we call \emph{Locally Linear Single-Index Regression} (LLSIR). A hallmark of this procedure is that it avoids the bi-level optimization problem in \eqref{bilevel optimization} through localization; in a neighborhood of an out-of-sample point $x_0$ the $g$ function is approximately linear, and one can therefore employ classical single-index techniques to estimate $\mu(x_0)$ and $\nabla g(x_0)$ up to scale. 

The motivation behind classical single-index or multi-index models stems from the idea that the conditional expectation $\mu(x) = \mathbb{E}[Y|X = x]$ depends on $x$ only through a low-dimensional linear function. This should allow one to estimate $\mu(\cdot)$ without a curse of dimensionality \citep{gaiffas2007optimal}. Our motivation in considering the non-linear extension is the same. In particular, given the compositional structure of our assumed $\mu(\cdot)$, we study whether it is possible to apply other more general methods to directly estimate $\mu(\cdot)$. For example,~\cite{Schmidt_Hieber_2020} proves the minimax rate in mean squared error (MSE) for estimating $\mu(x) = f(g(x))$ when $g:\mathbb{R}^d\to\mathbb{R}$ is $\beta$-smooth and $f:\mathbb{R}\to\mathbb{R}$ is $\alpha$-smooth is given by \[r_n = \max\left\{n^{-\frac{2\beta}{2\beta + d}},\, n^{-\frac{2\alpha}{2\alpha + 1}}\right\}.\] Hence, if the inner function $g$ is significantly smoother, i.e., $\beta \asymp d$, then the conditional expectation can be estimated without curse of dimensionality. Moreover,~\cite{Schmidt_Hieber_2020} proves that a ``well-designed'' and ``well-trained'' neural network achieves this rate adaptively. However, both these requirements are difficult to achieve in practice given the significant non-convexity of the optimization landscape. For this reason, we directly target the estimation of identifiable aspects of $g(\cdot)$ through which we also obtain an estimate of the conditional mean $\mu(\cdot)$.

While we may not achieve the minimax rate for estimating $\mu(\cdot)$ in full generality, this approach is particularly beneficial in the context of transfer learning. Imagine, instead of one sample, we have $K$ samples $(X_{i}^{(k)}, Y_{i}^{(k)}), 1\le i\le n_k$ for $1\le k\le K$ such that
\[
\mathbb{E}[Y_i^{(k)}|X_{i}^{(k)} = x] = f_k(g(x))\quad\mbox{for}\quad 1\le i\le n_k, 1\le k\le K
\]
where the inner function $g$ stays the same while the outer function $f_k$ can change between samples. In this scenario, an estimation procedure that targets $g$ can pool information from all samples while the estimation of outer function can be performed within each sample. See, for example,~\cite{chen2013estimation} where $g(x) = x^{\top}\theta$. While we do not pursue transfer learning in this paper, it plays a significant motivating role for our method. In particular, the procedure and analysis that follow can be extended to this transfer learing setting with minor modification.

\subsection{Related Work}

Our work relates to several areas of the statistical and computer science literature. We review each area here.
\medskip

\emph{Single-index regression.}
The classical single-index model assumes that the regression function takes the form $m(x)=f(\theta^\top x)$ for an unknown ``index'' direction $\theta$ and an unknown univariate link $f$. A substantial literature studies estimation of $\theta$ under different assumptions on the covariate distribution. Inverse regression approaches, including sliced inverse regression, SAVE, contour regression, and directional regression, recover the index direction from moments of $X \,|\, Y$ \citep{li1991sliced,dennis2000save,li2005contour,li2007directional}. These methods are closely connected to the linear conditional expectation condition exploited by \citet{li1989regression}. Conversely, an abundance of methods consider the forward-regression approach. For example, average-derivative methods recover an index direction from derivatives of the regression function \citep{hardle1989average,powell1989semiparametric}, while profile and semiparametric least-squares methods jointly estimate the index and unknown link \citep{ichimura1993semiparametric,hardle1993optimal}. Iterative refinements of average-derivative estimation can also recover single and multi-index structure at parametric rates under suitable conditions \citep{hristache2001direct,hristache2001structure}. Our construction, however, is most directly related to the density-weighted least-squares estimator of \citet{Newey_Ruud_2005}, which reweights observations by the ratio of an elliptically symmetric target density to the covariate density so that the limiting regression problem behaves as though the design satisfied the LCE property. In contrast to these aforementioned classical methods, however, we do not assume that our target direction $\nabla g$ is a fixed global index vector. Instead, the possibly nonlinear map $g$ yields a location-dependent direction $\nabla g(x_0)$ that generally varies throughout the covariate space.
\medskip 

\emph{Compositional function estimation.}
The literature regarding general models of the compositional form $m=f\circ g$ is quite rich. Most directly, \citet{juditsky2009nonparametric} study the estimation of a multivariate function represented as the composition of a univariate outer function and a multivariate inner function, and they characterize minimax estimation rates over smooth composition classes. Unfortunately, their estimator requires discretization of a $d-1$ dimensional sphere and therefore may not be suitable in practice. More recently, \citet{Schmidt_Hieber_2020} establishes minimax rates for substantially more general hierarchical composition classes using sparse deep ReLU networks. 
Finally, our closest-point projection example in \Cref{sec: single variable model} is directly related to the nonlinear single-variable model of \citet{wu2026conditional}, who consider $m(x)=f(\Pi (x))$ with $\Pi$ the closest-point coordinate associated with an unknown regular curve $\gamma$. Their estimator combines ideas from inverse regression and manifold learning, though their analysis restricts the covariate distribution to a neighborhood of the curve $\gamma$; under additional conditions on the link and variation normal to the curve, they obtain a one-dimensional minimax rate up to logarithmic and geometry-dependent effects. We remark that this setting is related to, but distinct from, regression under a manifold-support assumption. In this setting, local polynomial regression has been shown to adapt to intrinsic dimension when the covariates lie on or sufficiently near a lower-dimensional manifold \citep{bickel2007local}.  Furthermore, locally-linear methods have also been developed for simultaneous regression and gradient estimation on an unknown manifold \citep{cheng2013local}. 
\medskip

\emph{Sufficient dimension reduction.}
Our work is also intimately related to the vast literature on sufficient dimension reduction for regression. \citet{cook2002dimension} formalize sufficient dimension reduction for the conditional mean through the study of the so-called ``central mean subspace", while \citet{xia2002adaptive} propose the minimum average variance estimation method which combines local linear approximations of the regression function with estimation of an informative linear subspace, placing substantially weaker restrictions on the covariate distribution than classical inverse-regression methods. Closely related procedures from  \citet{xia2007constructive} likewise recover directions from local derivatives of the regression function. Kernel methods subsequently relaxed the linear-subspace formulation \citep{fukumizu2009kernel}, while \citet{lee2013nonlinear} developed a general theory of nonlinear sufficient dimension reduction in terms of sufficient and central classes of functions. More recently, \citet{chen2024deep} combine deep neural networks with generalized martingale difference divergence for nonlinear sufficient dimension reduction, while \citet{tang2026belted} introduce belted and ensembled neural networks that accommodate linear and nonlinear sufficient dimension reduction for both the conditional mean and the full conditional distribution. In contrast to this literature, which generally permits a reduction of dimension $k \geq 1$, we restrict attention throughout to the case $k = 1$ so that the latent index $g$ is scalar-valued. The multi-index analogue, in which $\mu(x) = f(g_1(x), \dots, g_k(x))$ and the object of interest at $x_0$ is the span of $\{\nabla g_j(x_0)\}_{j \le k}$, is a natural extension of our localization scheme that we leave to future work. 

\medskip

\emph{Local regression and gradient estimation.}
% The first stage of our procedure is also closely connected to the classical literature on local polynomial regression and nonparametric derivative estimation. Local polynomial estimators and their bias and variance expansions were developed by \citet{ruppert1994multivariate}, while \citet{lu1996partial} develops the asymptotic theory for multivariate partial derivative estimation. This literature is especially pertinent because $\nabla(f\circ g)(x_0)=f'(g(x_0))\nabla g(x_0)$, so local derivative estimation provides one route to recovering the index direction. However, conventional approaches such as local polynomial regression can become computationally demanding in high dimensions, as fitting a polynomial of degree $p$ in $d$ variables requires constructing and solving a weighted least-squares problem involving $\binom{d+p}{p}$ polynomial coefficients \citep{ruppert1994multivariate,lu1996multivariate,fan2018local}. This complexity grows as $O(d^p)$ for fixed $p$, motivating methods that estimate the local index direction without explicitly fitting a multivariate polynomial. Our approach instead extends the density-weighted linear least-squares (DWLLS) methodology of \citet{Newey_Ruud_2005} to nonlinear $g$, thereby retaining the computational tractability and single-index interpretation of DWLLS while allowing the index to vary spatially.
A natural forward-regression alternative is to estimate the gradient of the regression function directly by local polynomial regression \citep{ruppert1994multivariate,lu1996multivariate,fan2018local}, since under the model $\mu = f\circ g$ one has $\nabla \mu(x_0) = f'(g(x_0))\,\nabla g(x_0)$. This route, however, recovers the index direction only by differentiating through the link. The bias expansions and rates for local polynomial derivative estimation require $\mu$ (and hence $f$) to be at least $C^2$ in a neighborhood of $x_0$. Our procedure recovers $\nabla g(x_0)$ by the density reweighting of \citet{Newey_Ruud_2005}, which ensures that the population least-squares slope is proportional to the local index direction regardless of the form of the link, in the spirit of \citet{li1989regression}. As a consequence, our velocity results (\Cref{thm: population prop} \Cref{thm: decomposition for theta} and \Cref{corr: central limit for v}) require only $f \in BV_{\mathrm{loc}}(\mathbb{R})$ -- permitting kinks and even jump discontinuities in the link -- and still yield an $O(\epsilon^2)$ localization bias, asymptotic normality, and the rate of \Cref{thm: velocity risk} at $P_X$-almost every inference point. We also find that the asymptotic variance of our estimator is smaller in the L\"owner ordering than that of local polynomial regression when the density is locally constant.

\subsection{Overview}
In \Cref{notation} we review some important notation, in \Cref{sec: estimation procedure} we describe our Locally Linear Single Index Regression (LLSIR) procedure, and in \Cref{sec: theoretical results} we provide theoretical analyses of the LLSIR estimator. We first derive influence function expansions for the estimation errors $\widehat{\nabla g} - \nabla g$ and $\hat \mu(x_0) - \mu(x_0)$ (\Cref{thm: decomposition for theta}). These expansions readily yield central limit theorems for the local index direction estimate and the conditional mean estimate (\cref{corr: central limit for v}), as well as rates of convergence (\cref{thm: velocity risk}). Finally, we pull from the literature on estimation of composition of H\"older-smooth functions, where we find that our rates indeed match the mean squared error minimax rates when $g$ is $2$-H\"older.

In \Cref{sec: single variable model}, we instantiate the LLSIR method in the context of a related, recently studied model from \citet{wu2026conditional} called the \emph{single-variable model}. The single-variable model assumes the same compositional structure as the generalized single index model, but it instead stipulates that $g$ is the nearest-point projection onto a time-parameterized curve $(\gamma(t))_{t} \subset \mathbb R^d$. The estimator proposed by \citet{wu2026conditional} requires that the covariates be concentrated near the curve $\gamma$; this therefore requires that the marginal law of the covariates depends on the marginal law of the responses, which is an unnatural assumption for a regression model. In \Cref{sec: single variable model}  we instead show that the LLSIR procedure can be applied directly to the single-variable model, and it requires no such assumption on the covariates. Finally, in \Cref{sec: simulations} we provide simulation experiments to demonstrate the utility of the LLSIR method.

\medskip

\subsection{Notation}\label{notation}
For $x\in\mathbb{R}^d$ and $r>0$, $\mathcal{B}(x,r)$ denotes the closed Euclidean ball of radius $r$ centered at $x$, and $\mathbbm{1}_{\{\cdot\}}$ denotes an indicator. Throughout, $\epsilon$ denotes the localization radius of our procedure, whereas $\varepsilon$ denotes the regression noise. We write $P_X$ for the marginal law of $X$, with Lebesgue density $p$, and we write $\mathcal{L}^d$ for Lebesgue measure on $\mathbb{R}^d$ (and simply $\mathcal{L}$ when the dimension is clear from context). We denote  $\nu\ll\mu$ for absolute continuity of measures, $\operatorname{supp}(\cdot)$ and $\operatorname{int}(\cdot)$ for the support of a measure and the interior of a set, and $[d]=\{1,\dots,d\}$. The symbols $\|\cdot\|_2$, $\|\cdot\|_{\operatorname{op}}$ and $\|\cdot\|_\infty$ denote the Euclidean norm, the operator norm of a matrix and the supremum norm of a function, and $\langle\cdot,\cdot\rangle$ denotes the Euclidean inner product. For symmetric matrices, $A\succ0$ (resp.\ $A\succeq0$) means that $A$ is positive definite (resp.\ positive semidefinite), and $A\succeq B$ refers to the L\"owner order. We denote $I_d$ as the $d\times d$ identity matrix, $\lambda_{\min}(A)$ the smallest eigenvalue of $A$, $\operatorname{tr}(A)$ its trace and $\operatorname{diag}(\cdot)$ the (block-)diagonal matrix with the indicated blocks. We write $\operatorname{D}^k h$ for the $k$-th derivative of a function $h$, and $BV_{\mathrm{loc}}(\mathbb{R})$ for the class of functions of locally bounded variation, with $Df$ and $|Df|$ denoting the distributional derivative of $f$ and its total variation measure (see \Cref{sec: BV}). For positive quantities, $a\lesssim b$ means $a\le Cb$ for a constant $C$ that does not depend on $n$ or $\epsilon$, $a\gtrsim b$ means $b\lesssim a$, and $a\asymp b$ means that both hold. A subscript $x_0$, as in $\lesssim_{x_0}$, $O_{x_0}(\cdot)$ and $o_{x_0}(\cdot)$, indicates that the constant or the rate implicit in an $o(\cdot)$ statement may depend on the inference point $x_0$. When comparing bandwidth sequences we also write $a_n\ll b_n$ for $a_n/b_n\rightarrow0$. The symbols $O_P(\cdot)$ and $o_P(\cdot)$ are the usual stochastic order symbols, $\overset{d}{\rightarrow}$ denotes convergence in distribution and $\overset{d}{=}$ equality in distribution.

\subsection{Setting and assumptions}

In this section we reiterate and elaborate on our modeling assumptions throughout this work. In particular, we consider the problem where we are given i.i.d. samples $\{(X_1, Y_1), \dots, (X_n, Y_n)\} \sim P_{X,Y}$ from some probability measure $P_{X,Y} \in \mathcal{P}(\mathbb{R}^d\times \mathbb{R})$ with density $P_{X,Y}$, where our goal is to model the conditional mean of $Y$ given $X$. We assume that the covariate distribution $P_X$ has bounded support. Furthermore, we assume that the conditional mean of the responses $Y$ are related to the covariates $X$ through the following composition, 
\begin{equation}
    Y_i = f(g(X_i)) + \varepsilon_i,  \label{eq: the model}
\end{equation}
where $g$ is a $\beta$-H\"older smooth map from $\mathbb{R}^d \rightarrow \mathbb{R}$ (with $\beta \geq 2$) and $f$ is some function of bounded variation. Recall that a $\beta$-smooth function $h$ is differentiable up to order $k \triangleq\lfloor{\beta}\rfloor$, and all derivatives of order $k$ are $\alpha \triangleq \beta - \lfloor \beta \rfloor$ H\"older, i.e.
\[\sup_{x \neq y}\frac{|\operatorname{D}^{k}h(x) - \operatorname{D}^k h(y)|}{\|x - y\|^\alpha} < \infty.\]
Moreover, we assume $\varepsilon_i$ satisfies $\mathbb{E}[\varepsilon_i \, | \, X_i] = 0$ and some conditional moments assumptions that we consider to be quite standard for the regression setting.
\medskip 
% In the single-index setting, we know that the conditional expectation of $y$ given $x$ has the  compositional structure defined in \cref{eq: the model}, where $g:\mathbb{R}^d \rightarrow \mathbb{R}$ and $f: \mathbb{R}\rightarrow\mathbb{R}$. The classical single-index model takes $g(x) = b^{\top}x$ for some $b \in \mathbb{R}^d$, but in this work we take $g$ to be an arbitrary $\beta$-H\"older function from $\mathbb{R}^d$ to $\mathbb{R}$ in a neighborhood of the desired inference location $x_0$. 

\emph{The Linear Conditional Expectation (LCE) property.}
The Linear Conditional Expectation (LCE) property of the density of the covariates has played a prominent role in the study of single index models \citep{li1989regression, li1991sliced}, and it will be a key element of our approach as well. We reiterate that we do not assume the LCE property on the covariates. Similar to \citet{Newey_Ruud_2005}, we will merely leverage its properties via density reweighting. To recall its definition, the LCE assumption on a random variable $X \sim P$ guarantees that the conditional expectation of any linear transformation of a random variable with an LCE density conditioned on a \textit{secondary} linear transformation is linear in the second transformation.

\begin{restatable}[Linear Conditional Expectation (LCE) property, \cite{Newey_Ruud_2005,ruud1986consistent}]{defn}{} \label{def: lce}
    A random variable $X \sim P$ is said to satisfy the Linear Conditional Expectation property if for any $b \in \mathbb{R}^{d}$ and $B \in \mathbb{R}^{d\times k}$, there exists a $c\in \mathbb{R}^k, a \in \mathbb{R}$ such that $\mathbb{E}[b^{\top}X \,| \,B^{\top}X ] = a +c^{\top}B^{\top}X.$
\end{restatable}

While the LCE assumption may seem stringent, it turns out that many distributions satisfy it. A subset of such distributions include zero-mean \textit{elliptically symmetric distributions}, a family that subsumes the mean-zero Gaussian family. Contained in this subset are \textit{spherically symmetric distributions} -- probability measures that admit a density that is invariant under rotations.

\begin{restatable}[Elliptic and spherical symmetry]{defn}{}\label{def: spherical symmetry}
Let $m \in \mathbb{R}^d$ and let $\Lambda \in \mathbb{R}^{d\times d}$ with $\Lambda \succ 0$. For a Lebesgue density
\begin{equation}
     \rho(x) \propto |\Lambda|^{-1/2}\rho^*\left((x-m)^{\top}\Lambda^{-1}(x - m)\right)
\end{equation}
where $\rho^*:[0, \infty) \rightarrow[0, \infty)$ does not depend on $\mu$, $\Lambda$, then the $\rho$ density is said to be elliptically symmetric. If, furthermore, $\Lambda = \sigma^2I_d$, then the density is said to be spherically symmetric. Equivalently, a random variable $X$ has a spherically symmetric distribution about its mean $m$ if for every orthogonal matrix $Q$, we have \[Q(X - m) \overset{d}{=} X - m\] where $\overset{d}{=}$ denotes equivalence in distribution.
\end{restatable}

% \begin{restatable}[Spherically symmetric $\implies $ elliptically symmetric]{rmk}{}
%     Any spherically symmetric distribution is also zero mean and elliptically symmetric, and therefore satisfies the LCE property. 
% \end{restatable}

In this work we will leverage the properties of LCE densities, but we won't require the density of the covariates to satisfy it -- our only assumption on the covariate density is that it has sufficient smoothness and its support is contained in $\mathcal{B}(0, r)$ for some $r > 0$. We will use density ratios to re-weight observations as done in \citet{Newey_Ruud_2005}, which will allow us to use the following property of LCE densities when applied to the single index model.

\begin{restatable}[LCE Proportionality,
\citet{li1989regression}]{thm}{}
\label{thm: lce single index}
Suppose that $X \sim P$ admits a Lebesgue density, that
$\mathbb E[\|X\|^2_2]<\infty$, and that
$\mathbb E[Y^2]<\infty$. Suppose further that
$Y=f(b^\top X)+\varepsilon$ for some nonzero
$b\in\mathbb R^d$, with
$\mathbb E[\varepsilon\mid X]=0$, and assume that $P$ satisfies the LCE property
in the direction $b$.
Then the population-level optimization problem
\[
(\theta^*,\alpha^*)
=\arg\min_{\theta\in\mathbb R^d,\,
\alpha\in\mathbb R}
\mathbb E[(Y-\theta^\top X-\alpha)^2]
\]
admits a unique optimizer satisfying $\theta^*\in\operatorname{span}(b)$,
where $\theta^*$ is a nonzero scalar multiple of
$b$ whenever
$\operatorname{Cov}(Y,b^\top X)\neq 0$.
\end{restatable}

\begin{proof} We have
\begin{align*}
    \mathbb{E}\big(Y - \theta^{\top}X-\alpha\big)^2 &= \mathbb{E}\big(f(b^{\top}X) - \theta^{\top}X-\alpha  + \varepsilon\big)^2\\
    &= \mathbb{E}\Big[\big(f(b^{\top}X) - \theta^{\top}X-\alpha\big)^2\Big] + \mathbb{E}[\varepsilon^2]. \tag{$\mathbb{E}[\varepsilon \,|\,X]$ = 0}
\end{align*}
Analyzing the first term, 
\begin{align*}
    \mathbb{E}\Big[\big(f(b^{\top}X) - \theta^{\top}X-\alpha\big)^2\Big]&= \mathbb{E}\Big[ \mathbb{E}\Big[\big(f(b^{\top}X) - \theta^{\top}X-\alpha\big)^2 \, \Big | \, b^{\top}X\Big] \Big] \\
    &\geq \mathbb{E}\Big[ \Big(f(b^{\top}X) - \mathbb{E}\big[\theta^{\top}X |b^{\top}X\big]-\alpha\Big)^2 \Big] \\
    &\geq \mathbb{E}\Big[ \Big(f(b^{\top}X) - c_*b^{\top}X-(\alpha+a_\theta)\Big)^2 \Big]
\end{align*}
where the penultimate line uses Jensen's inequality and the last line uses the LCE property. Thus, every candidate $(\theta,\alpha)$ can be replaced
by one with slope proportional to $b$ without
increasing the population risk. Moreover, since $P$
admits a Lebesgue density and has finite second
moments, $\operatorname{Cov}(X)\succ0$, so the
least-squares objective admits a unique minimizer.
It follows that $\theta^*=c_*b$ for some
$c_*\in\mathbb R$.
\end{proof}

\Cref{thm: lce single index} tells us that the ordinary least squares estimator recovers the index vector $b$ of a single index model when the covariate density satisfies the LCE property. \citet{Newey_Ruud_2005} leverage this fact to construct an estimator of the index vector they call \emph{Density Weighted Linear Least Squares} (DWLLS) by using density reweighting to weight observations so they are asymptotically distributed according to a LCE density. In this work we combine this principle with localization to estimate the nonlinear single index model.

\subsection{Identifiability}

Before describing our procedure, we will briefly remark on the nuances of identifiability. In particular, we cannot expect that $f$ and $g$ can be uniquely identified in general. Indeed, taking $\tilde f(z) = f(z + c)$ and $\tilde g(x) = g(x) - c$ gives $(\tilde f \circ \tilde g)(x) = (f \circ g)(x) = \mu(x)$ yielding an immediate counterexample. In fact, this can be generalized to any invertible $h: \mathbb{R}\to \mathbb{R}$, taking $\tilde f = f \circ h^{-1},$ and $\tilde g = h \circ g$ gives $\mu(x) = (f\circ g)(x) = (\tilde f \circ \tilde g)(x)$. Despite this, one can show that $\nabla g$ can be identified up to scale pointwise almost surely under reasonable assumptions about $f, g$. Intuitively, the assumptions guarantee that the set of covariate values where either $\nabla g(x)=0$ or $f'(g(x))=0$ has Lebesgue measure zero. Since $P_X$ is absolutely continuous, these exceptional sets also have $P_X$-measure zero.

\begin{prop}[Identification of $\nabla g(x_0)$ up to scale]\label{identification}
    Suppose $X \sim P_X$  and define the critical sets of $f, g$ as
    \[\mathfrak{C}_f = \{t : f \text{ is not differentiable at } t, \text{ or } f'(t) = 0\}, \quad \mathfrak{C}_g = \{x : \nabla g(x) =0\}.\]
    Then if $f$ and $g$ satisfy $\mathcal{L}^1(\mathfrak{C}_f) = \mathcal{L}^d(\mathfrak{C}_g) = 0$ where $\mathcal{L}^d$ is the $d$-dimensional Lebesgue measure on $\mathbb{R}^d$, then $\operatorname{span}\{\nabla g(X)\} \neq \{0\}$ is identifiable $P_X$-almost-surely.
\end{prop}
\begin{proof}
    We seek to show that if $(\tilde f, \tilde g)$ is another pair satisfying the stated assumptions with $f \circ g \equiv \tilde f \circ \tilde g$, then $\operatorname{span}\{\nabla g(X)\} = \operatorname{span}\{\nabla \tilde g(X)\}$ $P_X$-a.s. To see this, note that since $P_X \ll \mathcal{L}^d$, we have $P_X(\mathfrak{C}_g) = 0.$ Moreover, since $g$ is continuously differentiable,
    the coarea formula
    \citep[Theorem 3.2.22]{federer1969geometric}
    gives
    \begin{align*}
    \int_{\{x:g(x)\in\mathfrak{C}_f,\,
    \|\nabla g(x)\|_2>0\}}p(x)\,dx=
    \int_{\mathfrak{C}_f}
    \int_{g^{-1}(t)}
    \frac{p(x)}{\|\nabla g(x)\|_2}
    \,d\mathcal H^{d-1}(x)\,dt=0,
    \end{align*}
    where the integrand is interpreted as zero wherever
    $\nabla g(x)=0$. Together with
    $P_X(\mathfrak{C}_g)=0$, this implies
    $P_X(g^{-1}(\mathfrak{C}_f))=0$. The same argument applied to $(\tilde f, \tilde g)$ gives $P_X(\mathfrak{C}_{\tilde g}) = P_X(\tilde g^{-1}(\mathfrak{C}_{\tilde f})) = 0$. Now fix an $x \in \mathbb{R}^d$ outside these four null sets. Then $f$ is differentiable at $g(x)$ and $\tilde f$ is differentiable at $\tilde g(x)$, so by the chain rule both sides of $\tilde f \circ \tilde g = f \circ g$ are differentiable at $x$, which gives
    \begin{align*}
        \tilde f'(\tilde g(x)) \cdot \nabla \tilde g(x) = f'(g(x)) \cdot \nabla g(x).
    \end{align*}
    The right side is nonzero since $f'(g(x)) \neq 0$ and $\nabla g(x) \neq 0$. As this holds for $P_X$-almost every $x$, we have $\operatorname{span}\{\nabla \tilde g(X)\} = \operatorname{span}\{\nabla g(X)\} \neq \{0\}$ almost surely, and the conclusion follows.
\end{proof}

We remark that the conditions on the critical sets of $f$ and $g$ in
\Cref{identification} are fairly mild. For instance, it suffices that each
of $f$ and $g$ have at most countably many stationary points. These
conditions are also weaker than those typically imposed in order to identify
a nonlinear inner function $g$. In the context of survival analysis, for example,
\citet[Section 4.2.1]{zeng2026rank} considers the case of a known link, for
which $g$ is identified outright, and otherwise requires the link to be
strictly monotone \citep[Assumption A2]{zeng2026rank} in order to guarantee
identifiability. \Cref{identification} shows that neither monotonicity nor
knowledge of $f$ is required, provided one targets the weaker object $\operatorname{span}\{\nabla
g(x_0)\}$ rather than $g$ itself. This is precisely the object our procedure
estimates, and it guided the theoretical results to follow. As we cannot
hope to recover $g$ directly, we instead prove convergence results of the
form $\hat v_{x_0, \epsilon} \to c_* \nabla g(x_0)$ in probability, where
$c_* \in \mathbb{R}$ is an unknown scalar.

\section{Estimation procedure}\label{sec: estimation procedure}

In this section we will describe our estimation procedure, which we call \emph{Locally Linear Single Index Regression} (LLSIR), while in the subsequent section we will analyze LLSIR by deriving limit theorems for $\hat{\mu}$ and $\nabla g$ as well as rates of convergence. Algorithm \ref{alg: LLSIR} describes the procedure in detail for a general query point $x_0$.  Without loss of generality, suppose $x_0 = 0$, and fix an $\epsilon > 0$ and consider a first-order Taylor expansion of $g$ on the $\epsilon$-ball $\mathcal{B}(0,\epsilon)$. One can informally write
\begin{equation} \label{eq: surrogate decomposition curve}
    f(g(x)) \approx f(g(0) + \nabla g(0)^\top x) \qquad \text{for all }x \in \mathcal{B}(0, \epsilon).
\end{equation}
This creates a local and, critically, \textit{linear} \say{surrogate} single-index problem with an index vector proportional to $\nabla g(0)$. This, in turn, allows us to apply an estimator similar in flavor to that of \citet{Newey_Ruud_2005},
\begin{align} \label{empirical estimator}
\widehat{\theta}_{x_0,\epsilon} &= \arg\min_{\theta \in \mathbb{R}^{d+1}} \sum_{i = 1}^{n}\hat{w}_i\left(Y_i - \boldsymbol{X}_i^{\top}\theta\right)^2 \qquad \text{with }\qquad \hat{w}_i = \mathbbm{1}_{\{X_i \in \mathcal{B}(0, \epsilon)\}}\frac{p^*(X_i)}{\hat{p}(X_i)}
\end{align} 
where $\boldsymbol{X}_i = \begin{bmatrix}
    1 & X_i/\epsilon
\end{bmatrix}^{\top}$, $\hat{p}$ is a consistent estimate of the density of $X$, and $p^*$ is some user-chosen spherically symmetric (thus LCE) density. We note that numerical instability arising from separately estimating and dividing density estimates may be mitigated by directly estimating the density ratio, for instance through $k$-nearest-neighbor methods \citep{kremer2015nearest} or probabilistic classification techniques \citep{menon2016linking}. We nevertheless employ kernel density estimation (KDE), as the resulting weighting scheme exhibits desirable bias-reducing properties. The population optimization problem corresponding to \eqref{empirical estimator} looks like
\begin{align}
\theta^*_{x_0, \epsilon} &= \arg\min_{\theta \in \mathbb{R}^{d+1}} \mathbb{E}_{p} \Big[w\Big(f(g(X)) - \boldsymbol{X}^{\top}\theta \Big)^2\Big] \qquad \text{with }\qquad w = \mathbbm{1}_{\{X \in \mathcal{B}(0, \epsilon)\}}\frac{p^*(X)}{p(X)}\label{eq: population estimator}
\end{align}
where $p$ is the Lebesgue density of the random variable $X$. We will show that the last $d$ entries of $\widehat{\theta}_{x_0,\epsilon}$ are a consistent estimate of $\nabla g(0)$ up to scale, and that the first entry is a consistent estimate of the conditional mean $\mu(0) = f(g(0))$. To formalize this, let $e_1 = \begin{bmatrix} 1 & 0 & \dots & 0 \end{bmatrix}^{\top} \in \mathbb{R}^{d+1}$ and $V = \begin{bmatrix} 0 & I_d\end{bmatrix}$, and define the population and sample approximations of $\mu(0)$ and $\nabla g(0)$ as
\begin{equation}
    \mu^*_{x_0, \epsilon} = e_1^{\top}\theta^*_{x_0, \epsilon}, \qquad \hat{\mu}_{x_0, \epsilon} = e_1^{\top}\hat{\theta}_{x_0, \epsilon}, \qquad v^*_{x_0, \epsilon} = V\theta^*_{x_0, \epsilon} \qquad \text{and }\qquad \hat{v}_{x_0, \epsilon} = V\hat{\theta}_{x_0, \epsilon}. \label{eq: velocity estimator}
\end{equation}
Since $\boldsymbol{X}_i$ carries a leading entry of one, the intercept of the local least-squares fit plays the role of a locally linear estimate of the regression function at $x_0$, so no separate regression stage is required once $\widehat{\theta}_{x_0,\epsilon}$ has been computed. It is worth noting that the optimization problems in \eqref{empirical estimator} and \eqref{eq: population estimator} admit closed form solutions
\begin{align}
    \widehat{\theta}_{x_0,\epsilon} &= \left(\sum_{i = 1}^n\hat{w}_i\boldsymbol{X}_i\boldsymbol{X}_i^{\top}\right)^{-1}\left(\sum_{i = 1}^n\hat{w}_i\boldsymbol{X}_iY_i\right) \label{eq: sample est closed form}
\end{align}
and
\begin{align}
    \theta^*_{x_0, \epsilon} = \mathbb{E}_{p^*}\left[\mathbbm{1}_{\{X \in \mathcal{B}(0,\epsilon)\}}\boldsymbol{X}\boldsymbol{X}^{\top}\right]^{-1}\mathbb{E}_{p^*}\left[\mathbbm{1}_{\{X \in \mathcal{B}(0,\epsilon)\}}\boldsymbol{X}f(g(X))\right]. \label{eq: population estimator closed form}
\end{align}
Note that the density ratios in \eqref{empirical estimator} and \eqref{eq: population estimator} re-weight each observation's contribution so that $X$ has a limiting spherically symmetric (and thus LCE) density. This means the solution to the population ordinary least squares problem $v^*_{x_0}$ is $O(\epsilon^2)$ away from the index vector of the linear surrogate problem with $\nabla g(0)$ up to scale. We prove this formally in \Cref{lemma: surrogate problem} and \Cref{thm: population prop}.

\begin{algorithm}[h]
    \begin{algorithmic}[1]
        \caption{Locally Linear Single Index Regression (LLSIR)}
        \label{alg: LLSIR}
        \Require New sample $x_0$, samples $\{(X_i, Y_i)\}_{i=1}^n \subseteq \mathbb{R}^d \times \mathbb{R}$, a density estimation kernel $K$ and bandwidth $h$, localization parameter $\epsilon$, LCE density $p^*$.\\
        Compute localized density ratios \[\hat{w}_i = \mathbbm{1}_{\{X_i - x_0 \in \mathcal{B}(0, \epsilon)\}}\frac{p^*(X_i - x_0)}{\hat p(X_i) } \qquad \text{where} \qquad \hat p(x) = \frac{1}{nh^d}\sum_{i = 1}^nK\left(\frac{x -X_i}{h}\right)\]for all $i \in \{1, \dots, n\}$. \\
        Center and augment data, $\boldsymbol{X}_i = \begin{bmatrix}
            1 & \frac{X_i - x_0}{\epsilon}
        \end{bmatrix}^{\top}$ for all $i \in \{1, \dots , n\}.$ \\
        Compute velocity estimate as 
        \[ \hat{v}_{x_0, \epsilon} = V\left(\sum_{i = 1}^n\hat{w}_i\boldsymbol{X}_i\boldsymbol{X}_i^{\top}\right)^{-1}\left(\sum_{i = 1}^n\hat{w}_i\boldsymbol{X}_iY_i\right) \qquad \text{with} \qquad V = \begin{bmatrix}
            0 & I_d
        \end{bmatrix}.\] \\
        Compute the conditional mean estimate as
        \begin{equation} \label{eq: kernel regressor}
            \hat{\mu}(x_0) =e_1^\top\left(\sum_{i = 1}^n\hat{w}_i\boldsymbol{X}_i\boldsymbol{X}_i^{\top}\right)^{-1}\left(\sum_{i = 1}^n\hat{w}_i\boldsymbol{X}_iY_i\right) \qquad \text{with} \qquad e_1 = \begin{bmatrix}
            1 & 0 & \dots & 0
        \end{bmatrix}^\top \in \mathbb R^{d+1}.
        \end{equation}
        \Return $\hat{\mu}(x_0), \hat v_{x_0, \epsilon}$
    \end{algorithmic}
\end{algorithm}

Our entire procedure for estimating the generalized single index model, which we call Locally Linear Single Index Regression (LLSIR), is described in \Cref{alg: LLSIR}. It is quire clear that the method is algorithmically simple. The estimation procedure produces $\hat v_{x_0, \epsilon}$ and $\hat \mu(x_0)$ jointly, and has a simple matrix-algebraic closed form. This is in stark contrast to the na\"ive bilevel optimization problem in \eqref{bilevel optimization} for estimating the generalized single index model.

\section{Asymptotics of the Estimator} \label{sec: theoretical results}

In this section we present theoretical results regarding the LLSIR procedure. To provide an overview, we prove a central limit theorem for both the gradient $\hat{v}_{x_0, \epsilon}$ and the conditional mean estimate $\hat \mu(x_0)$ (\Cref{corr: central limit for v}). The limiting normal for the former is indeed centered at its population counterpart, while the limiting normal for the latter has an asymptotic bias that can be eliminated with an undersmoothing condition. As an intermediate result, we derive influence function expansions for the gradient $\hat{v}_{x_0, \epsilon}$ and the response $\hat{\mu}(x_0)$ in \Cref{thm: decomposition for theta}. Before detailing these results, we will outline our main assumptions on the model, the reference density $p^*$, and the covariate law $p \, dx = P_X$.

\begin{restatable}{asmpt}{} \label{assumption: lce and covariate density}
    We require that $p^*(x)$ be a zero-mean spherically symmetric density with isotropic covariance. Moreover, we require the:
    \begin{enumerate}
        \item (LCE density): the density $p^*(x)$ satisfies $\mathbb{E}_{p^*}[\boldsymbol{X}\boldsymbol{X}^{\top}] \succ 0$, is continuous at $0$, and satisfies $p^*(0)>0$.
        \item (Covariate support): We assume that the support of the covariates is bounded, i.e.  for some $r > 0$ \[\overline{\{x : p(x) \neq 0\}} \subseteq \mathcal{B}(0, r).\]
        \item (Covariate density): there exists a non-negative integer $s > 2$ such that the $s^{\text{th}}$ derivative of the density $p(x)$ is uniformly continuous.
        \item (Noise regularity): at the interior point $x_0 =0$ without loss of generality, $\sigma^2(x)$ is continuous at $0$, $\sigma^2(0)>0$, and there exists $\delta>0$ such that
        \[
            \sup_{x\in\mathcal{B}(0,\delta)}
            \mathbb{E}[|\varepsilon|^3\mid X=x] < \infty.
        \]
        \item (Link and index smoothness):
        at the interior point $x_0 =0$ without loss of generality, we assume that
        $g \in C^2(U)$ for some open neighborhood
        $U \subseteq \mathbb R^d$ of $0$,
        $f \in BV_{\operatorname{loc}}(\mathbb R)$, and $\nabla g(0)\neq 0$.
    \end{enumerate}
\end{restatable}

Importantly, the reference density $p^*$ is freely
chosen by the practitioner, subject only to the
conditions above, and need not reflect the underlying
covariate distribution $p$. Moreover, our velocity
estimation results accommodate the weak assumption that
$f\in BV_{\mathrm{loc}}(\mathbb R)$, meaning that the
distributional derivative $Df$ is a signed Radon
measure taking a finite value on every compact
interval. Throughout the BV arguments, we use the
right-continuous representative of $f$, which satisfies
\[
|f(v)-f(u)|\leq |Df|((u,v]),\qquad u<v,
\]
by \citet[Theorem 3.28(a)]{ambrosio2000functions}. Occasionally, a stronger $C^2$ assumption is used for the sharper expansion of the proportionality coefficient $c_*$. We emphasize however that $C^2$ regularity is not needed for the limit theorems or other key results. For more details on functions of bounded variation, we refer the reader to \Cref{sec: BV}.

We note that for the convergence analysis and its appendix proofs, we work in coordinates centered at the inference point, so that $x_0=0$. We retain $x_0$ in estimator subscripts and in constants to indicate their dependence on the original query point. Moreover, almost-everywhere assertions and statements concerning arbitrary locations, including the geometric and local-regularity results, refer to that point before translation.

\medskip

% \tls{make sure to emphasize that $p^*$ is user-chosen. Also add an assumption about smoothness of $f,g$}

\emph{Density Estimation}. Our estimators $\hat v_{x_0, \epsilon}$ and $\hat \mu$ involve density estimation to re-weight observations so the covariate law asymptotically has an LCE density. In what follows, we estimate the covariate density $p$ appearing in the density ratio of $\hat v_{x_0, \epsilon}$ using Kernel Density Estimation (KDE). We impose sufficient conditions on the kernel to obtain a uniform rate of convergence in probability for $\hat p$ to $p$. We emphasize, however, that KDE is not essential for consistency, and alternative density estimators with suitable convergence properties may also be employed. Our choice of KDE is motivated by its tractability for establishing the subsequent central limit theorems. We begin with some standard definitions.

\begin{restatable}[Kernel order]{defn}{}
    A kernel $K$ is of order $s \in \mathbb N$ if 
    \[\int_{\mathbb{R}^d}\|u\|_2^jK(u)du < \infty, \quad \text{and}\quad \int_{\mathbb{R}^d}u_1^{j_1}\dots u_d^{j_d}K(u) du = 0, \text{where } \sum_kj_k = j\]
    for all $1 \leq j \leq s$ and 
    $\int_{\mathbb{R}^d}K(u)du = 1.$
\end{restatable}
A sufficiently high order kernel ensures that the bias $h^s$ of the density estimate has a dependence on the H\"older smoothness $s$ of the population density instead of the ambient dimension $d$ \citep[Proposition 1.2]{tsybakov2009introduction}. Let $K:\mathbb{R}^d \rightarrow \mathbb{R}$ denote an $s^{\text{th}}$ order kernel and define the kernel density estimate of $p$ as
\begin{equation}
    \hat{p}(x) = \frac{1}{nh^d}\sum_{i = 1}^nK\left(\frac{x -X_i}{h}\right). \label{eq: kde}
\end{equation}
We will impose the following further assumptions on the kernel in order to guarantee \emph{uniform} convergence in probability of $\hat{p}$ to the true density $p$.

\begin{restatable}[Further kernel assumptions, \cite{hansen2008uniform}]{asmpt}{} \label{further kernel assumptions}
    Suppose that the kernel $K$ is symmetric and satisfies $\|K\|_{\infty} < \infty$ and 
    \[\int_{\mathbb{R}^d}|K(u)|\,du_1\dots du_d < \infty.\]
    Moreover, suppose that the kernel $K$ satisfies $\int_{\mathbb{R}^d}\|u\|_2^s|K(u)|du  < \infty$ where $s$ is the order of the kernel, assume that the kernel is differentiable, assume that $\left|\frac{\partial}{\partial u}K(u)\right| \leq \Lambda_1$ for some $\Lambda_1 > 0$ and for some $\nu > 1$, and finally assume that $\left|\frac{\partial}{\partial u}K(u)\right| \leq \Lambda_1\|u\|_2^{-\nu}$ for $\|u\|_2$ sufficiently large.
\end{restatable}
Under these assumptions, we can invoke a uniform convergence in probability result from \citet{hansen2008uniform}. This will be used heavily in our influence function expansion in the proof of \Cref{thm: decomposition for theta}.
\begin{restatable}[Uniform convergence over a bounded set, \citet{hansen2008uniform}]{lemma}{} \label{lemma: uniform convergence of kde}
    Suppose that $\|p\|_{\infty}< \infty$ and the $s^{\text{th}}$ derivative of $p$ is uniformly continuous, where $s$ is the order of the kernel $K$. Moreover, suppose that \Cref{further kernel assumptions} holds. Then for any sequence $r_n  = O(1)$ we have
    \[\sup_{x \in \mathcal{B}(0, r_n)}\left|\hat{p}(x) - p(x)\right| = O_P\left(\sqrt{\frac{\log n}{nh^d}} +h^s \right)\]
    when $h \to 0$ and $nh^d/\log n \to \infty$.
\end{restatable}

Now we will analyze $\hat v_{x_0, \epsilon}$ the estimator of $\nabla g(0)$ (up to scale). To do so, we will decompose the difference as
\[\hat v_{x_0, \epsilon} - c_* \nabla g(0) = \underbrace{\hat v_{x_0, \epsilon} - v^*_{x_0, \epsilon}}_{\text{statistical error}} + \underbrace{v^*_{x_0, \epsilon} - c_* \nabla g(0)}_{\text{localization bias}}. \]
In particular, we will first show that the population quantity $v^*_{x_0, \epsilon}$ converges to the desired quantity $\nabla g(0)$ up to scale at an $\epsilon^2$ rate. This is formalized in \Cref{thm: population prop}.2 below.

\begin{restatable}[Bias shrinkage with localization]{thm}{}
\label{thm: population prop}
    Suppose \Cref{assumption: lce and covariate density}
    holds.
    Then, for $P_X$-almost every
    $x_0\in\operatorname{int}(\operatorname{supp}(P_X))$
    satisfying $p(x_0)>0$ and $\nabla g(x_0)\neq0$,
    and for all sufficiently small $\epsilon>0$, we have:
    \begin{enumerate}
        \item \textup{(Intercept):} The population intercept admits the closed form
        \[
            \mu^*_{x_0,\epsilon}
            =\frac{\mathbb{E}[w\,f(g(X))]}{\mathbb{E}[w]},
        \]
        and satisfies $\mu^*_{x_0,\epsilon}=\mu(0)+o_{x_0}(\epsilon)$.
        \item \textup{(Velocity):} There exists $c_*\in\mathbb R$ such that
        \[
            \|v^*_{x_0,\epsilon}
            -c_*\nabla g(0)\|_2
            \lesssim_{x_0}\epsilon^2,
        \]
        with $v^*_{x_0,\epsilon}$ defined in
        \eqref{eq: velocity estimator}. Moreover, $c_*=\epsilon f'(g(0))+o_{x_0}(\epsilon).$
    \end{enumerate}
    If, in addition, $f\in C^2(I)$, then both
    conclusions hold with the sharper expansions
    \[
        \mu^*_{x_0,\epsilon}
        =\mu(0)+\frac{\epsilon^2}{2(d+2)}
        \left[f''(g(0))\|\nabla g(0)\|_2^2+f'(g(0))\operatorname{tr}(\nabla^2g(0))\right]
        +o(\epsilon^2)
    \]
    and $c_*=\epsilon f'(g(0))+o_{x_0}(\epsilon^2)$.
\end{restatable}

The proof of \Cref{thm: population prop} is provided in \Cref{sec: proof of population vel prop}. This result establishes that the rate with which the bias induced by localization (or rather, lack thereof) shrinks as a function of the localization parameter $\epsilon$. Our next result instead considers the deviation between the population quantity $v^*_{x_0, \epsilon}$ and the sample estimator $\hat v_{x_0, \epsilon}$ -- in particular, we provide an influence function expansion of the difference, which will later allow us to establish a central limit theorem and rates of convergence. This influence function expansion is described in \Cref{thm: decomposition for theta}, the proof of which is similar in flavor to Theorem 2 of \citet{Newey_Ruud_2005}.

\begin{restatable}[Error expansion for $\hat{\theta}_{x_0, \epsilon}$]{lemma}{} \label{thm: decomposition for theta}
    Suppose that \cref{assumption: lce and covariate density} and \cref{further kernel assumptions} are satisfied. Let $\epsilon_0 > 0$ be such that $\mathcal{B}(0, \epsilon_0) \subseteq U$, $p$ is bounded away from zero and $p^*$ is bounded above on $\mathcal{B}(0, \epsilon_0)$, and $\epsilon_0 \leq \delta$. Let $\epsilon = \epsilon_n \in (0, \epsilon_0]$ be a sequence, either fixed or tending to zero, with $n\epsilon^d \rightarrow \infty$, and suppose that the density kernel bandwidth $h = h_n$ satisfies
    \[
        h \rightarrow 0, \qquad \frac{nh^d}{\log n} \rightarrow \infty, \qquad \sqrt{n\epsilon^d}\,h^s \rightarrow 0 \qquad \text{and} \qquad \sqrt{n\epsilon^d}\,\frac{\log n}{nh^d} \rightarrow 0.
    \]
    If $\epsilon_n\rightarrow0$, assume moreover that the right-continuous representative of $f$ is differentiable at $g(0)$. Then it holds that
    \[\hat{\theta}_{x_0, \epsilon} - \theta^*_{x_0, \epsilon} = \frac{1}{n}Q^{-1}\sum_{i = 1}^nw_i\boldsymbol{X}_i(u_i - \mathbb{E}[u_i\,|\,\boldsymbol{X}_i]) +o_P\left(\frac{1}{\sqrt{n\epsilon^{d}}}\right)\]
    where $u = Y - \boldsymbol{X}^{\top}\theta_{x_0, \epsilon}^*$ and $Q = \mathbb{E}[w\boldsymbol{X}\boldsymbol{X}^{\top}]$. In particular, 
    \begin{align*}
        \hat{\mu}_{x_0, \epsilon} - \mu^*_{x_0, \epsilon} &= \frac{1}{n}e_1^{\top}Q^{-1}\sum_{i = 1}^nw_i\boldsymbol{X}_i(u_i - \mathbb{E}[u_i\,|\,\boldsymbol{X}_i]) + o_P\left(\frac{1}{\sqrt{n\epsilon^{d}}}\right), \\
        \hat{v}_{x_0, \epsilon} - v^*_{x_0, \epsilon} &= \frac{1}{n}VQ^{-1}\sum_{i = 1}^nw_i\boldsymbol{X}_i(u_i - \mathbb{E}[u_i\,|\,\boldsymbol{X}_i]) + o_P\left(\frac{1}{\sqrt{n\epsilon^{d}}}\right),
    \end{align*}
    where $\hat{\mu}_{x_0, \epsilon} = e_1^{\top}\hat{\theta}_{x_0, \epsilon}$ and $\mu^*_{x_0, \epsilon} = e_1^{\top}\theta^*_{x_0, \epsilon}$ are the sample and population intercepts.
\end{restatable}

\begin{rmk}
For $f\in BV_{\operatorname{loc}}(\mathbb R)$ the differentiability hypothesis holds at $P_X$-almost every inference point $x_0$ with $\nabla g(x_0)\neq0$. In particular, the set $N_2$ of non-differentiability points of $f$ is Lebesgue-null, and the pullback argument in the proof of \Cref{thm: population prop} shows that $P_X(g^{-1}(N_2))=0$. If $f\in C^2(I)$ it holds at every such point.
\end{rmk}

The proof of \Cref{thm: decomposition for theta} is provided in \Cref{sec: proof of velocity decomp}. Having established this influence function expansion, we will now use it to characterize the limiting distribution of $\hat{v}_{x_0, \epsilon} - v^*_{x_0, \epsilon}.$ To do this, we will appeal to the multivariate Lyapunov central limit theorem, described in \Cref{thm: lyapunov clt}.

% \begin{restatable}[Central limit for $\hat{v}_{x_0, \epsilon}$]{thm}{} \label{corr: central limit for v}
%     Suppose that the hypotheses of
%     \Cref{thm: decomposition for theta} hold and suppose that $\epsilon/h \to0 $. Then we have asymptotic normality of the velocity vector estimate $\hat{v}_{x_0, \epsilon}$,
%     \[\sqrt{n\epsilon^{d}}(\hat{v}_{x_0, \epsilon} - v^*_{x_0, \epsilon}) \overset{d}{\rightarrow}N(0, \Sigma_0) \quad \text{where}\quad  \Sigma_0 = \frac{(d+2)\sigma^2(0)}{p(0)\mathcal{L}^d(\mathcal{B}(0,1))} I_d.\]
%     If, moreover, $n\epsilon^{d+4}\to 0$, then
%     \[
%     \sqrt{n\epsilon^d}
%     \left(\hat v_{x_0,\epsilon}-c_*\nabla g(0)\right)
%     \overset{d}{\to}N(0,\Sigma_0),
%     \]
%     while if we further have $f \in C^2(I)$ then we can deduce that
%     \[
%     \sqrt{n\epsilon^d}
%     \left(\hat v_{x_0,\epsilon}
%     -\epsilon f'(g(0))\nabla g(0)\right)
%     \overset{d}{\to}N(0,\Sigma_0).
%     \]
% \end{restatable}

\begin{restatable}[Central limit for $\hat{\theta}_{x_0, \epsilon}$]{thm}{} \label{corr: central limit for v}
    Suppose that the hypotheses of \Cref{thm: decomposition for theta} hold and let
    \[
        \Sigma_{\epsilon} = Q^{-1}\,\mathbb{E}\left[w^2\sigma^2(X)\boldsymbol{X}\boldsymbol{X}^{\top}\right]Q^{-1}.
    \]
    
    \noindent \textup{(Fixed $\epsilon$.)} If $\epsilon \in (0, \epsilon_0]$ is fixed, then $\Sigma_\epsilon\succ0$ and
        \[\sqrt{n}\left(\hat{\theta}_{x_0, \epsilon} - \theta^*_{x_0, \epsilon}\right) \overset{d}{\rightarrow} N(0, \Sigma_{\epsilon}).\]

    \noindent \textup{(Shrinking $\epsilon$.)} If $\epsilon \rightarrow 0$, then $\epsilon^d\Sigma_{\epsilon} \rightarrow \operatorname{diag}(\sigma_{\mu}^2, \Sigma_0)$ and
        \[\sqrt{n\epsilon^{d}}\left(\hat{\theta}_{x_0, \epsilon} - \theta^*_{x_0, \epsilon}\right) \overset{d}{\rightarrow}N\left(0, \operatorname{diag}(\sigma_{\mu}^2, \Sigma_0)\right)\]
        where
        \[
            \sigma_{\mu}^2 = \frac{\sigma^2(0)}{p(0)\mathcal{L}^d(\mathcal{B}(0,1))} \qquad \text{and} \qquad \Sigma_0 = \frac{(d+2)\sigma^2(0)}{p(0)\mathcal{L}^d(\mathcal{B}(0,1))} I_d.
        \]
        Moreover, for $P_X$-almost every inference point $x_0$ (and for \emph{every} inference point with $p(x_0)>0$ and $\nabla g(x_0)\neq0$ if $f\in C^2(I)$), \Cref{thm: population prop} yields the following centered statements. If $n\epsilon^{d+2}=O(1)$, then
        \[
        \sqrt{n\epsilon^d}
        \left(\hat{\mu}_{x_0,\epsilon}-\mu(0)\right)
        \overset{d}{\to}N(0,\sigma_{\mu}^2),
        \]
        and if $n\epsilon^{d+4} \rightarrow 0$, then
        \[
        \sqrt{n\epsilon^d}
        \left(\hat v_{x_0,\epsilon}-c_*\nabla g(0)\right)
        \overset{d}{\to}N(0,\Sigma_0).
        \]
        If $f \in C^2(I)$, then $n\epsilon^{d+4}\rightarrow0$ suffices for both statements, and $c_*$ may be replaced by $\epsilon f'(g(0))$.
\end{restatable}

% \begin{restatable}[Weaker regularity for velocity estimation]{rmk}{}
%     Although our standing assumptions require $f\in C^2(I)$
%     for the subsequent conditional mean estimation results,
%     the population localization-bias bound, velocity expansion,
%     and velocity central limit theorems centered at
%     $v^*_{x_0,\epsilon}$ and $c_*\nabla g(0)$  for $P_X$-almost every
%     inference point remain
%     valid under the weaker assumption
%     $f\in BV_{\mathrm{loc}}(I)$.
% \end{restatable}

We provide the proof of \Cref{corr: central limit for v} in \Cref{sec: proof of central limit for v}. See \Cref{rmk: bandwidth regimes} for a brief discussion on the bandwidth regimes required in \Cref{corr: central limit for v}. This result characterizes the limiting distribution of $\hat\theta_{x_0,\epsilon}$, i.e. of the intercept and velocity estimates jointly. In particular, it obtains limiting distributions centered at the biased population counterpart $\theta^*_{x_0,\epsilon}$ and, under the additional bias conditions, at the targets $\mu(0)$ and $c_*\nabla g(0)$. The proof involves verifying that the random variables in the influence function expansion derived in \Cref{thm: decomposition for theta} satisfy the Lyapunov condition of \Cref{thm: lyapunov clt}. 

\begin{restatable}[Comparison with local linear regression]{rmk}{}\label{rem: llr variance}
Fix $\epsilon \in (0,\epsilon_0]$ and take the boxcar kernel
$K(x) = \mathbbm{1}_{\{x \in \mathcal{B}(0,\epsilon)\}}\big/\mathcal{L}_d(\mathcal{B}(0,\epsilon))$,
which corresponds to choosing $p^*$ uniform on $\mathcal{B}(0,\epsilon)$ in \eqref{eq: population estimator}.
The local linear estimator with kernel $K$ and bandwidth $\epsilon$ \citep{ruppert1994multivariate,fan2018local} is
\[
    \hat{\theta}^{\mathrm{LL}}_{x_0,\epsilon} = \arg\min_{\theta \in \mathbb{R}^{d+1}} \sum_{i=1}^n K(X_i)\left(Y_i - \boldsymbol{X}_i^{\top}\theta\right)^2,
\]
that is, \eqref{empirical estimator} with the density ratio $1/\hat{p}$ removed from the weights. Its population target $\theta^{\mathrm{LL}}_{x_0,\epsilon}$ minimizes $\mathbb{E}\big[K(X)\big(\mu(X) - \boldsymbol{X}^{\top}\theta\big)^2\big]$. Since the weights $K(X_i)$ are fixed, the argument of \Cref{thm: decomposition for theta} without the density-estimation term gives an asymptotic variance of
\[
    \Sigma^{\mathrm{LL}}_{\epsilon} = \big(Q^{\mathrm{LL}}\big)^{-1}\,\mathbb{E}\!\left[K^2\big(\sigma^2(X) + r^{\mathrm{LL}}(X)^2\big)\boldsymbol{X}\boldsymbol{X}^{\top}\right]\big(Q^{\mathrm{LL}}\big)^{-1},
    \qquad Q^{\mathrm{LL}} = \mathbb{E}\!\left[K\boldsymbol{X}\boldsymbol{X}^{\top}\right],
\]
where $r^{\mathrm{LL}}(x) = \mu(x) - \boldsymbol{x}^{\top}\theta^{\mathrm{LL}}_{x_0,\epsilon}$ is the local linear approximation error. The residual $Y - \boldsymbol{X}^{\top}\theta^{\mathrm{LL}}_{x_0,\epsilon}$ has conditional second moment $\sigma^2(X) + r^{\mathrm{LL}}(X)^2$, whereas in $\Sigma_{\epsilon}$ of \Cref{corr: central limit for v} the estimated weights have replaced it by $Y - \mu(X)$, which has conditional second moment $\sigma^2(X)$ alone. Suppose that $p \equiv p_0 > 0$ on $\mathcal{B}(0,\epsilon)$. Then
\[
    \Sigma^{\mathrm{LL}}_{\epsilon} - \Sigma_{\epsilon} = Q^{-1}\,\mathbb{E}\!\left[w^2 r(X)^2\,\boldsymbol{X}\boldsymbol{X}^{\top}\right]Q^{-1} \succeq 0,
    \qquad r(x) = \mu(x) - \boldsymbol{x}^{\top}\theta^*_{x_0,\epsilon},
\]
and the difference is positive definite unless $\mu$ is $P_X$-a.e.\ affine on $\mathcal{B}(0,\epsilon)$, in which case it is zero. When $p$ is not constant on $\mathcal{B}(0,\epsilon)$ the two covariances are not ordered in general. The design matrix $Q^{\mathrm{LL}}$ then depends on $p$ while $Q$ does not, and in the normalization of \Cref{corr: central limit for v} the resulting discrepancy is of order $\epsilon^2$, dominating the $O(\epsilon^4)$ term above.
\end{restatable}

We provide the proof of \Cref{rem: llr variance} in \Cref{proof of llr variance remark}. The remark establishes that, in the special case where the density is constant on $\mathcal{B}(0, \epsilon)$, the asymptotic covariance of LLSIR is smaller in the L\"owner ordering than that of local polynomial (with $p = 1$) regression. Unfortunately, this cannot be stated in further generality, as nonconstant density introduces lower order asymptotic variance terms that prevent a general ordering of the two covariances. 

When one combines the influence function expansion of \Cref{thm: decomposition for theta} and the bias rate of \Cref{thm: population prop}, one immediately obtains a stochastic convergence rate for the estimator at the fixed inference point. 

\begin{restatable}[Error of $\hat{v}_{x_0, \epsilon}$]{corollary}{} \label{thm: velocity risk}
    Suppose that the hypotheses of \Cref{thm: decomposition for theta} hold with $\epsilon \asymp n^{-1/(d+4)}$, so that $f \in BV_{\operatorname{loc}}(\mathbb R)$. Then, for $P_X$-almost every inference point $x_0$,
    \[\left\|\hat{v}_{x_0, \epsilon} - c_*\nabla g(0)\right\|_{2}^2  =  O_P\left(n^{-\frac{4}{d+4}}\right). \]
    If, in addition, $f \in C^2(I)$, the bound holds at every inference point with $p(x_0) > 0$ and $\nabla g(x_0) \neq 0$.
\end{restatable}
\begin{proof} By the triangle inequality, the fact that $(a + b)^2 \leq 2(a^2 + b^2)$ and \Cref{thm: population prop} we have
\begin{align*}
    \|\hat{v}_{x_0, \epsilon} - c_*\nabla g(0)\|_{2}^2 &\leq \left(\|\hat{v}_{x_0, \epsilon} - v^*_{x_0, \epsilon}\|_{2} + \|v^*_{x_0, \epsilon}  - c_*\nabla g(0)\|_{2}\right)^2 \\
    &\lesssim_{x_0}  \|\hat{v}_{x_0, \epsilon} - v^*_{x_0, \epsilon} \|_{2}^2+  \epsilon^4.
\end{align*}
By \Cref{corr: central limit for v}, $\sqrt{n\epsilon^d}(\hat{v}_{x_0, \epsilon} - v^*_{x_0, \epsilon})$ converges in distribution and is therefore $O_P(1)$, so $\|\hat{v}_{x_0, \epsilon} - v^*_{x_0, \epsilon}\|_2^2 = O_P(\frac{1}{n\epsilon^d})$. Hence $\|\hat{v}_{x_0, \epsilon} - c_*\nabla g(0)\|_{2}^2 =  O_P(\frac{1}{n\epsilon^d} + \epsilon^4)$, and choosing $\epsilon \asymp n^{-1/(d+4)}$ proves the result.
\end{proof}

\begin{restatable}[Regularity of the link]{rmk}{} \label{rmk: link regularity}
    The rate $n^{-4/(d+4)}$ in \Cref{thm: velocity risk} is the rate attained by the local linear gradient estimator, whose standard analyses \citep{ruppert1994multivariate,lu1996partial} assume twice continuous differentiability of the regression function. \Cref{thm: velocity risk} shows that for the index direction this rate requires only $f \in BV_{\operatorname{loc}}(\mathbb R)$.  The additional regularity of $f$ enters only through the sharpening the expression $c_* - \epsilon f'(g(0)) = o_{x_0}(\epsilon)$ to $o(\epsilon^2)$.
\end{restatable}

\begin{restatable}[Error of $\hat{\mu}_{x_0, \epsilon}$]{corollary}{} \label{thm: regression risk}
    Suppose that the hypotheses of \Cref{thm: decomposition for theta} hold, so that $f \in BV_{\operatorname{loc}}(\mathbb R)$. Then, for $P_X$-almost every inference point $x_0$, choosing $\epsilon \asymp n^{-1/(d+2)}$ gives
    \[\left|\hat{\mu}_{x_0, \epsilon} - \mu(0)\right|^2  =  O_P\left(n^{-\frac{2}{d+2}}\right).\]
    If, in addition, $f$ is differentiable at $g(x_0)$ with a locally Lipschitz derivative, then choosing $\epsilon \asymp n^{-1/(d+4)}$ gives, at every inference point with $p(x_0) > 0$ and $\nabla g(x_0) \neq 0$,
    \[\left|\hat{\mu}_{x_0, \epsilon} - \mu(0)\right|^2  =  O_P\left(n^{-\frac{4}{d+4}}\right).\]
\end{restatable}
\begin{proof} As in the proof of \Cref{thm: velocity risk}, $|\hat{\mu}_{x_0, \epsilon} - \mu(0)|^2 \lesssim |\hat{\mu}_{x_0, \epsilon} - \mu^*_{x_0, \epsilon}|^2 + |\mu^*_{x_0, \epsilon} - \mu(0)|^2$. By \Cref{corr: central limit for v}, $\sqrt{n\epsilon^d}(\hat{\mu}_{x_0, \epsilon} - \mu^*_{x_0, \epsilon})$ converges in distribution and is therefore $O_P(1)$, so $|\hat{\mu}_{x_0, \epsilon} - \mu^*_{x_0, \epsilon}|^2 = O_P(1/n\epsilon^d)$, while \Cref{thm: population prop} gives $|\mu^*_{x_0, \epsilon} - \mu(0)|^2 = o_{x_0}(\epsilon^2)$ under $f \in BV_{\operatorname{loc}}(\mathbb R)$ and $O(\epsilon^4)$ when $f'$ is Lipschitz near $g(0)$. In this case, we would have $f(g(x))-f(g(0))=f'(g(0))(g(x)-g(0))+O(\epsilon^2)$ uniformly on $\mathcal B(0,\epsilon)$, and the argument for the differentiable case in \Cref{thm: population prop} goes through with $o(\epsilon)$ replaced by $O(\epsilon^2)$. The stated choices of $\epsilon$ balance the two terms in each case.
\end{proof}

Before moving on, we would like to remark on the quality of this rate of convergence. \citet{Schmidt_Hieber_2020} studies the minimax rates of convergence of estimating the composition of a sequence of functions in mean squared error. In particular, they derive that for a sequence of functions $\{g_1, \dots, g_m\}$ where each $g_i$ is $\beta_i$-H\"older and maps from $\mathbb{R}^{t_{i-1}}$ to $\mathbb{R}^{t_i}$, the minimax rate for estimating the composition $\mu = g_1 \circ \cdots \circ g_m$ in mean-squared-error is
\[ \inf_{\hat{\mu}_n } \sup_{\mu\in \mathcal{F}}  \mathbb{E}\left[(\hat{\mu}_n(X) - \mu(X))^2\right] \gtrsim \max_{i \in \{1, \dots, m\}} n^{-\frac{2\beta_i^*}{2\beta_i^* + t_i}}, \quad \text{where} \quad  \beta_i^* \triangleq \beta_i \prod_{\ell = i+1}^m(\beta_\ell \wedge 1)\]
where $\mathcal{F}$ is the model class induced by the composition of $\beta_i$-H\"older functions. When one instantiates this minimax bound with our generalized single-index model, we have $t_0 = d$, $t_1 = t_2 = 1$, $\beta_0 = \beta$ and $\beta_1 = 2$. Thus, we have a first stage effective smoothness of $\beta_0^* = \beta$ and $\beta_1^* = 2$, and one would therefore expect the following lower bound on the minimax error
\[\inf_{\hat{\mu}_n } \sup_{\mu\in \mathcal{F}}  \mathbb{E}\left[(\hat{\mu}_n(X) - \mu(X))^2\right] \gtrsim \max\left\{n^{-\frac{2\beta}{2\beta + d}}, n^{-4/5}\right\}. \quad \]
Indeed, when $\beta = 2$ the pointwise stochastic rate of convergence described in  \Cref{thm: regression risk} matches this minimax rate.
\begin{restatable}{rmk}{}
    In \citet{Schmidt_Hieber_2020} their minimax rates are global mean squared error rates, while our rates of convergence are pointwise. We expect that the same mean-squared error minimax rate holds pointwise, though we leave a rigorous proof to future work.
\end{restatable}

Having established an array of theoretical results regarding our generalized procedure, we will now explore a specific instantiation of that recovers the setting of \citet{wu2026conditional}. Thereafter, we will shift our attention to analyzing the performance of these estimators on simulated data. 

\section{Example: single index regression via projection onto a curve} \label{sec: single variable model}

In this section we discuss an instantiation of our framework that aligns with the setting explored by \citet{wu2026conditional}. Classical single-index methods operate in the setting where $g(x) = b^{\top}x$ for some $b \in \mathbb{R}^d$; up to rescaling, this can be interpreted as the orthogonal projection of $x$ onto a line in $\mathbb{R}^d$. \citet{wu2026conditional} propose an estimator for a generalization of this setting, which they call the \textit{nonlinear single variable model} (NSVM), where $g = \Pi$ is a nearest point projection onto a \textit{curve} instead of a line. To avoid projection regularity issues, their procedure and analysis assumes that the covariate law $X$ are concentrated near the curve $\gamma$, which is a highly restrictive assumption. While they prove that their estimator achieves an ambient dimension free rate, this assumption on the covariates turns the problem into one of manifold learning where minimax rates are known to be ambient dimension free \citep{genovese2012minimax}. Constructing and analyzing an estimator of their nonlinear single variable model under milder assumptions on the distribution of the covariates therefore remains an open problem. In this section, we address this open problem by instantiating our proposed procedure to estimate the nonlinear single variable model model under minimal assumptions about the covariate law.

In particular, we consider the setting where $\Pi$ is a closest-point projection onto a compact regular $C^3$ curve $\gamma$. We choose to represent $\gamma$ as a function $\gamma:I \rightarrow \Gamma \subset \mathbb{R}^d$ with $I = [0, 1]$, where $\gamma$ is a constant-speed curve through $\mathbb{R}^d$ with speed $\|\gamma'(t)\|_2 \equiv \mathfrak{s} > 0$.
\begin{restatable}[Further assumptions]{asmpt}{} \label{assumption: the curve gamma}
    For $\gamma: [0, 1]\rightarrow \Gamma \subset \mathbb{R}^d$ we assume that
    \begin{enumerate}
        \item $\Gamma$ is compact and the map $\gamma$ is injective.
        \item for all $x' \in \operatorname{supp}(P_X) \triangleq \Omega$ we have that $\Pi(x') \in \operatorname{Int}(I)$.
    \end{enumerate}
\end{restatable}

\Cref{assumption: the curve gamma} outlines our main assumptions on the curve $\gamma$. In particular, we assume that $\gamma$ is injective, which ensures that $\gamma^{-1}:\Gamma\rightarrow I$ is well defined and rules out any self-intersections. We also add the assumption that for all points in the support of the covariate density, the image under the projection of the curve is not on the boundary of the curve. This mild assumptions avoids some unnecessary proof technicalities later on. Finally, we define the closest-point projection map as
\begin{equation}\label{def: projection}
\Pi(x) = \arg\inf_{t\in I}\|x - \gamma(t)\|_2 = \arg\min_{t\in I}\|x - \gamma(t)\|_2.
\end{equation}
% \begin{restatable}[Infimum is attained]{rmk}{} \label{remark: infimum}
%     The infimum in \cref{def: projection} is attained by at least one point in the set $I$ for every $x \in \mathbb{R}^d$.
% \end{restatable}
% \noindent \textit{Proof.} Since the set $\Gamma$ is compact and the Euclidean distance $\|x-a\|_2$ is continuous, the infimum must be attained on some element in $\Gamma$. The result follows from the injectivity of $\gamma$. 

% \qed{}

Before instantiating our estimator, we will first prove some results regarding the regularity of the projection map $\Pi$. Since $\gamma$ is a regular compact curve, $\Pi$ can be seen as a nearest point projection onto a $1$-dimensional submanifold of $\mathbb{R}^d$. In the manifold learning literature, it is common to analyze algorithms that deal with data concentrated near a submanifold of $\mathbb{R}^d$, where \say{near} is usually taken to mean that the support of the data is contained within the so-called reach of the manifold \citep{genovese2012minimax, balakrishnan2012minimax, kokotnoisy, saidi2025recovering}.

\begin{restatable}[Medial axis and reach, \cite{federer1959curvature}]{defn}{}
    Given a closed subset $A \subset \mathbb{R}^d$, the medial axis $M(A)$ is the subset of $\mathbb{R}^d$ consisting of points with at least two nearest neighbors on $A$. Formally,
    \begin{equation}
    M(A) = \{x \in \mathbb{R}^d : |\{y \in A : \|x-y\|_2 = d(x,A)\}| > 1\}. \label{eq: medial axis definition}
    \end{equation}
    The reach of a closed subset $A \subset \mathbb{R}^d$ is defined as 
    \begin{equation}
        \tau_A = \inf_{p\in A} d(p,M(A)) = \inf_{z\in M(A)}d(z,A). \label{eq: reach definition}
    \end{equation}
\end{restatable}

Assuming the data is concentrated near the submanifold of interest guarantees uniqueness and regularity of the nearest point projection map. \citet{wu2026conditional} employ this assumption in the design and analysis of their estimator for the nonlinear single variable model for this exact reason. In particular, they stipulate that the support of the data is contained in the set $\bigcup_{I}\mathcal{B}^\circ(\gamma(t), d(\gamma(t), M(\Gamma)))$, where $d(\gamma(t), M(\Gamma))$ is a local analog of the global reach and $\mathcal{B}^\circ$ denotes the \emph{open} metric ball. Although this is standard in manifold learning, it is a very restrictive assumption for regression, as it forces the distribution of covariates $X$ to contain information about the responses $Y$. To circumvent this, we will show that our procedure leverages the almost-everywhere regularity of the projection map in a way that sidesteps the assumption of covariate concentration.

\subsection{Projection regularity}
We will now show that under the assumptions on $\gamma$ and $P_X$ described earlier, the projection map $\Pi$ and its first two derivatives have desirable properties on an open set of full $P_X$ measure. We emphasize that our assumptions on $P_X$ only require absolute continuity and and bounded support; no concentration near $\gamma$ is assumed.

\begin{restatable}[Differentiability of $\Pi$]{thm}{} \label{thm: differentiability of pi}
    Suppose that $\gamma$ and $P_X$ satisfy \Cref{assumption: the curve gamma}. Then there exists an \textit{open} set $R$ such that $P_X(R) = 1$, and for all $x \in R$:
    \begin{enumerate}
        \item $s = \Pi(x)$ is unique and well defined
        \item both $\nabla\Pi(x)$, $\nabla^2\Pi(x)$ exist and can be expressed as $\nabla\Pi(x)  = -\gamma'(s)/\ell(x)$ and 
        \begin{equation}
            \nabla^2\Pi(x) = \frac{1}{\ell^2(x)}\left(\gamma''(s)\gamma'(s)^{\top} + \gamma'(s)\gamma''(s)^{\top}\right)- \frac{\langle\gamma^{(3)}(s), x - \gamma(s)\rangle}{\ell^3(x)}\gamma'(s)\gamma'(s)^{\top}
        \end{equation}
        where $\ell(x) = -\mathfrak s^2 + \langle\gamma''(s), x-\gamma(s)\rangle \neq 0$. 
    \end{enumerate}
\end{restatable}
\begin{proof}[Proof of \Cref{thm: differentiability of pi}]
By \Cref{lemma: IFT degen closed}, we know that $R$ \eqref{R} is open with full $P_X$ measure. We also know that for any $x \in R$, $\Pi$ is $C^2$ at $x$ due to \Cref{prop: proj gradient} and \Cref{prop: proj hessian}, where the expressions for the gradient and Hessian of $\Pi$ are contained therein.
\end{proof}
For $P_X$-almost every $x_0$, there exists an $\epsilon >0$ such that $\Pi(x), \nabla\Pi(x)$ and $\nabla^2\Pi(x)$ are unique and well behaved for all $x \in \mathcal{B}(x_0, \epsilon).$ We will use this critical property to instantiate our estimator of $\nabla g$, which we will show is proportional to the tangent vector to the curve $\gamma$, as it holds that $\gamma'(\Pi(x))=\big(\mathfrak s^2-\langle\gamma''(\Pi(x)),x-\gamma(\Pi(x))\rangle\big)\nabla\Pi(x)\propto\nabla g(x)$ on $R$.

\subsection{Estimation}

To estimate the gradient of $\gamma$ we will estimate its tangent vector $\gamma'(t_0)$ evaluated at $t = \Pi(x_0)$ where $x_0\in \mathbb{R}^d$ is the point at which we want to estimate $\mu$. To produce this estimate we will employ the LLSIR procedure (\Cref{alg: LLSIR}). In particular, we will choose an $\epsilon > 0$ such that on $\mathcal{B}(x_0, \epsilon)$ the projection map $\Pi$ is twice differentiable -- note that \Cref{thm: differentiability of pi} indicates that for $P_X$-a.e. $x_0$, there exists an $\epsilon > 0$ satisfying such property. This allows us to directly apply the LLSIR procedure to estimate $\gamma'$ and $\mu$. In the subsequent content, we will instantiate our theoretical results from \Cref{sec: theoretical results} in the setting of the nonlinear single variable model. Again, we will operate under \Cref{assumption: lce and covariate density}. Note that, unlike prior work on estimating the nonlinear single variable model, we do not assume that the covariates are concentrated near the curve $\gamma$. Instead, we only require that the support is bounded and the density is sufficiently smooth. We consider this a key aspect of our approach, as any assumption about concentration near $\gamma$ reduces the problem to one of manifold learning.

\begin{restatable}[Bias shrinkage with localization]{corollary}{}
\label{corollary: population prop curve}
    Suppose that \Cref{assumption: the curve gamma} holds, together with items 1--4 of \Cref{assumption: lce and covariate density}, and set $g=\Pi$. Let $R$ be the projection regularity set from \Cref{thm: differentiability of pi} and set $s_0=\Pi(x_0)$ and $\ell_0=\mathfrak s^2-\langle\gamma''(s_0),x_0-\gamma(s_0)\rangle\neq0$. If $f\in BV_{\mathrm{loc}}(\operatorname{Int}(I))$, then for $P_X$-almost every $x_0\in R$ and all sufficiently small $\epsilon>0$ then
    \[\mu^*_{x_0,\epsilon}=\mathbb{E}[w\,f(\Pi(X))]/\mathbb{E}[w] = f(s_0)+o_{x_0}(\epsilon)\]
    and 
    \[
        \|v^*_{x_0,\epsilon}-c_*\gamma'(s_0)\|_2\lesssim_{x_0}\epsilon^2,
        \qquad \text{where} \qquad
        c_*=\frac{\epsilon f'(s_0)+o_{x_0}(\epsilon)}{\ell_0},
    \]
    and $v^*_{x_0,\epsilon}$ is defined in \eqref{eq: velocity estimator}. If, in addition, $f\in C^2(\operatorname{Int}(I))$, then both conclusions hold at every $x_0\in R$ with $p(x_0)>0$ with the sharper expansions
    \[
        \mu^*_{x_0,\epsilon}
        = f(s_0)+\frac{\epsilon^2\,\mathfrak s^2}{2(d+2)\,\ell_0^{2}}
        \left[f''(s_0)+\frac{\langle\gamma^{(3)}(s_0),x_0-\gamma(s_0)\rangle}{\ell_0}\,f'(s_0)\right]+o(\epsilon^2)  
    \]
    and $c_*=(\epsilon f'(s_0)+o(\epsilon^2))/\ell_0.$
\end{restatable}

The proof is provided in \Cref{proof of population prop curve}. We restate and reprove this result for the single variable model specifically as it is the only result that requires special treatment in this setting; this stems from the fact that our general setting considers any $\beta$-H\"older $g$, while the projection map $\Pi$ is only twice differentiable locally (almost surely). This nuance requires special attention but does not change any results. In particular, since $P_X(R) = 1$ by \Cref{thm: differentiability of pi}, restricting
inference points to $R$ does not affect the almost-everywhere qualifiers in
\Cref{thm: population prop} and the remark following
\Cref{thm: decomposition for theta}. On $R$, $\Pi$ is $C^2$ with $\nabla\Pi \neq 0$,
so the exceptional null set is handled exactly as in \Cref{sec: theoretical results}
(see the proof of \Cref{corollary: population prop curve}). Consequently, all limiting
distribution and rate of convergence results in \Cref{sec: theoretical results}
transfer directly to LLSIR in the nonlinear single variable model, at $P_X$-almost
every inference point.

\begin{restatable}{rmk}{}
    While we demonstrated that our procedure achieves a stochastic convergence rate matching the minimax MSE rate in the general case (\Cref{sec: theoretical results}), the same cannot be said for the nonlinear single variable model. The constraint that $g$ take the form of a closest point projection onto a curve might yield exploitable structure that is not available in the $\beta$-H\"older case. While the estimator constructed by \citet{wu2026conditional} attains a dimension free rate that coincides with the minimax rate for 1-dimensional function estimation, their procedure assumes that the covariates are concentrated near the curve $\gamma$. Dropping this assumption, as we have, likely changes the minimax rates. We consider a rigorous exploration of the minimax properties of estimating the nonlinear single variable model under mild assumptions about the distribution of the covariates an open question and a fruitful direction of future research. 
\end{restatable}

\section{Experiments}\label{sec: simulations}

\begin{figure}[htbp]
    \centering
    \includegraphics[width=\linewidth]{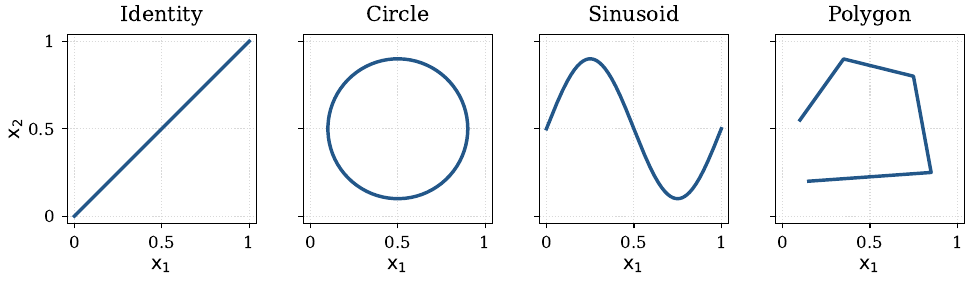}
    \caption{The four planar curves $\gamma([0,1])$ used in the synthetic experiments: Identity, Circle, Sinusoid, and Polygon, from left to right. All panels use the same coordinate scale and limits. For $d>2$, the curves are embedded by setting the remaining coordinates to zero.}
    \label{fig:simulation-curves}
\end{figure}

Having analyzed the statistical properties of the LLSIR procedure, we now evaluate it through simulation in several settings. We begin by evaluating the method in the \emph{single-variable model} setting described in \Cref{sec: single variable model}, where $\mu(x) = f \circ \Pi$ where $\Pi$ is the closest-point projection onto a compact and regular curve $\gamma$. We compare the method directly against the NSVM estimator from \citet{wu2026conditional} and against local polynomial regression with polynomial order $p = 1$ (denoted LLR) for four different choices of $\gamma$ in dimension $d = 2$ and $d = 3$. Moreover, for $\gamma = \{(x,y)  : x = y \}$ we also include the DWLLS squares estimator from \citet{Newey_Ruud_2005} as a baseline. The four choices of $\gamma$ (Identity, Circle, Sinusoid, Polygon) are shown in \Cref{fig:simulation-curves}.

\begin{figure}[H]
    \centering
    \includegraphics[width=\linewidth]{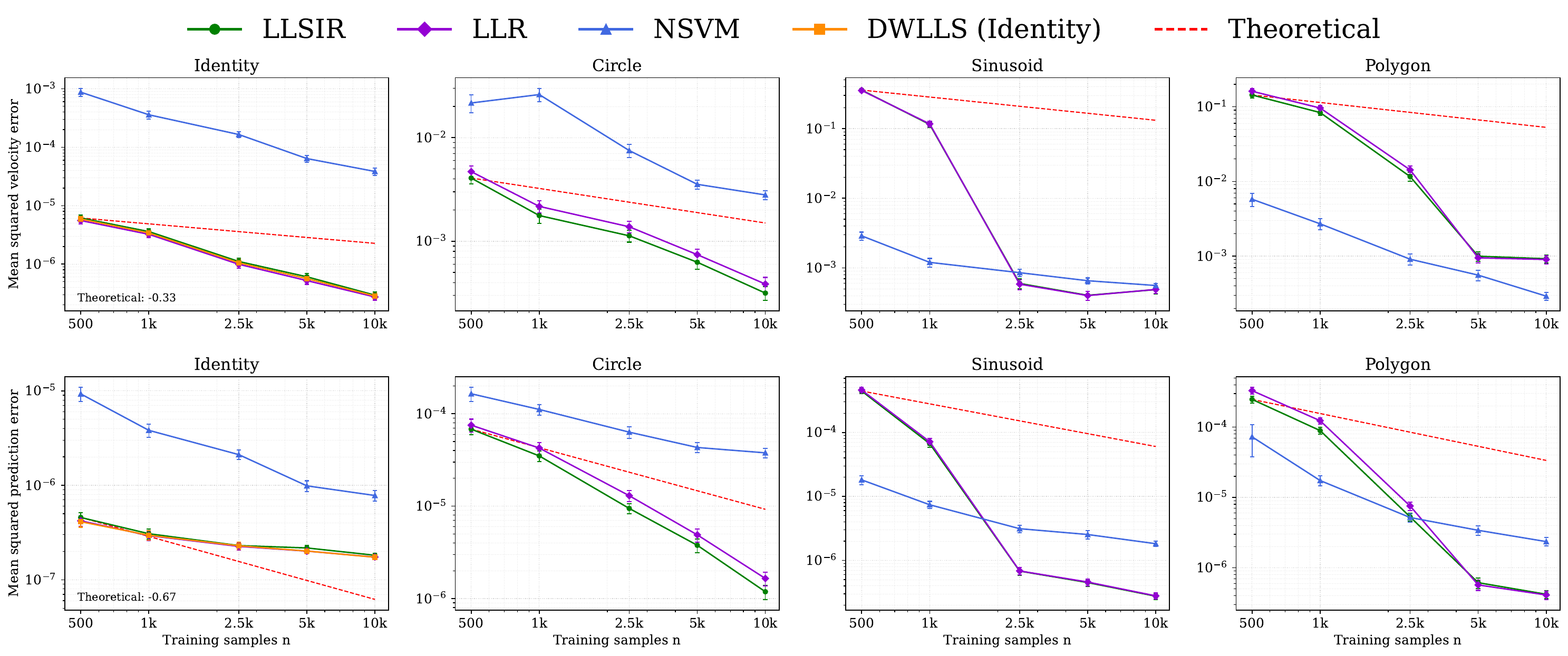}
    \caption{Simulation experiment for $d=2$. The top row shows mean squared velocity error between unit directions, (minimizing over signs), while the bottom row shows mean squared prediction error. Nodes average $B=100$ nested training trials, with one Monte Carlo standard error as error bars. Red dashed lines show the theoretical guides. More details for this experiment are provided in \Cref{experimental details}.}
    \label{fig:fixed-point-planar}
\end{figure}

\begin{figure}[H]
    \centering
    \includegraphics[width=\linewidth]{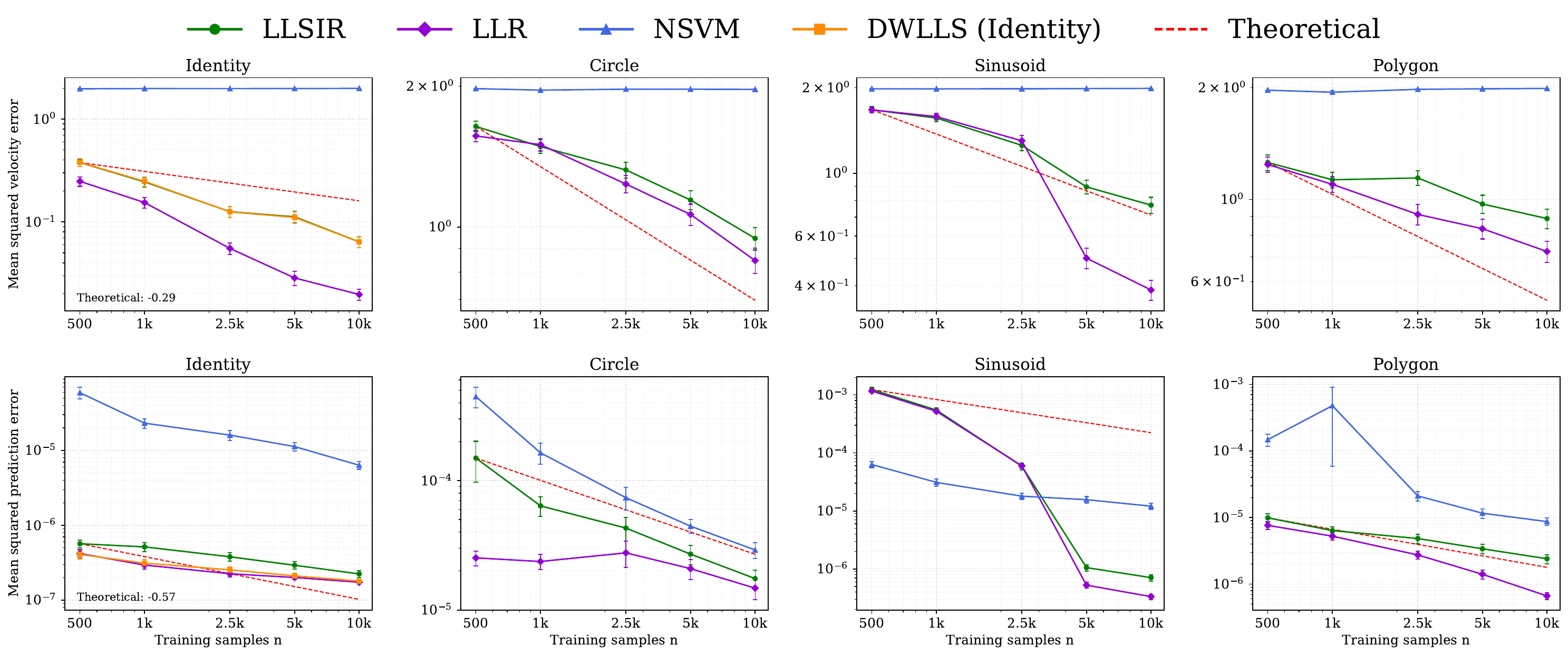}
    \caption{Lifted experiment ($d=3$), with rows and columns as in \Cref{fig:fixed-point-planar}. Covariates are embedded in three dimensions with independent $\mathcal{N}(0,10^{-6})$ noise only in the third coordinate, while curves and fixed inference locations remain at $z=0$. Nodes average $B=100$ nested training trials, with one Monte Carlo standard error as error bars. More details for this experiment are provided in \Cref{experimental details}.}
    \label{fig:fixed-point-lifted}
\end{figure}

We provide log-log plots of the mean squared velocity error and mean squared regression error in \Cref{fig:fixed-point-planar}. In it we see that LLSIR strictly outperforms NSVM on 4 of the 8 panels, while for three of the remaining panels we see that LLSIR outperforms NSVM in the high sample regime. We also unsurprisingly find that DWLLS and LLSIR exhibit nearly identical performance when $\gamma$ is linear, and we also see that LLSIR and LLR perform very similarly across all examples. Regarding comparison with the theoretical rates of convergence obtained in \Cref{sec: theoretical results}, we find that LLSIR typically exhibits slightly improved asymptotics than the theoretical results suggest. In \Cref{fig:fixed-point-lifted}, we provide the results for a nearly identical experiment, except we lift the covariates to $\mathbb R^3$ and add small Gaussian noise along the third axis. In this experiment we find that NSVM struggles to correctly identify the tangent velocity of the curve $\gamma$, as the geometry of the conditional measure $P_{X|Y}$ no longer identifies said velocity (a fact upon which NSVM relies).  On the other hand, we see that LLSIR still maintains strong performance and is typically consistent with the theoretical rate.

\begin{figure}[H]
    \centering
    \includegraphics[width=\linewidth]{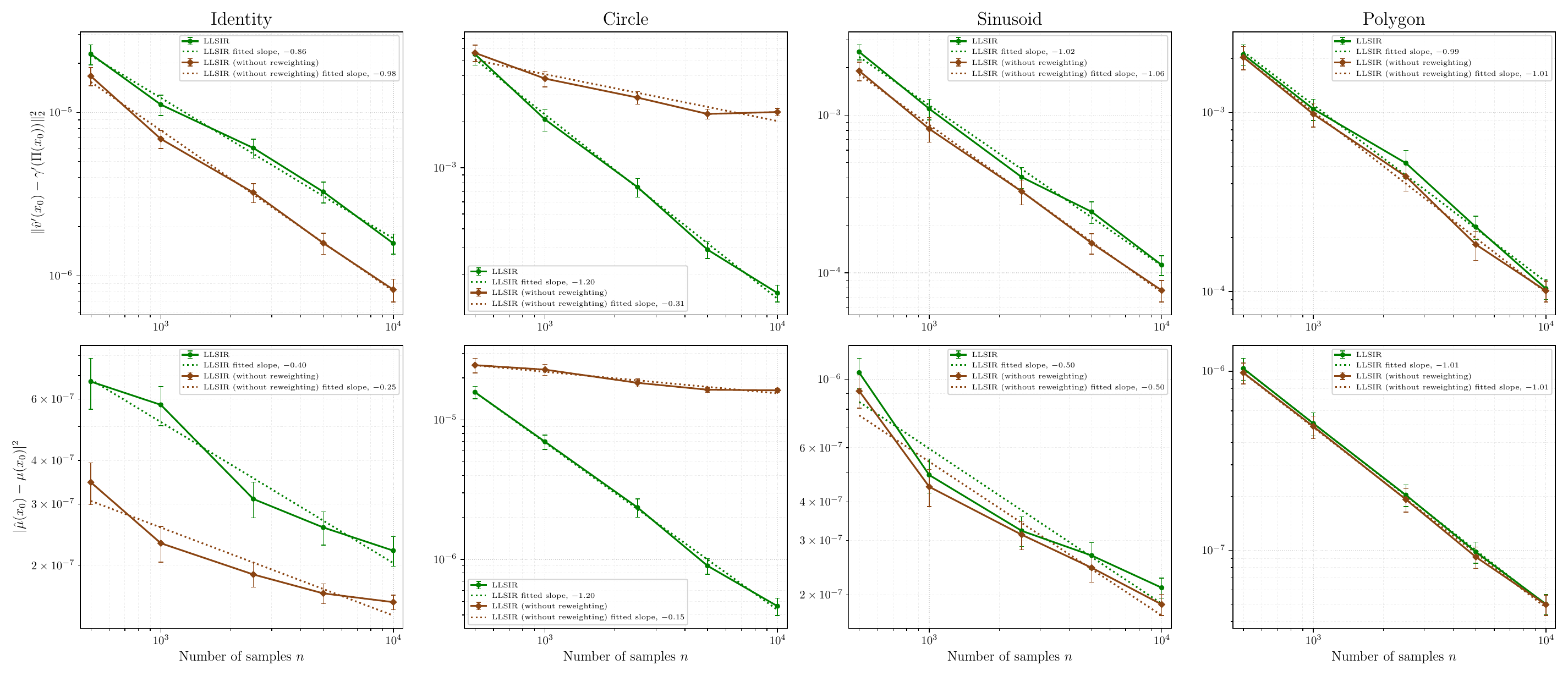}
    \caption{Density reweighting ablation ($d=2$), with localization radius fixed. Further details are provided in \Cref{experimental details}.}
    \label{fig:reweighting}
\end{figure}

In \Cref{fig:reweighting}, we consider the same datasets but ablate the density reweighting. In
particular, we compare LLSIR with weights
\[\hat w_i = \mathbbm{1}_{\{X_i\in \mathcal{B}(0,\epsilon)\}}\,p^*(X_i)/\hat p(X_i)\]
against the localization-only weights $\mathbbm{1}_{\{X_i\in \mathcal{B}(0,\epsilon)\}}$, where the localization parameter $\varepsilon$ fixed. For the
identity, sinusoid and polygon curves, the two variants attain comparable empirical rates for the
velocity error and for the conditional mean error on the sinusoid and polygon. On these curves the
unweighted fit often has slightly smaller error, reflecting the additional variance introduced by
dividing by $\hat p$. For the circle, however, the unweighted fit degrades substantially and its errors
plateau early.

\begin{figure}[H]
    \centering
    \includegraphics[width=\linewidth]{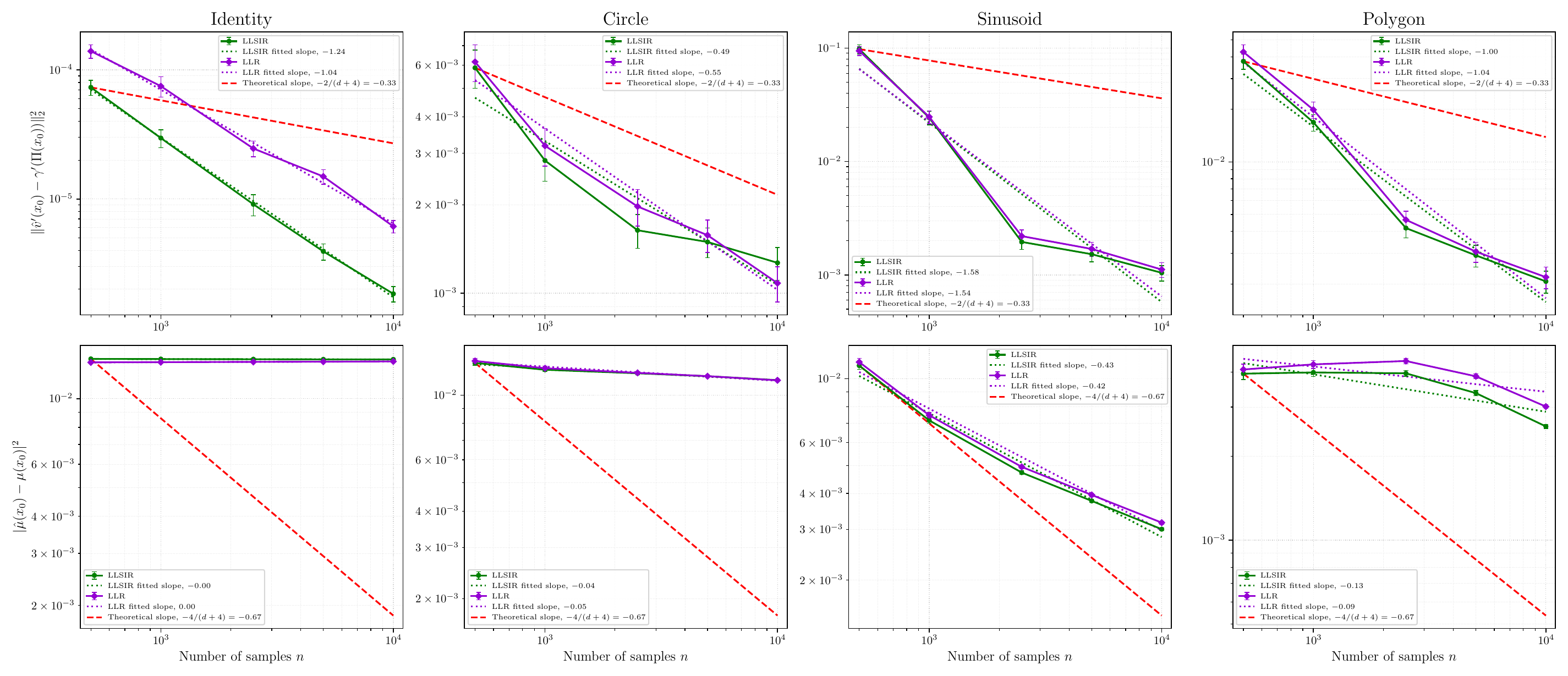}
    \caption{Velocity estimation with a bounded-variation link ($d=2$). The planar fixed-point experiment is repeated with the response link replaced. In particular, responses follow $Y = f(\Pi(X)) + \varepsilon$ with $f(t)=t+0.25\sum_j\mathbbm{1}\{t \geq s_j\}$. Further details are provided in \Cref{experimental details}.}
    \label{fig:reweighting-bv}
\end{figure}

Finally, in \Cref{fig:reweighting-bv} we compare the performance of LLSIR and local linear regression (LLR) when the link function $f$ is taken to be a discontinuous function of bounded variation. For velocity estimation (top row), LLSIR consistently attains a smaller error than LLR on nearly all settings. On the circle and sinusoid the two estimators are within one standard error of one another, with LLSIR slightly ahead at most sample sizes. In every panel the fitted slopes of both estimators match or exceed the $-2/(d+4)$ benchmark, so the discontinuity does not diminish the velocity rates. The conditional mean (bottom row) behaves differently, however. The two estimators nearly coincide, but consistently perform poorly. We attribute this to the small variation in the conditional mean relative to the jump discontinuity size over the localization.

\medskip

\emph{Dependence on $d$.} Now we describe experiments conducted to understand the empirical dimension dependence of the LLSIR procedure in the \emph{generalized single-index} setting. In particular, we generate data from the model $Y = f(g(X)) + \varepsilon$ with $f$ the identity link. We vary the ambient dimension over $d \in \{2, 3, 4, 5, 6\}$ and consider two families of index functions: the first family consists of the projections onto the four curves in \Cref{fig:simulation-curves}, while the second family consists of the functions
\[x \mapsto \sin(\pi x_1), \quad x\mapsto \|x\|_2^2, \quad x \mapsto x_1x_2.\]
For each $(\text{family}, \text{stage})$ pair, we provide a table detailing the log-log slope of the mean squared (stage) error where stage $\in \{\text{velocity}, \text{regression}\}.$ The tables corresponding to the curve-induced family are \Cref{tab:vel} and \Cref{tab:reg}, while the tables corresponding to the function family listed in the panel above are \Cref{tab:vel-general} and \Cref{tab:reg-general}. 

\begin{table}[H]
\small
\centering
\begin{tabular}{lcccc}
\toprule
 & Identity & Circle & Sinusoid & Polygon \\
\midrule
$d=2$ & $-0.32 \pm 0.49$ & $-0.35 \pm 0.53$ & $-0.77 \pm 0.18$ & $-0.23 \pm 0.10$ \\
$d=3$ & $-0.29 \pm 0.36$ & $-0.33 \pm 0.37$ & $-0.52 \pm 0.22$ & $-0.17 \pm 0.20$ \\
$d=4$ & $-0.27 \pm 0.32$ & $0.01 \pm 0.53$ & $-0.27 \pm 0.12$ & $-0.05 \pm 0.19$ \\
$d=5$ & $-0.30 \pm 0.28$ & $0.14 \pm 0.44$ & $-0.17 \pm 0.13$ & $-0.04 \pm 0.19$ \\
$d=6$ & $-0.31 \pm 0.29$ & $-0.07 \pm 0.32$ & $-0.16 \pm 0.11$ & $-0.14 \pm 0.26$ \\
\bottomrule
\end{tabular}
\caption{LLSIR log-log slopes for velocity error (projection) with mean $\pm$ standard deviation from $B=200$ Monte Carlo trials. Each trial's least-squares slope is fitted after averaging squared error over 5 fixed query points. For comparison, the theoretical slopes $-2/(d+4)$ are $(-0.333, -0.286, -0.250, -0.222, -0.200)$ for $d=(2,3,4,5,6)$, respectively. }
\label{tab:vel}
\end{table}

\begin{table}[H]
\small
\centering
\begin{tabular}{lcccc}
\toprule
 & Identity & Circle & Sinusoid & Polygon \\
\midrule
$d=2$ & $-0.67 \pm 0.49$ & $-0.61 \pm 0.55$ & $-0.91 \pm 0.18$ & $-0.53 \pm 0.16$ \\
$d=3$ & $-0.58 \pm 0.54$ & $-0.59 \pm 0.58$ & $-0.69 \pm 0.22$ & $-0.32 \pm 0.28$ \\
$d=4$ & $-0.55 \pm 0.54$ & $-0.51 \pm 0.71$ & $-0.42 \pm 0.12$ & $-0.21 \pm 0.26$ \\
$d=5$ & $-0.53 \pm 0.58$ & $-0.31 \pm 0.69$ & $-0.25 \pm 0.13$ & $-0.19 \pm 0.27$ \\
$d=6$ & $-0.56 \pm 0.67$ & $-0.41 \pm 0.54$ & $-0.16 \pm 0.15$ & $-0.32 \pm 0.54$ \\
\bottomrule
\end{tabular}
\caption{LLSIR log-log slopes for regression error (projection) with mean $\pm$ standard deviation from $B=200$ Monte Carlo trials. Each trial's least-squares slope is fitted after averaging squared error over 5 fixed query points. For comparison, the theoretical slopes $-4/(d+4)$ are $(-0.667, -0.571, -0.500, -0.444, -0.400)$ for $d=(2,3,4,5,6)$, respectively. }
\label{tab:reg}
\end{table}

\begin{table}[H]
\small
\centering
\begin{tabular}{lccc}
\toprule
 & $\sin(\pi x_1)$ & $\|x\|^2$ & $x_1x_2$ \\
\midrule
$d=2$ & $-0.51 \pm 0.60$ & $-0.30 \pm 0.60$ & $-0.26 \pm 0.63$ \\
$d=3$ & $-0.53 \pm 0.45$ & $-0.37 \pm 0.35$ & $-0.36 \pm 0.37$ \\
$d=4$ & $-0.72 \pm 0.41$ & $-0.53 \pm 0.32$ & $-0.39 \pm 0.32$ \\
$d=5$ & $-0.68 \pm 0.37$ & $-0.57 \pm 0.28$ & $-0.38 \pm 0.29$ \\
$d=6$ & $-0.68 \pm 0.31$ & $-0.60 \pm 0.23$ & $-0.49 \pm 0.29$ \\
\bottomrule
\end{tabular}
\caption{LLSIR log-log slopes for velocity error (general) with mean $\pm$ standard deviation from $B=200$ Monte Carlo trials. Each trial's least-squares slope is fitted after averaging squared error over 5 fixed query points. For comparison, the theoretical slopes $-2/(d+4)$ are $(-0.333, -0.286, -0.250, -0.222, -0.200)$ for $d=(2,3,4,5,6)$, respectively.}
\label{tab:vel-general}
\end{table}

\begin{table}[H]
\small
\centering
\begin{tabular}{lccc}
\toprule
 & $\sin(\pi x_1)$ & $\|x\|^2$ & $x_1x_2$ \\
\midrule
$d=2$ & $-0.67 \pm 0.10$ & $-0.66 \pm 0.16$ & $-0.65 \pm 0.45$ \\
$d=3$ & $-0.58 \pm 0.14$ & $-0.57 \pm 0.12$ & $-0.63 \pm 0.53$ \\
$d=4$ & $-0.49 \pm 0.11$ & $-0.51 \pm 0.06$ & $-0.60 \pm 0.54$ \\
$d=5$ & $-0.45 \pm 0.11$ & $-0.45 \pm 0.04$ & $-0.57 \pm 0.55$ \\
$d=6$ & $-0.39 \pm 0.12$ & $-0.40 \pm 0.03$ & $-0.66 \pm 0.53$ \\
\bottomrule
\end{tabular}
\caption{LLSIR log-log slopes for regression error (general) with mean $\pm$ standard deviation from $B=200$ Monte Carlo trials. Each trial's least-squares slope is fitted after averaging squared error over 5 fixed query points. For comparison, the theoretical slopes $-4/(d+4)$ are $(-0.667, -0.571, -0.500, -0.444, -0.400)$ for $d=(2,3,4,5,6)$, respectively. }
\label{tab:reg-general}
\end{table}

In \Cref{tab:vel} and \Cref{tab:reg}, we see that the dependence of the log-log slope typically tends to $0$ as $d$ increases, and the slopes roughly track those predicted by the theory. Some notable exceptions are the velocity log-log slope for the circle, where $d = 4$ and $5$ yield a positive dependence, while there are scattered instances of non-monotone empirical dependence on $d$. We note that both of these phenomena are likely due to the stochastic nature of the experiment. Moreover, the circle induces a discontinuity at $\theta = 2\pi$ which could also explain the artifact. For \Cref{tab:reg-general} we see that the tracking of the theoretical rates are very good, while \Cref{tab:vel-general} surprisingly exhibits improved empirical performance for increasing $d$. 
\medskip

\begin{figure}[htbp]
    \centering
    \includegraphics[width=\linewidth]{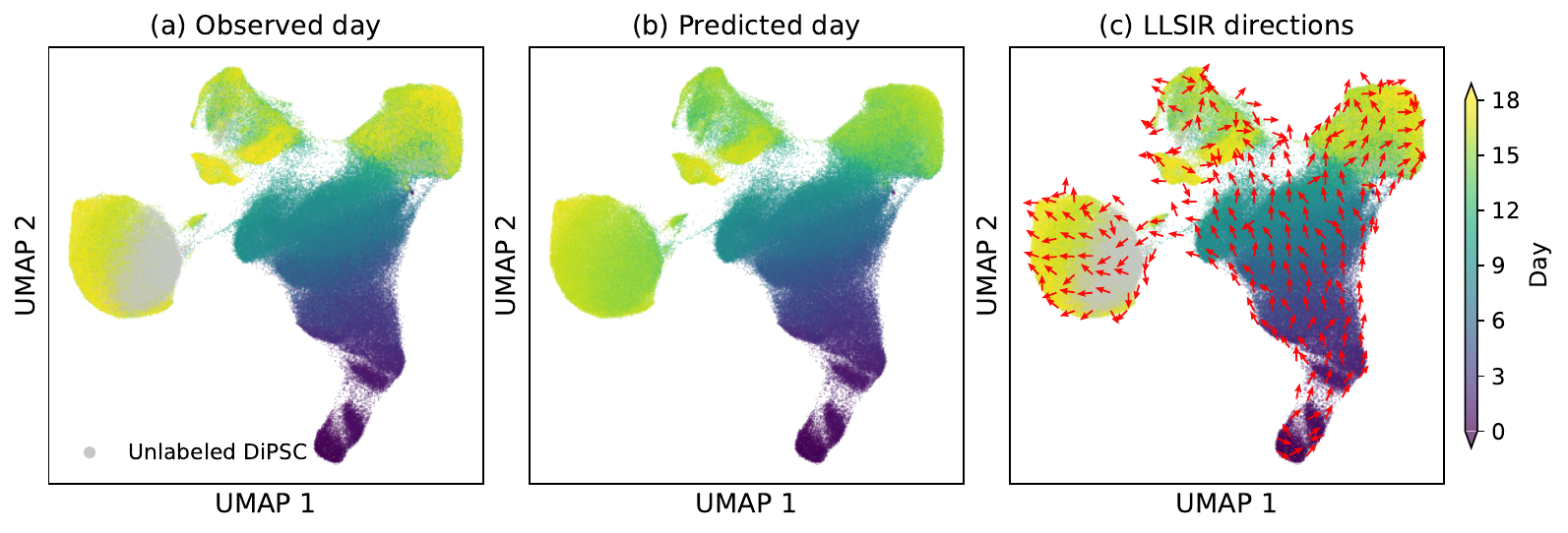}
    \caption{Cellular reprogramming with LLSIR on all 251,203 cells in the processed release. Panels show (a) observed collection day, (b) predicted day, and (c) observed day with approximately projected LLSIR directions. Gray points have no observed collection-day label. Red arrows point toward increasing estimated time and have equal display length.}
    \label{fig:reprogramming-umap}
\end{figure}

\emph{Example: Cellular reprogramming.} Finally, we apply LLSIR to the single-cell reprogramming data of \citet{schiebinger2019reprogramming}. The processed Waddington-OT release contains 251,203 cells and 1,479 log-normalized variable genes. We split the data by collection day, and assign 177,213 labeled cells to training and 59,072 to evaluation. LLSIR uses collection day as the response, a nominal radius of about $0.9$, expanded where necessary to retain at least 128 training neighbors, and Gaussian weight standard deviation $0.440$. We find that LLSIR obtains a held-out mean absolute prediction error of $0.805$ days, with root-mean-square error $1.183$ days (while the entire dataset has a range of approximately 20 days). In \Cref{fig:reprogramming-umap}, we visualize all cells using an unsupervised UMAP embedding \citep{mcinnes2018umap} of the saved PCA coordinates, and we project the LLSIR estimated tangent velocity $\hat v$ using local ridge regression between the original distances and the embedded distances. In particular, we find that LLSIR does well in recovering the underlying gradient field driving the temporal dependence. 

\section{Conclusion}\label{sec: conclusion}

In this work we proposed and analyzed LLSIR, a conceptually simple and computationally tractable procedure for generalized single-index regression. Through localization, the procedure jointly estimates the spatially varying direction $\nabla g(x_0)$ up to scale and the conditional mean $\mu(x_0)$, avoiding the need to jointly optimize over the link and index functions. We derived influence function expansions, central limit theorems, and pointwise rates of convergence for both stages. In particular, when $g$ is $2$-H\"older and $f$ has a locally Lipschitz derivative, the regression rate coincides with the minimax rate for the corresponding composition class. We also instantiated this procedure in the nonlinear single-variable model, where the index is a closest-point projection onto a curve. A key aspect of this application is that it does not require the covariates to be concentrated near the curve, allowing estimation under the bounded-support and density regularity assumptions of our general framework. Our simulations and cellular reprogramming example demonstrate the utility of the procedure for estimating both local directions and regression responses. 

\section{Acknowledgements and AI Declaration}
The authors are grateful to Yuxi (Eric) Chen for his assistance with simulations and for fruitful discussions regarding the method. ChatGPT 5.6 Sol and Claude Fable 5.2 were used to audit the proofs, aid in the literature review and assist with writing and phrasing. The development of the estimation procedure and proof the strategies for the limit theorems and decomposition involved little AI usage. 

\break

\begin{appendices}

\section{Preliminaries on the function class $BV_{\operatorname{loc}}(\mathbb R)$}\label{sec: BV}

In this section we will provide a brief overview of functions of bounded variation that will loosely follow Chapter 3 of \citet{ambrosio2000functions}. For any open interval $I = (a,b)$ and any function $f: I\to \mathbb R$, we define the pointwise variation $\operatorname{pV}(f,I)$ of $u$ in $I$ to be 
\begin{equation}
    \operatorname{pV}(f, I) = \sup\left\{\sum_{i = 1}^{N-1}\left|f(t_{i+1}) - f(t_i)\right|: N \geq 2, \, a < t_1< \cdots < t_n < b\right\}.
\end{equation}
Note that if $f \in C^1(I)$ then $\operatorname{pV}(f, I) = \int_a^b|f'(x)|\,dx = \|f\|_{\dot{W}^{1,1}}$, coinciding with the homogeneous $W^{1,1}$ seminorm. For $d > 1$, partitions have no natural analogue in $ \Omega\subset \mathbb R^d$. So one
instead defines the variation of $f$ by duality,
\[
    \operatorname{V}(f,\Omega) = \sup\left\{ \int_\Omega f\,\mathrm{div}\,\phi\,dx :
    \phi \in C_c^1(\Omega;\mathbb R^d),\ \|\phi\|_\infty \le 1 \right\},
\]
which reduces to $\int_\Omega \|\nabla f\|_2\,dx$ for smooth $f$.  A function $f \in L^1(\Omega)$ is a \textit{function of bounded variation} in $\Omega$ if its distributional derivative is representable by a finite Radon measure in $\Omega$, i.e. if 
\begin{equation}
    \int_\Omega f\, \nabla \phi(x) \, dx = -\int_\Omega \phi \, d(\mathbf{D}f) \quad \text{for all $\phi \in C^\infty_c(\Omega)$}
\end{equation}
for some $\mathbb R^d$-valued measure $\mathbf{D}f = \begin{bmatrix}
    D_1f & \cdots & D_df
\end{bmatrix}.$ Equivalently, $f \in L^1(\Omega)$ is a function of bounded variation precisely when $V(f,\Omega) < \infty$, in
which case $V(f,\Omega) = |\mathbf D f|(\Omega)$
\citep[Proposition~3.6]{ambrosio2000functions}. The vector space of all functions of bounded variation in $\Omega$ is denoted $BV(\Omega).$ Finally, we say that $f \in BV_{\mathrm{loc}}(\Omega)$ if $f \in L^1_{\mathrm{loc}}(\Omega)$ and
$f \in BV(\Omega')$ for every open set $\Omega' \subset \Omega$. Equivalently,
the distributional derivative $\mathbf D f$ is an $\mathbb R^d$-valued Radon measure on
$\Omega$, so that $|\mathbf D f|(K) < \infty$ for every compact $K \subset \Omega$,
although $|\mathbf D f|(\Omega)$ may be infinite. The measure $|\mathbf D f|$ is sometimes referred to as the \emph{total variation measure} of $f.$

When $d = 1$ and $I = (a,b)$ for some $a < b$, the measure $\operatorname{D}f$ can be related to $\operatorname{pV}(f, I)$ via the \textit{essential variation}
\[
    \operatorname{eV}(f, I) = \inf\left\{ \operatorname{pV}(g, I) : g = f \text{ a.e.\ in } I \right\}.
\]
The infimum is needed because $f \in L^1(I)$ is only an equivalence class, and
$\operatorname{pV}$ is sensitive to modifications on null sets. For
$f \in L^1(I)$, the essential variation coincides with the variation defined by
duality, $\operatorname{eV}(f, I) = \operatorname{V}(f, I)$, so that $f \in BV(I)$
if and only if $\operatorname{eV}(f, I) < \infty$, in which case
$|\mathbf D f|(I) = \operatorname{eV}(f, I)$
\citep[Theorem~3.27]{ambrosio2000functions}. Moreover, the infimum is attained, as we'll see below.

\begin{restatable}[Good representatives,  {\citep[Theorem~3.28]{ambrosio2000functions}}]{thm}{}
    \label{thm:good-representatives}
Let $I = (a,b) \subset \mathbb R$ be an interval and $f \in [BV(I)]^d$. Let $A$ be the
set of atoms of $\mathbf D f$, i.e.\ $t \in A$ if and only if
$|\mathbf D f|(\{t\}) \neq 0$. Then, the following statements hold.
\begin{enumerate}
    \item[(a)] There exists a unique $c \in \mathbb R^d$ such that
    \[
        f^l(t) := c + \mathbf D f\big((a,t)\big), \qquad
        f^r(t) := c + \mathbf D f\big((a,t]\big) \qquad t \in I
    \]
    are good representatives of $f$, the left continuous one and the right continuous
    one. Any other function $\bar f : I \to \mathbb R^d$ is a good representative of
    $f$ if and only if
    \begin{equation}
        \bar f(t) \in \left\{ \theta f^l(t) + (1-\theta) f^r(t) : \theta \in [0,1] \right\}
        \qquad \forall\, t \in I. 
    \end{equation}
    \item[(b)] Any good representative $\bar f$ is continuous in $I \setminus A$ and has
    a jump discontinuity at any point of $A$:
    \[
        \bar f(t_-) = f^l(t) = f^r(t_-), \qquad
        \bar f(t_+) = f^l(t_+) = f^r(t) \qquad \forall\, t \in A.
    \]
    \item[(c)] Any good representative $\bar f$ is differentiable at
    $\mathcal L^1$-a.e.\ point of $I$. The derivative $\bar f'$ is the density of
    $\mathbf D f$ with respect to $\mathcal L^1$.
\end{enumerate}
\end{restatable}
    
Thus, every $f \in BV(I)$ has a \textit{good representative} $\bar f$, equal to $f$ almost
everywhere, with $\operatorname{pV}(\bar f, I) = |\mathbf D f|(I)$. For instance, for
a suitable constant $c$, the function $\bar f(t) = c + \mathbf D f((a, t))$ is such a
representative. Any good representative is continuous away from the atoms of
$\mathbf D f$ and has a jump discontinuity at each atom
\citep[Theorem~3.28]{ambrosio2000functions}. Note that the theorem extends to $f \in BV_{\mathrm{loc}}(J)$ for any open interval
$J \subseteq \mathbb R$ by applying it to the restriction of $f$ to each bounded open
interval $I \subset J$. Since any two right-continuous functions that agree
almost everywhere agree everywhere, the right-continuous good representatives obtained
on nested intervals coincide on their common domain. They therefore define a single
right-continuous representative of $f$ on $J$, for which (a)--(c) hold on every bounded
open subinterval $I \subset J$.

\section{Proofs of key results}

\subsection{Proof of \Cref{thm: population prop}} \label{sec: proof of population vel prop}

The proof of \Cref{thm: population prop} follows from  \Cref{intercept error BV}, \Cref{velocity error BV} and \Cref{velocity and intercept error C2}. \qed{}

\begin{restatable}{lemma}{} \label{intercept error BV}
    Suppose \Cref{assumption: lce and covariate density}
    holds.
    Then, for $P_X$-almost every
    $x_0\in\operatorname{int}(\operatorname{supp}(P_X))$
    satisfying $p(x_0)>0$, $\nabla g(x_0)\neq0$,and for all sufficiently small $\epsilon>0$ the population intercept admits the closed form
    \[
        \mu^*_{x_0,\epsilon}
        =\frac{\mathbb{E}[w\,f(g(X))]}{\mathbb{E}[w]},
    \]
    and satisfies $\mu^*_{x_0,\epsilon}=\mu(0)+o_{x_0}(\epsilon)$.
\end{restatable}
\begin{proof}
    Since $p$ is continuous at $0$ with
    $p(0)>0$, we have $p>0$ on $\mathcal B(0,\epsilon)$ for all
    sufficiently small $\epsilon$, so that
    $\mathbb E_p[w\,h(X)] = \int_{\mathcal B(0,\epsilon)} h(x)p^*(x)\,dx$. In particular this applies to
    $h = f\circ g$, which is bounded on $\mathcal B(0,\epsilon)$ because
    $f\in BV_{\operatorname{loc}}(\mathbb R)$ is bounded on compact
    intervals. Writing
    \[
    A(\epsilon) := \mathbb E[w]
    = \int_{\mathcal B(0,\epsilon)} p^*(x)\,dx,
    \]
    continuity of $p^*$ at $0$ and $p^*(0)>0$ give
    $A(\epsilon)\asymp\epsilon^d$. By spherical
    symmetry $\mathbb E[wX]=0$, so \eqref{eq: Q closed form} gives
    $e_1^\top Q^{-1} = A(\epsilon)^{-1}e_1^\top$, and
    \eqref{eq: population estimator closed form} yields
    \[
    \mu^*_{x_0,\epsilon}
    = e_1^\top Q^{-1}\,\mathbb E\big[w\boldsymbol X f(g(X))\big]
    = \frac{\mathbb E[w\,f(g(X))]}{\mathbb E[w]},
    \]
    which is the claimed closed form. Consequently, recalling that
    $\mu(0)=f(g(0))$ and $\xi(x) = f(g(x)) - f(g(0) + \nabla g(0)^{\top}x)$,
    \begin{align*}
    \mu^*_{x_0,\epsilon}-\mu(0)
    &= \frac{1}{A(\epsilon)}\int_{\mathcal B(0,\epsilon)}
       \big[f(g(x))-f(g(0))\big]\,p^*(x)\,dx\\
    &= \underbrace{\frac{1}{A(\epsilon)}\int_{\mathcal B(0,\epsilon)}
       \xi(x)\,p^*(x)\,dx}_{T_1}
     + \underbrace{\frac{1}{A(\epsilon)}\int_{\mathcal B(0,\epsilon)}
       \big[f(g(0)+\nabla g(0)^\top x)-f(g(0))\big]\,p^*(x)\,dx}_{T_2}.
    \end{align*}
    For $T_1$, the bound
    $\int_{\mathcal B(0,\epsilon)}|\xi(x)|p^*(x)\,dx
    \lesssim_{x_0} \epsilon^{d+1}|Df|(J_\epsilon)$
    established in \eqref{function integrated error bound}, together with $A(\epsilon)\asymp\epsilon^d$, gives
    \[
    |T_1| \lesssim_{x_0} \epsilon\,|Df|(J_\epsilon) = O_{x_0}(\epsilon^2),
    \]
    since $|Df|(J_\epsilon)=O_{x_0}(\epsilon)$ at the inference points under
    consideration (\Cref{tv measure bound}). For $T_2$, define
    \[
    \rho(t) = f(g(0)+t)-f(g(0))-f'(g(0))\,t,
    \qquad \text{and} \qquad
    \eta(r) = \sup_{0<|t|\leq r}\frac{|\rho(t)|}{|t|},
    \]
    so that $\eta(r)\to0$ as $r\downarrow0$ by differentiability of $f$ at
    $g(0)$. Since
    $\int_{\mathcal B(0,\epsilon)}(\nabla g(0)^\top x)\,p^*(x)\,dx = 0$ by spherical
    symmetry and $|\nabla g(0)^\top x|\leq\|\nabla g(0)\|_2\epsilon$ on $\mathcal B(0,\epsilon)$,
    \begin{multline*}
    |T_2|
    = \left|\frac{1}{A(\epsilon)}\int_{\mathcal B(0,\epsilon)}
      \rho(\nabla g(0)^\top x)\,p^*(x)\,dx\right|
    \\ \leq \eta(\|\nabla g(0)\|_2\epsilon)\,
      \frac{1}{A(\epsilon)}\int_{\mathcal B(0,\epsilon)}|\nabla g(0)^\top x|\,p^*(x)\,dx
    \\ \leq \eta(\|\nabla g(0)\|_2\epsilon)\,\|\nabla g(0)\|_2\,\epsilon
    = o_{x_0}(\epsilon).
    \end{multline*}
    Combining the two bounds gives
    $\mu^*_{x_0,\epsilon} = \mu(0) + o_{x_0}(\epsilon)$, which completes the
    proof.
\end{proof}

\begin{restatable}{prop}{}\label{velocity error BV}
    Suppose \Cref{assumption: lce and covariate density}
    holds, except that the assumptions on $f$ are
    replaced by $f\in BV_{\operatorname{loc}}(\mathbb R)$.
    Then, for $P_X$-almost every
    $x_0\in\operatorname{int}(\operatorname{supp}(P_X))$
    satisfying $p(x_0)>0$ and $\nabla g(x_0)\neq0$,
    for all sufficiently small $\epsilon>0$ there exists $c_*\in\mathbb R$ such that
    \[
        \|v^*_{x_0,\epsilon}
        -c_*\nabla g(0)\|_2
        \lesssim_{x_0}\epsilon^2,
    \]
    with $v^*_{x_0,\epsilon}$ defined in
    \eqref{eq: velocity estimator}. Moreover, $c_*=\epsilon f'(g(0))+o_{x_0}(\epsilon).$
\end{restatable}
\begin{proof}
    Let $\tilde\theta^*_{x_0,\epsilon}$ denote the
    population optimizer of the linearized surrogate
    problem in \Cref{lemma: surrogate problem}.
    By that lemma, there exist
    $c_0,c_*\in\mathbb R$ such that
    \[
    \tilde\theta^*_{x_0,\epsilon}
    =\begin{bmatrix}
    c_0 & c_*\nabla g(0)^\top
    \end{bmatrix}^{\top}.
    \]
    Thus, it suffices to establish that
    $\|v^*_{x_0,\epsilon}
    -V\tilde\theta^*_{x_0,\epsilon}\|_2
    \lesssim\epsilon^2$. To show this, we'll define the \say{surrogate} target
    \[m(x) = f\Big(g(0) + \nabla g(0)^{\top}x\Big)\]
    as described in \eqref{eq: surrogate decomposition curve} and the surrogate residual $\xi(x) = f(g(x)) - m(x).$
    Since $g\in C^2(U)$, for sufficiently small
    $\epsilon>0$ and all $x\in\mathcal B(0,\epsilon)$,
    \[
    g(x)=g(0)+\nabla g(0)^\top x+R(x),
    \qquad |R(x)|\le L\epsilon^2.
    \]
    Writing $|Df|$ for the total variation measure
    of the distributional derivative of $f$, the
    right continuous representative $f$ satisfies $f(v) - f(u) = Df((u,v])$ (\Cref{thm:good-representatives}) and thus
    \[
    |\xi(x)|
    \le
    |Df|\left(
    [g(0)+\nabla g(0)^\top x-L\epsilon^2,\,
     g(0)+\nabla g(0)^\top x+L\epsilon^2]
    \right).
    \]
    Since $\nabla g(0)\neq 0$, we have
    \[
    J_\epsilon =
    [g(0)-\|\nabla g(0)\|_2\epsilon-L\epsilon^2,\,
     g(0)+\|\nabla g(0)\|_2\epsilon+L\epsilon^2].
    \]
    By continuity of $p^*$ at $0$, it is uniformly bounded
    on $\mathcal B(0,\epsilon)$ for sufficiently small
    $\epsilon$. Rotating coordinates so that
    $\nabla g(0)=\|\nabla g(0)\|_2e_1$, the ball becomes
    \[
    \mathcal B(0,\epsilon)
    =\{(u,z) \in \mathbb R \times \mathbb R^{d-1}:|u|\leq\epsilon,\,
    \|z\|_2\leq\sqrt{\epsilon^2-u^2}\}.
    \]
    Since \(p^*\) is bounded above by some \(M<\infty\) on the ball for sufficiently small \(\epsilon\), writing $\omega_{d-1}$ for
    the volume of the unit ball in $\mathbb R^{d-1}$
    (with $\omega_0=1$), writing $J_\epsilon(u) = [g(0) + \|\nabla g(0)\|_2u - L\epsilon^2, g(0) + \|\nabla g(0)\|_2]u + L\epsilon^2]$ and applying Fubini's theorem gives
    \begin{align*}
    \int_{\mathcal B(0,\epsilon)}|\xi(x)|p^*(x)\,dx
    &\leq M\int_{-\epsilon}^{\epsilon}
    \int_{\|z\|_2\leq\sqrt{\epsilon^2-u^2}}
    |Df|(J_\epsilon(u))\,dz\,du\\
    &=M\omega_{d-1}\int_{-\epsilon}^{\epsilon}
    (\epsilon^2-u^2)^{(d-1)/2}
    |Df|(J_\epsilon(u))\,du\\
    &\leq M\omega_{d-1}\epsilon^{d-1}
    \int_{-\epsilon}^{\epsilon}
    |Df|(J_\epsilon(u))\,du.
    \end{align*}
    Thus, applying a change of variables $s = g(0) + \|\nabla g(0)\|_2 u$ (such that $du = ds/\|\nabla g(0)\|_2$) yields
    \begin{align}
    \int_{\mathcal B(0,\epsilon)}
    |\xi(x)|p^*(x)\,dx
    &\lesssim_{x_0}
    \frac{\omega_{d-1}\epsilon^{d-1}}{\|\nabla g(0)\|_2}
    \int_{g(0)-\|\nabla g(0)\|_2\epsilon}^{g(0)+\|\nabla g(0)\|_2\epsilon}
    |Df|([s-L\epsilon^2,s+L\epsilon^2])\,ds\nonumber\\
    &= \frac{\omega_{d-1}\epsilon^{d-1}}{\|\nabla g(0)\|_2}
    \int_{\mathbb R} \int_{g(0)-\|\nabla g(0)\|_2\epsilon}^{g(0)+\|\nabla g(0)\|_2\epsilon}
    \mathbbm{1}_{\{|t-s|\leq L\epsilon^2\}}\,ds \,|Df|(dt)\nonumber\\
    &\le
    \frac{2L\omega_{d-1}\epsilon^{d+1}
    }{\|\nabla g(0)\|_2}|Df|(J_\epsilon) \nonumber\\
    &\lesssim_{x_0}
    \epsilon^{d+1}|Df|(J_\epsilon). \label{function integrated error bound}
    \end{align}
    The second inequality follows from Tonelli's theorem,
    since each point integrated against $|Df|$ belongs to
    $[s-L\epsilon^2,s+L\epsilon^2]$ for a set of values of $s$
    having Lebesgue measure at most $2L\epsilon^2$. Let
    \[
    \nu^2
    :=\int_{\mathcal B(0,\epsilon)}x_1^2p^*(x)\,dx.
    \]
    By spherical symmetry, continuity of $p^*$ at $0$,
    and $p^*(0)>0$, we have
    $\nu^2\asymp\epsilon^{d+2}$.
    Consequently, by \eqref{eq: population estimator closed form}
    and \eqref{theta tilde simplification}, we have
    \[
    v^*_{x_0,\epsilon}
    -V\tilde\theta^*_{x_0,\epsilon}
    =
    \frac{\epsilon}{\nu^2}
    \int_{\mathcal B(0,\epsilon)}
    x\,\xi(x)p^*(x)\,dx,
    \]
    where
    $\nu^2\asymp\epsilon^{d+2}$.
    Taking norms and applying the preceding bound gives
    \begin{align*}
    \|v^*_{x_0,\epsilon}
    -V\tilde\theta^*_{x_0,\epsilon}\|_2
    &\leq
    \frac{\epsilon^2}{\nu^2}
    \int_{\mathcal B(0,\epsilon)}
    |\xi(x)|p^*(x)\,dx\\
    &\lesssim_{x_0}
    \epsilon\,|Df|(J_\epsilon).
    \end{align*}
    By \Cref{tv measure bound}, we have that $|Df|(J_\epsilon) = O_{x_0}(\epsilon)$, which gives the bound
    \[
    \|v^*_{x_0,\epsilon}
    -c_*\nabla g(0)\|_2
    \lesssim_{x_0}\epsilon^2.
    \]
    It remains to characterize $c_*$. Recall that
    $\nabla g(0)\neq0$. By
    \eqref{theta tilde simplification} and
    \Cref{lemma: surrogate problem},
    \[
    c_*=
    \frac{\epsilon}{\nu^2\|\nabla g(0)\|_2^2}
    \int_{\mathcal B(0,\epsilon)}
    (\nabla g(0)^\top x)f(g(0)+\nabla g(0)^\top x)p^*(x)\,dx.
    \]
    Since $f$ is differentiable at $g(0)$, we have
    \[
    f(g(0)+t)
    =f(g(0))+f'(g(0))t+o(|t|).
    \]
    Moreover, this expansion holds uniformly over
    $|t|\leq\|\nabla g(0)\|_2\epsilon$, with remainder bounded
    by $o(1)|t|$. By spherical symmetry,
    \[
    \int_{\mathcal B(0,\epsilon)}
    (\nabla g(0)^\top x)p^*(x)\,dx=0,
    \qquad \text{and}\qquad
    \int_{\mathcal B(0,\epsilon)}
    (\nabla g(0)^\top x)^2p^*(x)\,dx
    =\nu^2\|\nabla g(0)\|_2^2.
    \]
    Substituting into the expression for $c_*$ gives
    \[
    |c_*-\epsilon f'(g(0))|
    \leq\epsilon\,o(1)=o_{x_0}(\epsilon),
    \]
    establishing the first-order expansion.
\end{proof}

\begin{restatable}{prop}{}\label{velocity and intercept error C2}
    Suppose \Cref{assumption: lce and covariate density}
    holds, with the assumptions that $f \in C^2(I)$. Then, for $P_X$-almost every
    $x_0\in\operatorname{int}(\operatorname{supp}(P_X))$
    satisfying $p(x_0)>0$, $\nabla g(x_0)\neq0$,
    and for all sufficiently small $\epsilon>0$ we have
    \[
        \mu^*_{x_0,\epsilon}
        =\mu(0)+\frac{\epsilon^2}{2(d+2)}
        \left[f''(g(0))\|\nabla g(0)\|_2^2+f'(g(0))\operatorname{tr}(\nabla^2g(0))\right]
        +o(\epsilon^2)
    \]
    and $c_*=\epsilon f'(g(0))+o(\epsilon^2)$.
\end{restatable} 
\begin{proof}
For all sufficiently small
$\epsilon$ we have $J_\epsilon\subseteq I$, so $f$ is Lipschitz on
$J_\epsilon$ and
\[|Df|(J_\epsilon)
\leq \operatorname{Lip}(f|_{J_\epsilon})\,\mathcal L_1(J_\epsilon)
= O_{x_0}(\epsilon)\]
at every inference point with $p(x_0)>0$ and $\nabla g(x_0)\neq0$.
Moreover $f$ is differentiable at $g(0)$. Hence the quadratic
localization-bias bound and the first-order expansion of $c_*$ hold
without the almost-everywhere restriction, and
\Cref{lemma: constant of prop} sharpens the latter to
\[
c_* = \epsilon f'(g(0)) + o(\epsilon^2).
\]
Now we will address the intercept. Since $g\in C^2(U)$, a
second-order Taylor expansion gives
\[
t(x) = g(x)-g(0) = \nabla g(0)^\top x + \tfrac12 x^\top \nabla^2g(0) x + r_g(x),
\qquad
\sup_{x\in\mathcal B(0,\epsilon)}|r_g(x)| = o(\epsilon^2),
\]
and in particular
$\sup_{x\in\mathcal B(0,\epsilon)}|t(x)|\leq 2\|\nabla g(0)\|_2\epsilon$ for all
sufficiently small $\epsilon$. Likewise, since $f\in C^2(I)$,
\[
f(g(0)+t) = f(g(0)) + f'(g(0))\,t + \tfrac12 f''(g(0))\,t^2 + r_f(t),
\qquad
|r_f(t)| \leq \omega(|t|)\,t^2,
\]
with $\omega(s)\to0$ as $s\downarrow0$. Substituting $t=t(x)$ and
collecting terms, we obtain, uniformly over $x\in\mathcal B(0,\epsilon)$,
\[
f(g(x))-f(g(0))
= f'(g(0))\Big(a^\top x+\tfrac12 x^\top \nabla ^2g(0)x\Big)
 + \tfrac12 f''(g(0))\,(\nabla g(0)^\top x)^2 + o(\epsilon^2),
\]
where we used $f'(g(0))\,r_g(x) = o(\epsilon^2)$,
$t(x)^2 = (\nabla g(0)^\top x)^2 + O(\epsilon^3)$, and
\[|r_f(t(x))|\leq \omega(2\|\nabla g(0)\|_2\epsilon)\,4\|\nabla g(0)\|_2^2\epsilon^2 = o(\epsilon^2).\]
Integrating against $p^*(x)/A(\epsilon)$ over $\mathcal B(0,\epsilon)$,
the linear term vanishes by spherical symmetry, while
\[
\frac{1}{A(\epsilon)}\int_{\mathcal B(0,\epsilon)} xx^\top p^*(x)\,dx
= \frac{\nu^2}{A(\epsilon)}\,I_d
\]
gives
$A(\epsilon)^{-1}\int_{\mathcal B(0,\epsilon)} x^\top \nabla^2g(0)x\,p^*(x)\,dx
= \operatorname{tr}(\nabla ^2g(0))\,\nu^2/A(\epsilon)$
and
$A(\epsilon)^{-1}\int_{\mathcal B(0,\epsilon)} (\nabla g(0)^\top x)^2p^*(x)\,dx
= \|\nabla g(0)\|_2^2\,\nu^2/A(\epsilon)$.
Therefore
\[
\mu^*_{x_0,\epsilon}-\mu(0)
= \frac{\nu^2}{2A(\epsilon)}
  \Big[f''(g(0))\|\nabla g(0)\|_2^2
     + f'(g(0))\operatorname{tr}\big(\nabla^2 g(0)\big)\Big]
 + o(\epsilon^2).
\]
It remains to identify $\nu^2/A(\epsilon)$. Substituting $x=\epsilon z$,
\[
\frac{\nu^2}{A(\epsilon)}
= \epsilon^2\,
  \frac{\int_{\mathcal B(0,1)} z_1^2\,p^*(\epsilon z)\,dz}
       {\int_{\mathcal B(0,1)} p^*(\epsilon z)\,dz}.
\]
By continuity of $p^*$ at $0$ we have $p^*\leq 2p^*(0)$ on
$\mathcal B(0,\epsilon)$ for all sufficiently small $\epsilon$, so
dominated convergence gives
$\int_{\mathcal B(0,1)} z_1^2\,p^*(\epsilon z)\,dz
\to p^*(0)\int_{\mathcal B(0,1)} z_1^2\,dz
= p^*(0)\,\mathcal{L}^d(\mathcal B(0,1))/(d+2)$
by \Cref{prop: spherical cov}, and
$\int_{\mathcal B(0,1)} p^*(\epsilon z)\,dz
\to p^*(0)\,\mathcal{L}^d(\mathcal B(0,1))$.
Hence $\nu^2/A(\epsilon) = \epsilon^2/(d+2) + o(\epsilon^2)$, and
\[
\mu^*_{x_0,\epsilon}
= \mu(0)
 + \frac{\epsilon^2}{2(d+2)}
   \Big[f''(g(0))\|\nabla g(0)\|_2^2
      + f'(g(0))\operatorname{tr}\big(\nabla^2 g(0)\big)\Big]
 + o(\epsilon^2),
\]
as claimed. Since this argument uses neither $N_1$ nor $N_2$, it holds at
every inference point with $p(x_0)>0$ and $\nabla g(x_0)\neq0$.
\end{proof}

\begin{lemma}\label{tv measure bound}
    Let $J_\epsilon := [g(0) - \|\nabla g(0)\|_2\epsilon - L\epsilon^2, g(0) + \|\nabla g(0)\|_2\epsilon + L\epsilon^2]$ and let $|Df|$ be the total variation measure of the distributional derivative of $f.$ Then we have $|Df|(J_\epsilon) = O_{x_0}(\epsilon).$
\end{lemma}
\begin{proof}
    We will apply the Besicovitch derivation theorem
    \citep[Theorem 2.22]{ambrosio2000functions} with $\Omega = \mathbb R$, $\mu = \mathcal{L}^1$ and $\nu = |Df|$, the latter of which is a positive Radon measure by our assumption that $f \in BV_{\operatorname{loc}}(\mathbb R)$. Thus, the derivation theorem guarantees the existence of a Lebesgue-null set $N_1\subset\mathbb R$
    such that, for every $t\notin N_1$,
    \[
    |Df|([t-r,t+r])=O_t(r)
    \qquad\text{as }r\downarrow0.
    \]
    Applying this with $J_\epsilon$ proves the statement for any inference point satisfying the hypotheses. We will now show that this occurs $P_X$ almost surely. 
    
    By \citet[Theorem 3.28(c)]{ambrosio2000functions},
    the chosen BV representative of $f$ is differentiable
    outside a Lebesgue-null set $N_2\subset\mathbb R$.
    Let $N=N_1\cup N_2$, and let $\bar x \in \{x: \nabla g(x) \neq 0\}.$ Then there exists an index $j \in \{1, \dots, d\}$ such that $(\nabla g(\bar x))_j \neq 0.$ For such a $j$, the map \[\Phi(x) = (x_1,\ldots,x_{j-1},g(x), x_{j+1},\ldots,x_d)\]
    has an invertible Jacobian, since $\operatorname{det}(\nabla \Phi(x)) = \pm \partial_jg(\bar x) \neq 0.$ Now we can apply the inverse function theorem \citep[Theorem 9.24]{rudin1976principles} by taking $\mathbf{f} = \Phi$, $\mathbf{a} = \bar x$ and $E$ any open set containing $\mathbf{a}$ (since $g$ is $C^2$) to conclude the existence of an open    $U \ni \bar x$ and $V = \Phi(U)$ with $\Phi: U \to V$ as a bijection, and  $\Phi^{-1} \in C^1(V)$.
    We now claim that $\mathcal L^d(g^{-1}(N) \cap U) = 0$. First observe that for
    $x \in U$, we have $g(x) \in N$ if and only if the $j$-th coordinate of $\Phi(x)$
    lies in $N$ (by definition), so that
    \[
        g^{-1}(N) \cap U = \left\{x \in U : \Phi(x) \in V \cap S_N\right\} = \Phi^{-1}\big(V \cap S_N\big),
        \qquad
        S_N := \mathbb R^{j-1} \times N \times \mathbb R^{d-j}.
    \]
    Since $N$ is Lebesgue-null in $\mathbb R$, Fubini's theorem gives
    \[
        \mathcal L^d(V \cap S_N)
        = \int_{\mathbb R^d}\mathbbm{1}_{V\cap S_N}(x) \, dx =  \int_{\mathbb R^{d-1}} \int_{\mathbb R}
        \mathbbm 1_V(z_1, \ldots, z_{j-1}, t, z_{j+1}, \ldots, z_d)\,
        \mathbbm 1_N(t)\, dt\, dz = 0.
    \]
    Since $\Phi^{-1}$ is differentiable at every point of $V$ and
    $\mathcal L^d(V \cap S_N) = 0$, \citet[Lemma 7.25]{rudin1987real} gives
    $\mathcal L^d\big(g^{-1}(N) \cap U\big)
    = \mathcal L^d\big(\Phi^{-1}(V \cap S_N)\big) = 0$. Finally, $\{\nabla g \neq 0\}$ is an open subset of
    $\mathbb R^d$ and hence can be covered by countably many neighborhoods $U$ of the
    above form. Since a countable union of null sets is null, it follows that
    $\mathcal L^d\big(g^{-1}(N) \cap \{\nabla g \neq 0\}\big) = 0$. Since
    $P_X \ll \mathcal L^d$, we conclude that $g(x_0) \notin N$ for $P_X$-almost every
    regular inference point $x_0$. 
\end{proof}

\subsection{Proof of \Cref{thm: decomposition for theta}} \label{sec: proof of velocity decomp}

This proof will follow closely the argument from \citet{Newey_Ruud_2005}. Let's define the matrices
\begin{equation*}
    Q = \mathbb{E}_{p}\left[w\boldsymbol{X}\boldsymbol{X}^{\top}\right] \qquad \text{and} \qquad \hat{Q} = \frac{1}{n}\sum_{i = 1}^n\hat{w}_i\boldsymbol{X}_i\boldsymbol{X}_i^{\top}.
\end{equation*}
We have
\begin{align*}
    \hat{\theta}_{x_0, \epsilon} - \theta^*_{x_0, \epsilon} &= \hat{Q}^{-1}\left(\frac{1}{n}\sum_{i = 1}^n\hat{w}_i\boldsymbol{X}_iY_i\right) - \theta^*_{x_0, \epsilon} \\
    &= \hat{Q}^{-1}\left(\frac{1}{n}\sum_{i = 1}^n\hat{w}_i\boldsymbol{X}_iY_i - \hat{Q}\theta^*_{x_0, \epsilon}\right) \\
    &= \hat{Q}^{-1}\left(\frac{1}{n}\sum_{i = 1}^n\hat{w}_i\boldsymbol{X}_i\underbrace{\left(Y_i - \boldsymbol{X}_i^{\top}\theta^*_{x_0, \epsilon}\right)}_{u_i}\right) \\
    &= \hat{Q}^{-1}\left(\frac{1}{n}\sum_{i = 1}^n\hat{w}_i\boldsymbol{X}_iu_i\right).
\end{align*}
Now we'll add and subtract out a term with the true weights $w_i$ to obtain
\begin{align*}
    \hat{\theta}_{x_0, \epsilon} - \theta^*_{x_0, \epsilon} &= \hat{Q}^{-1}\left(\frac{1}{n}\sum_{i = 1}^nw_i\boldsymbol{X}_iu_i + \frac{1}{n}\sum_{i = 1}^n(\hat{w}_i - w_i)\boldsymbol{X}_iu_i\right) \\
    &= \hat{Q}^{-1}\Bigg(\frac{1}{n}\sum_{i = 1}^n\frac{\mathbbm{1}_{\{X_i \in \mathcal{B}(0, \epsilon)\}}p^*(X_i)\boldsymbol{X}_iu_i}{p(X_i)} \\
    &\hspace{40mm}+ \frac{1}{n}\sum_{i = 1}^n\left[\frac{1}{\hat{p}(X_i)} - \frac{1}{p(X_i)}\right]\mathbbm{1}_{\{X_i \in \mathcal{B}(0, \epsilon)\}}p^*(X_i)\boldsymbol{X}_iu_i\Bigg).
\end{align*}
Let $\mathcal Z_i=(X_i,Y_i)$ and
$\mathcal Z=(X,Y)$. Define
$a(\mathcal Z_i)=
\mathbbm{1}_{\{X_i\in\mathcal B(0,\epsilon)\}}
p^*(X_i)\boldsymbol X_i u_i$.
Then we can write the preceding expression as
\begin{align*}
    \hat{\theta}_{x_0, \epsilon} - \theta^*_{x_0, \epsilon} &= \hat{Q}^{-1}\left(\frac{1}{n}\sum_{i = 1}^n\frac{a(\mathcal Z_i)}{p(X_i)} + \frac{1}{n}\sum_{i = 1}^n\left[\frac{1}{\hat{p}(X_i)} - \frac{1}{p(X_i)}\right]a(\mathcal Z_i)\right).
\end{align*}
Applying a second-order Taylor expansion of
$t\mapsto t^{-1}$, with
$\Delta(X_i)=\hat p(X_i)-p(X_i)$, gives
\begin{align}
    \hat{\theta}_{x_0, \epsilon} - \theta^*_{x_0, \epsilon} &= \hat{Q}^{-1}\bigg(\frac{1}{n}\sum_{i = 1}^n\frac{a(\mathcal Z_i)}{p(X_i)} - \underbrace{\frac{1}{n}\sum_{i = 1}^n\frac{a(\mathcal Z_i)\Delta(X_i)}{p^2(X_i)}}_{R_1} + \underbrace{\frac{1}{n}\sum_{i = 1}^na(\mathcal Z_i) \frac{\Delta(X_i)^2}{\xi_i^3}}_{R_2}\bigg) \label{eq: expansion with remainder}
\end{align}
where $\xi_i$ is some value between $p(X_i)$ and $\hat{p}(X_i)$. Now note that $\mathbb{E}[a(\mathcal Z)/p(X)] = \mathbb{E}[w\boldsymbol{X}u] = 0$ by the population first order conditions. 
We write $\mathbb P F=\mathbb E[F(X,Y)]$
and $\mathbb P_nF=n^{-1}\sum_{i=1}^nF(X_i,Y_i)$,
allowing $F$ to depend on the full observation.
 Thus,
\begin{align*}
    \frac{1}{n}\sum_{i = 1}^n\frac{a(\mathcal Z_i)}{p(X_i)} &= \frac{1}{n}\sum_{i = 1}^n\frac{a(\mathcal Z_i)}{p(X_i)} - \mathbb{E}\left[\frac{a(\mathcal Z)}{p(X)}\right] = (\mathbb{P}_n -\mathbb{P})\left[\frac{a(\mathcal Z)}{p(X)}\right] = \frac{1}{n}\sum_{i = 1}^n w_i \boldsymbol{X}_i u_i.
\end{align*}
Now we'll consider the first remainder term. By \Cref{lemma: remainder term},
\[
    R_1 = \frac{1}{n}\sum_{i = 1}^n \frac{\mathbb{E}[a(\mathcal Z)\,|\,X=X_i]}{p(X_i)} + o_P\left(\sqrt{\frac{\epsilon^d}{n}}\right).
\]
For the second remainder term, \Cref{lemma: uniform convergence of kde} applies since $h\rightarrow0$ and $nh^d/\log n\rightarrow\infty$, and gives
\[
    \sup_{x\in \mathcal{B}(0, \epsilon)} \left|\Delta(x)\right| =O_P\left(\sqrt{\frac{\log n}{nh^d}} +h^s \right) =o_P(1).
\]
Thus,
\begin{align*}
    \|R_2\|_2 &= \left\|\mathbb{P}_n\left[\frac{a(\mathcal Z)}{p(X)}p(X)\frac{\Delta(X)^2}{\xi^3}\right]\right\|_2\\
    &\leq \mathbb{P}_n[\left\|w\boldsymbol{X}u\right\|_2 \xi_i^{-3}]\sup_{\mathcal{B}(0, \epsilon)}|p(x)|\cdot \sup_{\mathcal{B}(0, \epsilon)}|\Delta(x)|^2.
\end{align*}
Since $p\geq m>0$ on $\mathcal{B}(0,\epsilon)$ and $\sup_{x\in \mathcal{B}(0,\epsilon)}|\Delta(x)|=o_P(1)$, we know that $\hat{p}(x)>m/2$ uniformly on $\mathcal{B}(0,\epsilon)$ with probability approaching $1$. It follows that $\xi_i>m/2$ with probability approaching $1$ as well. Moreover, the standing moment assumptions and local boundedness of the density ratio imply $\mathbb{E}\|w\boldsymbol{X}u\|_2=O(\epsilon^d)$, and hence $\mathbb{P}_n\|w\boldsymbol{X}u\|_2=O_P(\epsilon^d)$. Therefore,
\begin{align*}
    \|R_2\|_2 &\leq \mathbb{P}_n\left\|w\boldsymbol{X}u\right\|_2\cdot (8/m^3)\cdot \sup_{\mathcal{B}(0, \epsilon)}|p(x)|\cdot \sup_{\mathcal{B}(0, \epsilon)}|\Delta(x)|^2 \\
    &= O_P\left(\epsilon^d\left\{\frac{\log n}{nh^d} +h^{2s}\right\}\right) \\
    &= o_P\left(\sqrt{\frac{\epsilon^d}{n}}\right).
\end{align*}
The final line follows from the rate assumptions. In particular, dividing by $\sqrt{\epsilon^d/n}$, the two contributions become
\[
    \sqrt{n\epsilon^d}\,\frac{\log n}{nh^d}\rightarrow0
    \qquad\text{and}\qquad
    \sqrt{n\epsilon^d}\,h^{2s}
    = \big(\sqrt{n\epsilon^d}\,h^s\big)\,h^s\rightarrow0,
\]
the first by assumption and the second because $\sqrt{n\epsilon^d}\,h^s\rightarrow0$ and $h\rightarrow0$. Combining the bounds for $R_1$ and $R_2$ in \eqref{eq: expansion with remainder} gives
\begin{align*}
    \hat{\theta}_{x_0, \epsilon} - \theta^*_{x_0, \epsilon}
    &= \hat{Q}^{-1}\left(\frac{1}{n}\sum_{i = 1}^n\frac{a(\mathcal Z_i) - \mathbb{E}[a(\mathcal Z)\,|\,X=X_i]}{p(X_i)}
    + o_P\left(\sqrt{\frac{\epsilon^d}{n}}\right)\right).
\end{align*}
Now we will introduce $Q$ as follows,
\begin{equation*}
    \hat{\theta}_{x_0, \epsilon} - \theta^*_{x_0, \epsilon} = Q^{-1}\psi_n + (\hat{Q}^{-1} - Q^{-1})\psi_n
\end{equation*}
with
\begin{align*}
    \psi_n &= \frac{1}{n}\sum_{i = 1}^n\frac{a(\mathcal Z_i) -\mathbb{E}[a(\mathcal Z)\,|\,X=X_i]}{p(X_i)}
    + o_P\left(\sqrt{\frac{\epsilon^d}{n}}\right) \\
    &= O_P\left(\sqrt{\frac{\epsilon^d}{n}}\right).
\end{align*}
Indeed, the summands in the first term are mean zero and the standing moment and density assumptions imply
$\mathbb{E}\|(a(\mathcal Z)-\mathbb{E}[a(\mathcal Z)\mid X])/p(X)\|_2^2=O(\epsilon^d)$. By \Cref{lemma: Q convergence}, which applies since $h\rightarrow0$, $nh^d/\log n\rightarrow\infty$ and $n\epsilon^d\rightarrow\infty$, we know that
\[
    \|\hat{Q}^{-1} - Q^{-1}\|_{\operatorname{op}}
    =
    O_P\left(\frac{1}{\epsilon^{3d/2}\sqrt{n}} + \frac{1}{n\epsilon^{2d}}\right) + o_P\left(\frac{1}{\epsilon^d}\right)
    =
    o_P(\epsilon^{-d}),
\]
where the last equality uses $n\epsilon^d\rightarrow\infty$. It follows that
\[
    \|(\hat Q^{-1}-Q^{-1})\psi_n\|_2
    =
    o_P(\epsilon^{-d})O_P\left(\sqrt{\frac{\epsilon^d}{n}}\right)
    =
    o_P\left(\frac{1}{\sqrt{n\epsilon^d}}\right).
\]
Furthermore, $\|Q^{-1}\|_{\operatorname{op}} \lesssim \epsilon^{-d}$ as derived in \Cref{lemma: Q op norm}, so
\[
    Q^{-1}o_P\left(\sqrt{\frac{\epsilon^d}{n}}\right)
    =
    o_P\left(\frac{1}{\sqrt{n\epsilon^d}}\right).
\]
Finally, observe that
\[
    \frac{a(\mathcal Z_i)-\mathbb{E}[a(\mathcal Z)\mid X=X_i]}{p(X_i)}
    =
    w_i\boldsymbol{X}_i\left(u_i-\mathbb{E}[u_i\mid \boldsymbol{X}_i]\right).
\]
Putting everything together, we obtain
\begin{equation}
    \hat{\theta}_{x_0, \epsilon} - \theta^*_{x_0, \epsilon}
    =
    \frac{1}{n}Q^{-1}\sum_{i = 1}^nw_i\boldsymbol{X}_i(u_i - \mathbb{E}[u_i\,|\,\boldsymbol{X}_i])
    + o_P\left(\frac{1}{\sqrt{n\epsilon^{d}}}\right),
    \label{eq: theta decomposition final}
\end{equation}
which is the first claim. Note that $u_i-\mathbb{E}[u_i\mid\boldsymbol X_i] = Y_i - f(g(X_i)) = \varepsilon_i$. Since $\hat{\mu}_{x_0,\epsilon}-\mu^*_{x_0,\epsilon}=e_1^\top(\hat{\theta}_{x_0,\epsilon}-\theta^*_{x_0,\epsilon})$ and $\hat{v}_{x_0,\epsilon}-v^*_{x_0,\epsilon}=V(\hat{\theta}_{x_0,\epsilon}-\theta^*_{x_0,\epsilon})$, the intercept and velocity expansions follow from \eqref{eq: theta decomposition final} by left-multiplying by $e_1^\top$ and by $V$, respectively.
\qed{}

\subsection{Proof of \Cref{corr: central limit for v}} \label{sec: proof of central limit for v}
Let $u = Y - \boldsymbol{X}^{\top}\theta_{x_0, \epsilon}^*$ and $\Lambda = \operatorname{Cov}\left(w\boldsymbol{X}(u - \mathbb{E}[u\,|\,\boldsymbol{X}])\right)$. Since $\mathbb{E}[u\mid\boldsymbol X]=f(g(X))-\boldsymbol X^\top\theta^*_{x_0,\epsilon}$, we have $u-\mathbb{E}[u\mid\boldsymbol X]=\varepsilon$, and therefore, as $\mathbb{E}[\varepsilon\mid X]=0$ and $\mathbb{E}[\varepsilon^2\mid X]=\sigma^2(X)$, with 
\[
    \Lambda = \mathbb{E}\left[w^2\sigma^2(X)\boldsymbol{X}\boldsymbol{X}^{\top}\right]
    \qquad\text{and}\qquad
    \Sigma_\epsilon = Q^{-1}\Lambda Q^{-1}.
\]
From \Cref{thm: decomposition for theta}, in both regimes,
\begin{align*}
    \sqrt{n\epsilon^{d}}\left(\hat{\theta}_{x_0, \epsilon} - \theta^*_{x_0, \epsilon}\right) = \sqrt{\frac{\epsilon^{d}}{n}}\,Q^{-1}\sum_{i = 1}^nw_i\boldsymbol{X}_i\varepsilon_i
    + o_P\left(1\right).
\end{align*}
We will apply the Lyapunov central limit theorem of \Cref{thm: lyapunov clt} to the triangular array
\[U_{n, i} = \sqrt{\frac{\epsilon^{d}}{n}}\,Q^{-1}w_i\boldsymbol{X}_i\varepsilon_i, \qquad 1\leq i\leq n,\]
whose entries are independent and identically distributed for each fixed $n$, and have mean zero since $\mathbb E[\varepsilon\mid X]=0$. Let
\[
    V_n = \sum_{i=1}^n\operatorname{Cov}(U_{n,i}) = \epsilon^d Q^{-1}\Lambda Q^{-1} = \epsilon^d\Sigma_\epsilon
    \qquad\text{and}\qquad
    v_n^2 = \lambda_{\min}(V_n).
\]

\medskip
\noindent\textit{Lyapunov condition.} We verify the condition of \Cref{thm: lyapunov clt} with $\delta=1$, assuming for the moment that $v_n$ is bounded away from zero for all sufficiently large $n$; this is established separately in each regime below. We have
\begin{align*}
    \frac{1}{v_n^3}\sum_{i=1}^n\mathbb{E}\|U_{n,i}\|_2^3
    &=
    \frac{\epsilon^{3d/2}}{v_n^3\sqrt{n}}
    \mathbb{E}\left\|Q^{-1}w\boldsymbol{X}\varepsilon\right\|_2^3
    \leq
    \frac{\epsilon^{3d/2}}{v_n^3\sqrt{n}}
    \|Q^{-1}\|_{\operatorname{op}}^3\,
    \mathbb{E}\left\|w\boldsymbol{X}\varepsilon\right\|_2^3.
\end{align*}
By \Cref{lemma: Q op norm}, $\|Q^{-1}\|_{\operatorname{op}}\lesssim\epsilon^{-d}$ (for fixed $\epsilon$ this says only that $\|Q^{-1}\|_{\operatorname{op}}$ is a finite constant, which also follows from \Cref{lemma: pd}). Moreover, using $\|\boldsymbol{X}\|_2\leq\sqrt{2}$ on $\mathcal{B}(0,\epsilon)$ and $w\leq w_{\max}:=\sup_{x\in\mathcal B(0,\epsilon_0)}p^*(x)/p(x)<\infty$,
\begin{align*}
    \mathbb{E}\|w\boldsymbol{X}\varepsilon\|_2^3
    \leq 2^{3/2}w_{\max}^2\,\mathbb{E}\left[w|\varepsilon|^3\right]
    = 2^{3/2}w_{\max}^2\int_{\mathcal{B}(0,\epsilon)}
    p^*(x)\,\mathbb{E}[|\varepsilon|^3\mid X=x]\,dx
    \lesssim \epsilon^d,
\end{align*}
where the last step uses $\epsilon\leq\epsilon_0\leq\delta$ and the local third-moment bound in \Cref{assumption: lce and covariate density}. It follows that
\begin{align*}
    \frac{1}{v_n^3}\sum_{i=1}^n\mathbb{E}\|U_{n,i}\|_2^3
    \lesssim
    \frac{\epsilon^{3d/2}}{v_n^3\sqrt{n}}\,\epsilon^{-3d}\epsilon^d
    =
    \frac{1}{v_n^3\sqrt{n\epsilon^d}}
    =o(1),
\end{align*}
since $n\epsilon^d\rightarrow\infty$ and $v_n$ is bounded away from zero. Thus \Cref{thm: lyapunov clt} gives
\begin{equation}
    V_n^{-1/2}\sum_{i=1}^nU_{n,i}\overset{d}{\rightarrow}N(0,I_{d+1}).
    \label{eq: standardized clt}
\end{equation}

\medskip
\noindent\textit{Part 1: fixed $\epsilon$.} Here $V_n=\epsilon^d\Sigma_\epsilon$ does not depend on $n$, so it suffices to show that $\Sigma_\epsilon\succ0$, i.e. that $\Lambda\succ0$. By continuity of $p^*$ and $\sigma^2$ at $0$, with $p^*(0)>0$ and $\sigma^2(0)>0$, there exists $\epsilon_1\in(0,\epsilon]$ such that $p^*\geq p^*(0)/2$ and $\sigma^2\geq\sigma^2(0)/2$ on $\mathcal B(0,\epsilon_1)$. Hence, for every $v\in\mathbb R^{d+1}\setminus\{0\}$,
\[
    v^\top\Lambda v
    = \int_{\mathcal B(0,\epsilon)}\frac{p^*(x)^2}{p(x)}\,\sigma^2(x)\,(\boldsymbol x^\top v)^2\,dx
    \geq \frac{p^*(0)^2\,\sigma^2(0)}{8\,\|p\|_\infty}\int_{\mathcal B(0,\epsilon_1)}(\boldsymbol x^\top v)^2\,dx
    >0,
\]
and the final inequality holds because $x \mapsto \boldsymbol x^\top v
= v_1 + v_{2:d+1}^\top x/\epsilon$ is a nonzero affine (hence real analytic)
function, and the zero set of a real analytic function that is not identically
zero has Lebesgue measure zero \citep{mityagin2020zero}. Thus $v_n^2=\lambda_{\min}(\epsilon^d\Sigma_\epsilon)$ is a positive constant, the Lyapunov condition holds, and \eqref{eq: standardized clt} together with
\[
    \sqrt n\left(\hat\theta_{x_0,\epsilon}-\theta^*_{x_0,\epsilon}\right)
    = \epsilon^{-d/2}\sum_{i=1}^nU_{n,i}+o_P(1)
    = \epsilon^{-d/2}V_n^{1/2}\left(V_n^{-1/2}\sum_{i=1}^nU_{n,i}\right)+o_P(1)
\]
yields $\sqrt n(\hat\theta_{x_0,\epsilon}-\theta^*_{x_0,\epsilon})\overset{d}{\rightarrow}N(0,\epsilon^{-d}V_n)=N(0,\Sigma_\epsilon)$ by Slutsky's theorem.

\medskip

\noindent\textit{Part 2: shrinking $\epsilon$.} The hypotheses of \Cref{lemma: asymptotic var bound} hold under those of \Cref{thm: decomposition for theta}. In particular, $\sigma^2$ is bounded on $\mathcal B(0,\epsilon_0)$ because $\epsilon_0\leq\delta$ and $\sigma^2(x)\leq\mathbb E[|\varepsilon|^3\mid X=x]^{2/3}$ by Lyapunov's inequality. Hence
\[
    V_n=\epsilon^d\Sigma_\epsilon\rightarrow\Sigma_\infty:=\operatorname{diag}(\sigma_\mu^2,\Sigma_0),
\]
which is positive definite since $\sigma^2(0)>0$ and $p(0)>0$. Therefore $v_n^2\rightarrow\lambda_{\min}(\Sigma_\infty)>0$, so $v_n$ is bounded away from zero for all large $n$ and the Lyapunov condition holds. The convergence of $V_n \to \Sigma_\infty$ then implies $V_n^{1/2}\rightarrow\Sigma_\infty^{1/2}$ (by \citet{bhatia2013matrix}[Corollary 2] with $p = \infty$, yielding operator norm convergence) and Slutsky's theorem applied to
\[
    \sqrt{n\epsilon^d}\left(\hat\theta_{x_0,\epsilon}-\theta^*_{x_0,\epsilon}\right)
    = V_n^{1/2}\left(V_n^{-1/2}\sum_{i=1}^nU_{n,i}\right)+o_P(1)
\]
gives $\sqrt{n\epsilon^d}(\hat\theta_{x_0,\epsilon}-\theta^*_{x_0,\epsilon})\overset{d}{\rightarrow}N(0,\Sigma_\infty)$. Left-multiplying by $e_1^\top$ and by $V=\begin{bmatrix}0 & I_d\end{bmatrix}$ gives, respectively,
\[
    \sqrt{n\epsilon^d}\left(\hat\mu_{x_0,\epsilon}-\mu^*_{x_0,\epsilon}\right)\overset{d}{\rightarrow}N(0,\sigma_\mu^2)
    \qquad\text{and}\qquad
    \sqrt{n\epsilon^d}\left(\hat v_{x_0,\epsilon}-v^*_{x_0,\epsilon}\right)\overset{d}{\rightarrow}N(0,\Sigma_0).
\]
For the centered statements, fix an inference point at which the conclusions of \Cref{thm: population prop} hold. Again, this is $P_X$-almost every inference point under $f\in BV_{\operatorname{loc}}(\mathbb R)$, and every inference point with $p(x_0)>0$ and $\nabla g(x_0)\neq0$ under $f\in C^2(I)$. By \Cref{thm: population prop}, $\mu^*_{x_0,\epsilon}-\mu(0)=o_{x_0}(\epsilon)$ and $\|v^*_{x_0,\epsilon}-c_*\nabla g(0)\|_2\lesssim_{x_0}\epsilon^2$. Hence, if $n\epsilon^{d+2}=O(1)$,
\[
    \sqrt{n\epsilon^d}\,\left|\mu^*_{x_0,\epsilon}-\mu(0)\right|
    = \sqrt{n\epsilon^{d+2}}\cdot o_{x_0}(1)\rightarrow0,
\]
and if $n\epsilon^{d+4}\rightarrow0$,
\[
    \sqrt{n\epsilon^d}\,
    \|v^*_{x_0,\epsilon}-c_*\nabla g(0)\|_2
    \lesssim_{x_0}\sqrt{n\epsilon^{d+4}}\rightarrow0,
\]
so the two centered limits follow from the uncentered ones by Slutsky's theorem. Finally, if $f\in C^2(I)$, then \Cref{thm: population prop} gives $\mu^*_{x_0,\epsilon}-\mu(0)=O(\epsilon^2)$ and $c_*=\epsilon f'(g(0))+o(\epsilon^2)$, so that under $n\epsilon^{d+4}\rightarrow0$ alone
\[
    \sqrt{n\epsilon^d}\,\left|\mu^*_{x_0,\epsilon}-\mu(0)\right|\lesssim\sqrt{n\epsilon^{d+4}}\rightarrow0
    \qquad\text{and}\qquad
    \sqrt{n\epsilon^d}\,\left|c_*-\epsilon f'(g(0))\right|\,\|\nabla g(0)\|_2 = o\left(\sqrt{n\epsilon^{d+4}}\right)\rightarrow0,
\]
which gives the intercept statement under $n\epsilon^{d+4}\rightarrow0$ and allows $c_*$ to be replaced by $\epsilon f'(g(0))$ in the velocity statement. \qed{}

\begin{restatable}[Bandwidth conditions in the two regimes]{rmk}{} \label{rmk: bandwidth regimes}
    The conditions on $h$ in \Cref{thm: decomposition for theta} say that the bias and the variance of the KDE are both negligible at the scale $1/\sqrt{n\epsilon^d}$ of the leading term. Writing $r_n=\sqrt{n\epsilon^d}$, they define the window
    \[
        \left(\frac{r_n\log n}{n}\right)^{1/d}\ll h\ll r_n^{-1/s}.
    \]
    For $\epsilon\asymp n^{-1/(d+4)}$ (\Cref{thm: velocity risk}) this reads \[n^{-(d+2)/(d(d+4))}(\log n)^{1/d}\ll h\ll n^{-2/(s(d+4))},\] which is nonempty since $s>2>2d/(d+2)$. For $\epsilon\asymp n^{-1/(d+2)}$ (\Cref{thm: regression risk}) it reads \[n^{-(d+1)/(d(d+2))}(\log n)^{1/d}\ll h\ll n^{-1/(s(d+2))},\] which is always nonempty. When $\epsilon$ is fixed it reads
    \[
        n^{-1/(2d)}(\log n)^{1/d}\ll h\ll n^{-1/(2s)},
    \]
    so that $\hat p$ must be undersmoothed relative to the MSE-optimal bandwidth, and the window is nonempty precisely when $s>d$.
\end{restatable}

\subsection{Proof of \Cref{rem: llr variance}}\label{proof of llr variance remark}

Only the displayed identity requires justification. Split $\Sigma^{\mathrm{LL}}_{\epsilon}$ into a noise term and an approximation-error term,
\begin{equation}\label{eq: llr variance split}
    \Sigma^{\mathrm{LL}}_{\epsilon}
    = \underbrace{\big(Q^{\mathrm{LL}}\big)^{-1}\,\mathbb{E}\!\left[K^2\sigma^2(X)\,\boldsymbol{X}\boldsymbol{X}^{\top}\right]\big(Q^{\mathrm{LL}}\big)^{-1}}_{N}
    \;+\;
    \underbrace{\big(Q^{\mathrm{LL}}\big)^{-1}\,\mathbb{E}\!\left[K^2 r^{\mathrm{LL}}(X)^2\,\boldsymbol{X}\boldsymbol{X}^{\top}\right]\big(Q^{\mathrm{LL}}\big)^{-1}}_{P},
\end{equation}
so that $\Sigma^{\mathrm{LL}}_{\epsilon} - \Sigma_{\epsilon} = (N - \Sigma_{\epsilon}) + P$. On $\mathcal{B}(0,\epsilon)$ we have $w = K/p_0$, which gives three identities. First, the weighted least squares problems defining $\theta^*_{x_0,\epsilon}$ and $\theta^{\mathrm{LL}}_{x_0,\epsilon}$ differ by the constant factor $p_0$ and therefore share their minimizer, so $r^{\mathrm{LL}} = r$. Second, $Q^{\mathrm{LL}} = \mathbb{E}[K\boldsymbol{X}\boldsymbol{X}^{\top}] = p_0\,\mathbb{E}[w\boldsymbol{X}\boldsymbol{X}^{\top}] = p_0\,Q$. Third, $K^2 = p_0^2 w^2$. Substituting into \eqref{eq: llr variance split},
\[
    N = (p_0 Q)^{-1}\,p_0^2\,\mathbb{E}\!\left[w^2\sigma^2(X)\boldsymbol{X}\boldsymbol{X}^{\top}\right](p_0 Q)^{-1}
      = Q^{-1}\,\mathbb{E}\!\left[w^2\sigma^2(X)\boldsymbol{X}\boldsymbol{X}^{\top}\right]Q^{-1}
      = \Sigma_{\epsilon},
\]
and in the same way $P = Q^{-1}\,\mathbb{E}\!\left[w^2 r(X)^2\boldsymbol{X}\boldsymbol{X}^{\top}\right]Q^{-1}$. Hence $\Sigma^{\mathrm{LL}}_{\epsilon} - \Sigma_{\epsilon} = P$, which is positive semidefinite. Without the constant density hypothesis, \eqref{eq: llr variance split} still gives $\Sigma^{\mathrm{LL}}_{\epsilon} - \Sigma_{\epsilon} = P + (N - \Sigma_{\epsilon})$. Write $p = p(0)(1+O(\epsilon))$ on $\mathcal{B}(0,\epsilon)$, which gives
\[\Sigma^{\operatorname{LL}}_\epsilon - \Sigma_\epsilon = P + O(\epsilon^2)\Sigma_\epsilon.\]
The second term is indefinite with asymptotic magnitude $O(\epsilon^2)$ which dominates the positive semidefinite term $P = O(\epsilon^4)\,\Sigma_{\epsilon}$ since $r^{\mathrm{LL}} = O(\epsilon^2)$ on the ball.
\qed{}

\subsection{Proof of \Cref{corollary: population prop curve}}
\label{proof of population prop curve}

\begin{proof}
We verify that \Cref{thm: population prop} applies with $g=\Pi$, and then evaluate its conclusions. Fix $x_0\in R$ and write $s_0=\Pi(x_0)$. By \Cref{lemma: IFT degen closed}, $R$ is open and $\Pi$ is $C^2$ on $R$, so $g=\Pi$ belongs to $C^2(U)$ for the open neighborhood $U=R$ of $x_0$. By \Cref{prop: proj gradient},
\[
\nabla\Pi(x_0)=\frac{\gamma'(s_0)}{\ell_0},
\qquad
\ell_0=\mathfrak{s}^2-\langle\gamma''(s_0),x_0-\gamma(s_0)\rangle,
\]
and $\ell_0\neq0$ by the definition of $R$ in \eqref{R}, so $\nabla\Pi(x_0)\neq0$ since $\gamma$ is regular. Moreover, $\Pi(R)\subseteq\operatorname{Int}(I)$ by the definition of $R$, so the regularity of $f$ is only required on the open interval $\operatorname{Int}(I)\supseteq g(U)$. In particular, all arguments in the proof of \Cref{thm: population prop} involving $f$ take place in a neighborhood of $g(x_0)=s_0\in\operatorname{Int}(I)$, so $f\in BV_{\mathrm{loc}}(\operatorname{Int}(I))$ suffices in place of $f\in BV_{\mathrm{loc}}(\mathbb R)$. Finally, $P_X$-almost every $x_0\in R$ satisfies $p(x_0)>0$ and lies in $\operatorname{int}(\operatorname{supp}(P_X))$, because $p$ is continuous and $P_X\ll\mathcal{L}^d$.

One point requires care. The almost-everywhere qualifier in \Cref{thm: population prop} is obtained in \Cref{tv measure bound} by pulling back a Lebesgue-null set $N\subset\mathbb R$, determined by $f$ alone, through $g$ on the regular set $\{\nabla g\neq0\}$. That argument uses only that $g$ is $C^1$ with nonvanishing gradient on an open subset of $\mathbb R^d$, together with $P_X\ll\mathcal{L}^d$. The same argument therefore applies verbatim with $\{\nabla g\neq0\}$ replaced by the open set $R$, on which $\Pi$ is $C^2$ with $\nabla\Pi\neq0$. Hence $P_X(\Pi^{-1}(N)\cap R)=0$, and the conclusions of \Cref{thm: population prop} hold for $P_X$-almost every $x_0\in R$.

The intercept conclusion is now the intercept conclusion of \Cref{thm: population prop} with $\mu(0)=f(\Pi(x_0))=f(s_0)$. For the velocity, \Cref{thm: population prop} yields a scalar $a_\epsilon\in\mathbb R$ with
\[
\|v^*_{x_0,\epsilon}-a_\epsilon\nabla\Pi(x_0)\|_2\lesssim_{x_0}\epsilon^2,
\qquad
a_\epsilon=\epsilon f'(s_0)+o_{x_0}(\epsilon).
\]
Substituting the expression for $\nabla\Pi(x_0)$ and setting $c_*=a_\epsilon/\ell_0$ gives the velocity conclusion.

When $f\in C^2(\operatorname{Int}(I))$, \Cref{thm: population prop} states that both conclusions hold at every inference point with $p(x_0)>0$ and $\nabla g(x_0)\neq0$, hence at every $x_0\in R$ with $p(x_0)>0$, and that $a_\epsilon=\epsilon f'(s_0)+o(\epsilon^2)$ and
\[
\mu^*_{x_0,\epsilon}
=f(s_0)+\frac{\epsilon^2}{2(d+2)}
\left[f''(s_0)\,\|\nabla\Pi(x_0)\|_2^2+f'(s_0)\operatorname{tr}\big(\nabla^2\Pi(x_0)\big)\right]+o(\epsilon^2).
\]
It remains to evaluate the two quantities. From the gradient formula, $\|\nabla\Pi(x_0)\|_2^2=\mathfrak s^2/\ell_0^{2}$. For the Hessian, \Cref{prop: proj hessian} with $\ell(x_0)=-\ell_0$ gives
\[
\operatorname{tr}\big(\nabla^2\Pi(x_0)\big)
=\frac{2\langle\gamma''(s_0),\gamma'(s_0)\rangle}{\ell_0^{2}}
+\frac{\mathfrak s^2\,\langle\gamma^{(3)}(s_0),x_0-\gamma(s_0)\rangle}{\ell_0^{3}}
=\frac{\mathfrak s^2\,\langle\gamma^{(3)}(s_0),x_0-\gamma(s_0)\rangle}{\ell_0^{3}},
\]
since $\langle\gamma''(t),\gamma'(t)\rangle=\tfrac12\tfrac{d}{dt}\|\gamma'(t)\|_2^2\equiv0$ for the constant-speed parametrization. Substituting both expressions into the display above gives the stated expansion of $\mu^*_{x_0,\epsilon}$, and $c_*=a_\epsilon/\ell_0$ gives the stated expansion of $c_*$.
\end{proof}

\section{Supplementary results}

\subsection{Supplementary results for \Cref{thm: differentiability of pi}}

For all proofs in this section, we will assume that \Cref{assumption: the curve gamma} holds for the $x$-marginal $P_X$ and $\gamma$. As a reminder, this assumption guarantees that for all $x \in \Omega$,  $\Pi$ projects to the interior of the parameter space of $\gamma$. This, coupled with the fact that $\gamma$ is $C^3$, allows us to differentiate $\gamma$ at all projected points in the support of $p$.

\begin{restatable}[Medial axis has zero Lebesgue measure]{lemma}{} \label{lemma: singular set measure}
    Let
    \[
        M(\Gamma)
        =
        \left\{
        x \in \mathbb{R}^d :
        \left|\arg\min_{t\in I}\|x-\gamma(t)\|_2\right| > 1
        \right\}
    \]
    denote the medial axis of $\Gamma$. Then $\mathcal{L}\big(M(\Gamma)\big)=0,$ where $\mathcal{L}$ denotes the $d$-dimensional Lebesgue measure on $\mathbb{R}^d$.
\end{restatable}

\begin{proof}[Proof of \Cref{lemma: singular set measure}]
Let
\[
    D(x) = d^2(x,\Gamma)
    =
    \inf_{t\in I}\|x-\gamma(t)\|_2^2.
\]
Since $\Gamma$ is compact, it is closed, and the map $x\mapsto d(x,\Gamma)$ is $1$-Lipschitz. It follows that $D$ is locally Lipschitz on $\mathbb{R}^d$. Thus, by Rademacher's theorem, $D$ is differentiable at $\mathcal{L}^d$-almost every $x\in\mathbb{R}^d$.

We will show that
\[
    M(\Gamma)
    \subseteq
    \left\{x\in\mathbb{R}^d:D\text{ is not differentiable at }x\right\}.
\]
Suppose, toward a contradiction, that $x\in M(\Gamma)$ and $D$ is differentiable at $x$. Let $p\in\Gamma$ be any nearest point to $x$, so that $\|x-p\|_2^2=D(x).$ Then for any $h\in\mathbb{R}^d$, the definition of $D$ gives
\begin{align*}
    D(x+h)
    &\leq
    \|x+h-p\|_2^2 =
    D(x)+2\langle x-p,h\rangle+\|h\|_2^2.
\end{align*}
On the other hand, differentiability of $D$ at $x$ implies
\[
    D(x+h)
    =
    D(x)+\langle\nabla D(x),h\rangle+o(\|h\|_2).
\]
Combining the two displays yields
\[
    \langle\nabla D(x)-2(x-p),h\rangle
    \leq
    \|h\|_2^2+o(\|h\|_2).
\]
Fixing $v\in\mathbb{R}^d$, taking $h=tv$ and then $h=-tv$ for $t\downarrow0$ gives
\[
    \langle\nabla D(x)-2(x-p),v\rangle=0.
\]
Since $v$ was arbitrary,
\[
    \nabla D(x)=2(x-p).
\]
This identity must hold for every nearest point $p$ to $x$. Thus, if $p,q\in\Gamma$ are both nearest points to $x$, then
\[
    2(x-p)=\nabla D(x)=2(x-q),
\]
which implies $p=q$. This contradicts $x\in M(\Gamma)$. It follows that $M(\Gamma)$ is contained in the set of nondifferentiability points of the locally Lipschitz function $D$. By Rademacher's theorem, this latter set has Lebesgue measure zero. Hence $\mathcal{L}\big(M(\Gamma)\big)=0$ as desired.
\end{proof}

\begin{restatable}[IFT degeneracy]{lemma}{} \label{lemma: IFT degen}
    Let
    \begin{equation} 
    B= \{x \in \Omega : \exists \,t \in \arg\inf_t \|x-\gamma(t)\|_2 \text{ such that } \mathfrak s^2 - \langle\gamma''(t), x-\gamma(t)\rangle = 0 \} \label{eq: ift degeneracy set}
    \end{equation}
    and let $\mathcal{L}^d$ denote the $d$-dimensional Lebesgue measure on $\mathbb{R}^d$. Then $\mathcal{L}^d(B) = 0.$
\end{restatable}
\noindent 
We prove the result by showing that $B$ is contained in a countable union of images of locally $C^1$
maps from subsets of $\mathbb R^{d-1}$ into
$\mathbb R^d$, each of which has zero
$d$-dimensional Lebesgue measure.

\begin{proof}[Proof of \Cref{lemma: IFT degen}] Let $N\gamma = \{(p, v) : p \in \gamma((0, 1)), v \in (T_p \gamma)^{\bot}\}$ denote the normal bundle of the interior of the curve $\gamma$, where $T_p\gamma$ denotes the tangent space of $\gamma$ at point $p$. Observe that $N\gamma$ forms a $d$-dimensional boundaryless manifold -- explicitly, one has the canonical local chart $(t,w) \rightarrow (\gamma(t), w_1n_1(t) + \dots+w_{d-1}n_{d-1}(t))$ where $\{n_i(t)\}_{i\in [d-1]}$ form an orthonormal basis of $N_p\gamma$. Since $\gamma\in C^3$ is regular, the normal frame
can be chosen locally $C^2$. Consequently,
$N\gamma$ is a $C^2$ manifold and $\Phi$ is $C^1$.

Now let $B'= \{(p,v) \in N\gamma : \langle v,  \gamma''(\gamma^{-1}(p))\rangle = \mathfrak s^2\}$, and let $\Phi:N\gamma \rightarrow \mathbb{R}$ be the function
\[\Phi(p,v) = \langle v, \gamma''(\gamma^{-1}(p))\rangle - \mathfrak s^2.\]
By \Cref{assumption: the curve gamma}, we know that $\arg \min_t\|x - \gamma(t)\|_2 \subseteq \text{Int}(I)$. This implies (by first order optimality conditions) that $x' - \gamma(t') \in N_{\gamma(t')}\gamma$. Now observe that for all $x' \in B$, there exists a $t' \in \arg\min_t \|x-\gamma(t)\|_2$ such that \[\langle \underbrace{x' - \gamma(t')}_{v'}, \gamma''(t')\rangle = \mathfrak s^2\] where $v' \in N_{\gamma(t')}\gamma$. Thus $x' \in \exp(B')$ where $\exp(p,v) = p + v$. It follows that $B\subset \exp(B')$.

We will now show that $0$ is a regular value of $\Phi$, allowing us to conclude that the level set $\Phi^{-1}(0) = B'$ is an embedded submanifold of $N\gamma$ of codimension 1. To show that $0$ is a regular value of $\Phi$, we will need to show that the differential of $\Phi$, $d\Phi$, is surjective for all $(p,v)$ such that $\Phi(p,v) = 0$. Since $d\Phi_{(p,v)}: T_{(p,v)}N\gamma \rightarrow T_{\Phi(p,v)}\mathbb{R} \simeq\mathbb{R}$, one just needs to show that the differential is not the zero map, as in this case the output spans all of $\mathbb{R}$. To prove this, observe that $\Phi$ can be expressed in local coordinates as
\[\Phi(t,w) = \sum_{i = 1}^{d-1}w_i\langle n_i(t), \gamma''(t)\rangle - \mathfrak s^2.\]
Taking the partial derivative with respect to $w_i$ we have
\[\frac{\partial}{\partial w_i}\Phi(t,w) = \langle n_i(t), \gamma''(t)\rangle.\]
To show the regularity of $\Phi^{-1}(0)$, we need to establish that for all $(t,w)$ such that $\Phi(t,w) = 0$, the gradient is nonzero. Observe that for all such $(t,w)$, $\langle v(t,w), \gamma''(t)\rangle = \mathfrak s^2$, where $v(t,w) = \sum_{i = 1}^{d-1}w_in_{i}(t)$ -- therefore $\gamma''(t) \neq 0$. Now we need to guarantee that $\gamma''(t)$ has a component in $N_{\gamma(t)}\gamma$. We will show this by contradiction: suppose $\gamma''(t) \in T_{\gamma(t)}\gamma$. Then $\langle\gamma''(t), v\rangle = 0$ for all $v$ normal to $\gamma$ at $t$, which contradicts $\langle \gamma''(t), v(t,w) \rangle = \mathfrak s^2$.

Since $\Phi(t,w)=0$ implies
$\sum_{i=1}^{d-1}w_i\langle n_i(t),\gamma''(t)\rangle
=\mathfrak{s}^2>0$, there exists an index $j$
such that $\langle n_j(t),\gamma''(t)\rangle\neq0$.
Restricting to a sufficiently small neighborhood,
we may therefore solve $\Phi(t,w)=0$ for $w_j$:
\[
w_j=
\frac{\mathfrak{s}^2-
\sum_{i\neq j}w_i\langle n_i(t),\gamma''(t)\rangle}
{\langle n_j(t),\gamma''(t)\rangle}.
\]
Since $\gamma\in C^3$ and the normal frame is
locally $C^2$, this expression is $C^1$ in
$(t,w_{-j})$. Thus, locally, $\exp(B')$ is
contained in the image of a $C^1$ map from
a subset of $\mathbb R^{d-1}$ into $\mathbb R^d$. Since $d-1<d$, every point in the domain of
this $C^1$ map is critical. Thus, by the
Morse--Sard theorem
\citep[Chapter 3, Theorem 1.3]{hirsch1976differential},
its image has zero $d$-dimensional Lebesgue measure.
By second countability, $\exp(B')$ is covered by
countably many such images and therefore has
Lebesgue measure zero. Since $B\subseteq\exp(B')$,
we conclude that $\mathcal{L}^d(B)=0$.
\end{proof}

\begin{restatable}[Open full-measure regularity set]{lemma}{} \label{lemma: IFT degen closed}
    Define
    \begin{equation}\label{R}
        R
        =
        \left\{
        x\in\mathbb{R}^d:
        \begin{array}{l}
        \arg\min_{t\in I}\|x-\gamma(t)\|_2=\{s\}
        \text{ for some }s\in\operatorname{Int}(I),\\[1mm]
        \mathfrak{s}^2
        -
        \langle\gamma''(s),x-\gamma(s)\rangle
        \neq0
        \end{array}
        \right\}.
    \end{equation}
    Then $R$ is an open subset of $\mathbb{R}^d$, $\Pi$ is uniquely defined and $C^2$ on $R$, and $P_X(R)=1.$
\end{restatable}

\begin{proof}[Proof of \Cref{lemma: IFT degen closed}]
We will first show that every $x\in R$ has an open neighborhood on which the closest-point projection is uniquely defined and $C^2$. Fix $x\in R$ and let $s=\Pi(x)\in\operatorname{Int}(I).$
Define
\[
    F(y,t)
    =
    \left\langle y-\gamma(t),\gamma'(t)\right\rangle.
\]
Since $s$ is an interior minimizer of $t\mapsto\|x-\gamma(t)\|_2^2$, the first-order optimality condition gives $F(x,s)=0.$ Moreover,
\begin{align*}
    \partial_tF(x,s)
    &=
    -\|\gamma'(s)\|_2^2
    +
    \left\langle x-\gamma(s),\gamma''(s)\right\rangle =
    -\mathfrak{s}^2
    +
    \left\langle x-\gamma(s),\gamma''(s)\right\rangle
    \neq0,
\end{align*}
where the last inequality follows from $x\in R$. Since $\gamma\in C^3(I)$, the map $F$ is $C^2$. The implicit function theorem \citet[Theorem C.40]{lee2003smooth} therefore guarantees open neighborhoods $U_0$ of $x$ and $J_0$ of $s$, together with a unique $C^2$ map $\pi:U_0\rightarrow J_0$ such that $F(y,\pi(y))=0$ for every $y\in U_0$, and such that $\pi(x)=s$.

We now show that, after possibly shrinking $U_0$, the implicit-function solution $\pi(y)$ is in fact the unique global nearest-point projection of $y$ onto $\Gamma$. To see this, choose an open interval $J$ satisfying
\[
    s\in J,
    \qquad
    \overline{J}\subseteq J_0\cap\operatorname{Int}(I).
\]
Since $s$ is the unique global minimizer of $t\mapsto\|x-\gamma(t)\|_2^2$ and $I\setminus J$ is compact, we have
\[
    \eta
    \triangleq
    \min_{t\in I\setminus J}
    \left\{
    \|x-\gamma(t)\|_2^2-\|x-\gamma(s)\|_2^2
    \right\}
    >0.
\]
By continuity, after shrinking $U_0$ to an open neighborhood $U$ of $x$, we may guarantee simultaneously that $\pi(y)\in J$ and
$\|y-\gamma(t)\|_2^2 > \|y-\gamma(\pi(y))\|_2^2$ for every $y\in U$ and every $t\in I\setminus J$. Thus every global minimizer of $t\mapsto\|y-\gamma(t)\|_2^2$ must lie in $J$. But $J\subseteq\operatorname{Int}(I)$, so any such minimizer must satisfy the first-order condition $F(y,t)=0.$ By the uniqueness statement supplied by the implicit function theorem on $J_0$, the only such point in $J$ is $t=\pi(y)$. Hence $\Pi(y)=\pi(y)$ for every $y\in U$. In particular, the closest-point projection is unique and $C^2$ on $U$. Furthermore, by continuity of
\[
    y
    \mapsto 
    \mathfrak{s}^2
    -
    \left\langle
    \gamma''(\Pi(y)),
    y-\gamma(\Pi(y))
    \right\rangle,
\]
we may shrink $U$ once more so that this quantity remains nonzero throughout $U$. Therefore $U\subseteq R.$ Since every $x\in R$ admits such an open neighborhood, $R$ is open. The same argument shows that $\Pi$ is uniquely defined and $C^2$ on $R$.

It remains to show that $R$ has full $P_X$ measure. By \Cref{assumption: the curve gamma}, for every $x\in\Omega$ every nearest-point projection lies in $\operatorname{Int}(I)$. Thus, if $x\in\Omega\setminus R$, then either the nearest-point projection is not unique, in which case $x\in M(\Gamma),$ or the projection is unique but
\[
    \mathfrak{s}^2
    -
    \left\langle
    \gamma''(\Pi(x)),
    x-\gamma(\Pi(x))
    \right\rangle
    =0,
\]
in which case $x\in B$. Therefore $\Omega\setminus R \subseteq
    M(\Gamma)\cup B.$ By \Cref{lemma: singular set measure} and \Cref{lemma: IFT degen},
\[
    \mathcal{L}\big(M(\Gamma)\cup B\big)=0.
\]
Since $P_X$ is absolutely continuous with respect to Lebesgue measure, $ P_X(\Omega\setminus R)=0.$ 
Since $P_X(\Omega)=1$, we conclude that $P_X(R)=1$ as desired.
\end{proof}

\begin{restatable}[Projection gradient]{prop}{} \label{prop: proj gradient}
    Let $\Pi: \mathbb{R}^d \rightarrow [0, 1]$ be the closest-point projection onto $\gamma$ as defined in \eqref{def: projection}. Then for every $x \in R$ \eqref{R} 
    we have 
    \[\nabla\Pi(x)  = -\frac{\gamma'(\Pi(x))}{-\mathfrak s^2 + \left\langle\gamma''(\Pi(x)), x - \gamma(\Pi(x))\right\rangle}.\]
\end{restatable}

\begin{proof}[Proof of \Cref{prop: proj gradient}] Let $f(x,t) = \|x - \gamma(t)\|_2^2$. By non-negativity of $\|\cdot\|_2$ and monotonicity of $h(x) = x^2$ for $x \geq 0$, we have $\Pi(x) = \arg\min_t \|x - \gamma(t)\|_2 = \arg \min_t \|x - \gamma(t)\|_2^2 = \arg\min_t f(x, t)$. We have the first-order condition that
\[ \partial_t f(x,t) = 0\]
at the optimum. Now, if we define $F(x, t) = \partial_t f(x,t)$ the first order condition amounts to the requirement that $F(x,t) = 0$. Expanding $F$, we have
\begin{align}
    F(x, t) &= \partial_t \|x - \gamma(t)\|_2^2 \\ \label{eq: ift}
    &= -2\langle x - \gamma(t)), \gamma'(t)\rangle.
\end{align}
Observe that the first order condition implies that the projection displacement $x - \gamma(t)$ lies normal to the curve at optimality. 
Now we'll apply the implicit function theorem to $F(x, t)$. Because $x \in R$, we have that $\Pi(x)$ is unique and $\Pi$ is differentiable in a neighborhood about $x$ -- its not hard to verify that this yields continuous differentiability of $F(x,t)$ in a neighborhood about $(x, \Pi(x))$. This, coupled with the fact that $\langle\gamma''(\Pi(x)), x - \gamma(\Pi(x))\rangle \neq \mathfrak s^2$ for $x\in R$ allows us apply the implicit function theorem to conclude that 
\begin{align}
    \nabla\Pi(x) &= - \frac{\nabla_x F\big(x, t\big)}{\partial_t F\big(x, t\big)} \Bigg|_{t = \Pi(x)}\\
    &= - \frac{\nabla_x\left(-2\langle x - \gamma(t), \gamma'(t)\rangle\right)}{\partial_t \left(-2\langle x - \gamma(t), \gamma'(t)\rangle\right)} \Bigg|_{t = \Pi(x)}\\
    &= -\frac{\gamma'(t)}{-\|\gamma'(t)\|_2^2 + 
    \langle\gamma''(t), x - \gamma(t)\rangle}\Bigg|_{t = \Pi(x)} \label{eq: projection grad} \\ 
    &= -\frac{\gamma'(\Pi(x))}{-\mathfrak s^2 + \left\langle\gamma''(\Pi(x)), x - \gamma(\Pi(x))\right\rangle}.
\end{align}
\end{proof}

\begin{restatable}[Projection hessian]{corollary}{} \label{prop: proj hessian}
    Let $\Pi: \mathbb{R}^d \rightarrow [0, 1]$ be the closest-point projection onto $\gamma$ as defined in \eqref{def: projection}. Then for every $x \in R$ \eqref{R}
    we have
    \[\nabla^2\Pi(x) = \frac{1}{\ell^2(x)}\left(\gamma''(s)\gamma'(s)^{\top} + \gamma'(s)\gamma''(s)^{\top}\right)- \frac{\langle\gamma^{(3)}(s), x - \gamma(s)\rangle}{\ell^3(x)}\gamma'(s)\gamma'(s)^{\top}.\]
    where $s = \Pi(x)$ and $\ell(x) = -\mathfrak s^2 + \langle\gamma''(s), x-\gamma(s)\rangle$.
\end{restatable}
\begin{proof} By \cref{prop: proj gradient} we know that for all $x \in R$, we have
\begin{align}
    \nabla\Pi(x) &= -\frac{\gamma'(s)}{-\mathfrak s^2 + \langle\gamma''(s), x - \gamma(s)\rangle} \\
    &= -\frac{\gamma'(s)}{\ell(x)}.
\end{align}
Taking the gradient again yields
\begin{align}
    \nabla^2\Pi(x) &= -\frac{\ell(x)\nabla \gamma'(s) - \gamma'(s)\nabla \ell(x)^{\top}}{\ell(x)^2} \\
    &= -\frac{\nabla \gamma'(s)}{\ell(x)} + \frac{1}{\ell^2(x)}\gamma'(s)\nabla \ell(x)^{\top}.
\end{align}
By the chain rule, we have $\nabla\gamma'(s) = \gamma''(s)\nabla\Pi(x)^{\top}$ and 
\begin{align}
    \nabla \ell(x) &= \nabla\langle\gamma''(s), x\rangle -\nabla\langle \gamma''(s), \gamma(s)\rangle \\
    &= \langle \gamma^{(3)}(s), x\rangle \nabla\Pi(x) + \gamma''(s) - \langle\gamma^{(3)}(s), \gamma(s)\rangle\nabla\Pi(x) - \underbrace{\langle\gamma''(s), \gamma'(s)\rangle}_{= 0}\nabla\Pi(x) \\
    &= \langle\gamma^{(3)}(s), x - \gamma(s)\rangle\nabla\Pi(x) + \gamma''(s).
\end{align}
Thus,
\begin{align*}
    \nabla^2\Pi(x) &= -\frac{1}{\ell(x)}\gamma''(s)\nabla\Pi(x)^{\top} + \frac{\langle\gamma^{(3)}(s), x - \gamma(s)\rangle}{\ell^2(x)}\gamma'(s)\nabla\Pi(x)^{\top} + \frac{1}{\ell^2(x)}\gamma'(s)\gamma''(s)^{\top} \\
    &= \frac{1}{\ell^2(x)}\gamma''(s)\gamma'(s)^{\top} - \frac{\langle\gamma^{(3)}(s), x - \gamma(s)\rangle}{\ell^3(x)}\gamma'(s)\gamma'(s)^{\top} + \frac{1}{\ell^2(x)}\gamma'(s)\gamma''(s)^{\top} \\
    &= \frac{1}{\ell^2(x)}\left(\gamma''(s)\gamma'(s)^{\top} + \gamma'(s)\gamma''(s)^{\top}\right)- \frac{\langle\gamma^{(3)}(s), x - \gamma(s)\rangle}{\ell^3(x)}\gamma'(s)\gamma'(s)^{\top}.
\end{align*}
\end{proof}

\begin{restatable}[Operator norm bound]{corollary}{} \label{lemma: operator norm bound}
    Suppose $x_0 \in R$ with $R$ defined in \eqref{R}. Then there exist finite $\epsilon, \eta > 0$ such that 
    \[ \sup_{x\in \mathcal{B}(x_0,\epsilon)}\|\nabla^2\Pi(x)\|_{\operatorname{op}} \leq 2\mathfrak sM_2\eta^{2}+2r\mathfrak s^2M_3\eta^{3}\]
    where $M_i$ are defined in \Cref{remark: deriv bound}.
\end{restatable}

\begin{proof} By \Cref{thm: differentiability of pi} we have that for every $x \in R$,
\[\nabla^2\Pi(x) = \frac{1}{\ell^2(x)}\left(\gamma''(s)\gamma'(s)^{\top} + \gamma'(s)\gamma''(s)^{\top}\right)- \frac{\langle\gamma^{(3)}(s), x - \gamma(s)\rangle}{\ell^3(x)}\gamma'(s)\gamma'(s)^{\top}.\]
where $s = \Pi(x)$ and $\ell(x) = -\mathfrak s^2 + \langle\gamma''(s), x-\gamma(s)\rangle$. Now let $x_0 \in R$ -- since $R$ is an open set, there exists an $\epsilon > 0$ such that this expression holds for all $x$ in the closed ball $\mathcal{B}(x_0, \epsilon)$. Bounding the operator norm pointwise yields
\begin{align*}
    \|\nabla^2\Pi(x)\|_{\operatorname{op}} &= \max_{v\in \mathbb{S}^{d-1}} \|\nabla^2\Pi(x)v\|_2 \\
    &\leq \max_{v\in \mathbb{S}^{d-1}}\left(\frac{\|\gamma''(s)\gamma'(s)^{\top}v + \gamma'(s)\gamma''(s)^{\top}v\|_2}{|\ell(x)|^2} + \frac{|\langle\gamma^{(3)}(s), x - \gamma(s)\rangle|}{|\ell(x)|^3} \|\gamma'(s)\gamma'(s)^{\top}v\|_2\right) \\
    &\leq \max_{v\in \mathbb{S}^{d-1}}\left(\frac{\|\gamma''(s)\gamma'(s)^{\top}v\|_2 + \|\gamma'(s)\gamma''(s)^{\top}v\|_2}{|\ell(x)|^2} + \frac{|\langle\gamma^{(3)}(s), x - \gamma(s)\rangle|}{|\ell(x)|^3} \|\gamma'(s)\gamma'(s)^{\top}v\|_2\right) \\
    &\leq \frac{2\mathfrak sM_2}{|\ell(x)|^2} + \frac{\mathfrak s^2|\langle\gamma^{(3)}(s), x - \gamma(s)\rangle|}{|\ell(x)|^3} \\
    &\leq \frac{2\mathfrak sM_2}{|\ell(x)|^2} + \frac{2r\mathfrak s^2 M_3}{|\ell(x)|^3}
\end{align*}
where the final line follows from the application of triangle inequality and the assumptions that $\Gamma = \text{Im}(\gamma) \subset \mathcal{B}(0, r)$ and $x \in  \mathcal{B}(0, r)$. Now we will establish that $1/|\ell(x)|$ is continuous everywhere in $\mathcal{B}(x_0,\epsilon)$. This property is a consequence of 
\begin{enumerate}
    \item the continuity of $|\ell(x)|$ in $\mathcal{B}(x_0, \epsilon)$. This follows from (a) the fact that $\Pi(x) \in \text{Int}(I)$ for all $x \in \Omega$ (\cref{assumption: the curve gamma}) in conjunction with the fact that $\gamma \in C^3$, and (b) the fact that $\Pi$ is twice-differentiable for all points in $\mathcal{B}(x_0, \epsilon)$ due to \Cref{thm: differentiability of pi}.
    \item the fact that $|\ell(x)| \neq 0$ for all $x \in R$ -- this follows from the definition of $R$.
\end{enumerate}

From these two properties, we can deduce that $1/|\ell(x)|$ is continuous everywhere in $\mathcal{B}(x_0,\epsilon)$. Since $\mathcal{B}(x_0,\epsilon)$ is a closed set contained entirely in $R$ and $1/|\ell(x)|$ is continuous everywhere in it, we know it must be bounded by some constant $\eta$. The statement follows. \end{proof}

\subsection{Supplementary results for \Cref{thm: population prop}}

% \begin{restatable}[Approximation]{lemma}{} \label{lemma: proj approx general}
%     For $P_X$-almost every $x_0$ there exists a $C_{x_0} > 0$ such that for all $x \in \mathcal{B}(x_0, \epsilon)$
%     \[\bigg|f\big(g(x)\big) - f\big(g(x_0) + \nabla g(x_0)^{\top}(x-x_0) \big)\bigg| \le C_{x_0}\epsilon^2. \]
% \end{restatable}

% \noindent \textit{Proof.} By the $\beta$-H\"older smoothness of $g$ (where $\beta \geq 2$), we know that for all $x \in \mathcal{B}(x_0, \epsilon)$
% \[\|\nabla^2g(x)\|_{\operatorname{op}} \leq C_{x_0}.\]
% We can employ an expansion of $g$ as follows,
% \begin{align*}
%     g(x) &= g(x_0) + \nabla g(x_0)^{\top}(x-x_0) + \underbrace{\int_{0}^{1}(1-t)(x-x_0)^{\top}\nabla^2g(x+t(x-x_0))(x-x_0) dt}_{= R(x)}.
% \end{align*}
% By the operator norm bound, we have that
% \[\|R(x)\|_2 \leq \int_{0}^{1}\|x - x_0\|_2^2\|\|\nabla^2g(x+t(x-x_0))\|_{\operatorname{op}} \,dt \leq C_{x_0, \eta}\epsilon^2.\]
% It follows that 
% \begin{align}
%     \left|\,g(x) - \left(g(x_0) + \nabla g(x_0)^{\top}(x-x_0)\right)\right| &\leq C_{x_0}\epsilon^2. \label{projection second order bound}
% \end{align}
% The lemma follows from the Lipschitz continuity of $f$. 

% \qed{}

% Having established the $O(\epsilon^2)$ deviation between the true target and the linearized surrogate, we will now show that the population index vector using the linearized surrogate recovers something proportional to $\nabla g(0)$.

We now show that the population regression coefficient of the linearized surrogate is proportional to $\nabla g(0)$.

\begin{restatable}[Surrogate]{lemma}{} \label{lemma: surrogate problem} Define the linearized surrogate response as
\[m(x) = f\Big(g(0) + \nabla g(0)^{\top}x\Big)\]
and define the population regression solution using the linearized surrogate as 
\[\tilde{\theta}^*_{x_0, \epsilon} = \arg\min_{\theta \in \mathbb{R}^{d+1}}\mathbb{E}_{ p}\Big[\mathbbm{1}_{\{X \in \mathcal{B}(0, \epsilon)\}}\frac{p^*(X)}{p(X)}\Big(m(X) - \boldsymbol{X}^{\top}\theta \Big)^2\Big].\]
If $f \in BV_{\operatorname{loc}}(\mathbb R)$ then for some $c_0, c_* \in \mathbb{R}$ we have $\tilde{\theta}^*_{x_0, \epsilon} = \begin{bmatrix}c_0  & c_*\nabla g(0) \end{bmatrix}^{\top}$.
\end{restatable}
\begin{proof} Define
\[R_{\epsilon}(\theta; x) = \mathbb{E}_{ p}\Big[\mathbbm{1}_{\{X \in \mathcal{B}(0, \epsilon)\}}\frac{p^*(X)}{p(X)}\Big(m(X) - \boldsymbol{X}^{\top}\theta\Big)^2\Big].\]
Then we have,
\begin{align*}
    R_{\epsilon}(\theta; X)&= \int_{\mathbb{R}^d}\mathbbm{1}_{\{x \in \mathcal{B}(0, \epsilon)\}}\frac{p^*(x)}{p(x)}\Big(m(x) - \boldsymbol{x}^{\top}\theta\Big)^2p(x) dx_1 \dots dx_d \\
    &= \int_{\mathbb{R}^d}\mathbbm{1}_{\{x \in \mathcal{B}(0, \epsilon)\}}p^*(x)\Big(m(x) - \boldsymbol{x}^{\top}\theta \Big)^2 dx_1 \dots dx_d.
\end{align*}
Now let 
\begin{equation}
    A(\epsilon) = \int_{\mathbb{R}^d} \mathbbm{1}_{\{x \in \mathcal{B}(0, \epsilon)\}}p^*(x) dx_1 \dots dx_d.
\end{equation}
Then we have,
\begin{align*}
    R_{\epsilon}(\theta; X)&= A(\epsilon) \bigintsss_{\mathbb{R}^d}\underbrace{\frac{\mathbbm{1}_{\{x \in \mathcal{B}(0, \epsilon)\}}p^*(x)}{A(\epsilon)}}_{\triangleq \tilde{p}(x)}\Big(m(x) - \boldsymbol{x}^{\top}\theta \Big)^2 dx_1 \dots dx_d \\
    &= A(\epsilon)\, \mathbb{E}_{ \tilde{p}}\left(m(X)-\boldsymbol{X}^{\top}\theta\right)^2.
\end{align*}
Applying the law of iterated expectation yields
\begin{align*}
    R_{\epsilon}(\theta; \boldsymbol{X})&= A(\epsilon)\,\mathbb{E}\left[\mathbb{E}\left[\left(m(X)-\boldsymbol{X}^{\top}\theta\right)^2 \, \Big|
    \, \nabla g(0)^{\top}X\right]\right]\\
    &= A(\epsilon)\,\mathbb{E}\left[\mathbb{E}\left[\left(f\left(g(0) + \nabla g(0)^{\top}X\right)-\boldsymbol{X}^{\top}\theta\right)^2 \, \Big|
    \, \nabla g(0)^{\top}X\right]\right].
\end{align*}
Now we'll apply Jensen's inequality to say,
\begin{align*}
    R_{\epsilon}(\theta; X) &\geq A( \epsilon)\,\mathbb{E}\left[\left(f\Big(g(0) + \nabla g(0)^{\top}X\Big)-\mathbb{E}\left[\boldsymbol{X}^{\top}\theta \, \Big|
    \, \nabla g(0)^{\top}X\right]\right)^2\right] \\
    &= A( \epsilon)\,\mathbb{E}\left[\left(f\Big( g(0) + \nabla g(0)^{\top}X\Big)-(1/\epsilon)\mathbb{E}\left[X^{\top}\theta' \, \Big|
    \, \nabla g(0)^{\top}X\right]- \theta_0\right)^2\right]
\end{align*}
where $\theta'$ is the last $d$ entries of $\theta$. By \Cref{prop: conditional spherical symmetry} we know that $\tilde{p}$ is also a spherically symmetric distribution, and it therefore satisfies the LCE property (\cref{def: lce}). Thus, there exist $a_0, a_* \in \mathbb{R}$ such that
\begin{align}
    R_{\epsilon}(\theta; X) & \geq A( \epsilon)\,\mathbb{E}\left[\left(f\Big(g(0) + \nabla g(0)^{\top}X\Big) - a_*\nabla g(0)^{\top}X - \underbrace{(a_0 + \theta_0)}_{\triangleq c_0}\right)^2\right]\label{eq: surrogate risk lower bound}
\end{align}
Observe that the inequality becomes an equality when $\theta = \begin{bmatrix}c_0  & c_*\nabla g(0) \end{bmatrix}^{\top}$ with $c_* := \epsilon a_*$, which completes the proof.
\end{proof}

\begin{restatable}[Constant of proportionality]{lemma}{} \label{lemma: constant of prop}
    When $f\in C^2(I)$, the constants of proportionality
    $c_0,c_*$ in \Cref{lemma: surrogate problem} satisfy
    \[
    c_0=
    \frac{\mathbb E_{p^*}[
    \mathbbm{1}_{\{X\in\mathcal B(0,\epsilon)\}}m(X)]}
    {\mathbb P_{p^*}(X\in\mathcal B(0,\epsilon))},
    \qquad
    c_*=\epsilon f'(g(0))+o(\epsilon^2).
    \]
\end{restatable}
\begin{proof} One can derive the following closed form expression for $\tilde \theta^*_{x_0, \epsilon}$ from \Cref{lemma: surrogate problem} as
\begin{align*}
    \tilde \theta^*_{x_0, \epsilon} &= \mathbb{E}[\mathbbm{1}_{\{X \in \mathcal{B}(0, \epsilon)\}}\boldsymbol{X}\boldsymbol{X}^{\top}]^{-1}\mathbb{E}[\mathbbm{1}_{\{X \in \mathcal{B}(0, \epsilon)\}}\boldsymbol{X}m(X)] \\
    &= Q^{-1}\mathbb{E}\left[\mathbbm{1}_{\{X \in \mathcal{B}(0, \epsilon)\}}\begin{pmatrix}
        1 \\ X/\epsilon
    \end{pmatrix}m(X)\right]
\end{align*}
where the expectation is taken with respect to $p^*$ and $Q = \mathbb{E}[\mathbbm{1}_{\{X \in \mathcal{B}(0, \epsilon)\}}\boldsymbol{X}\boldsymbol{X}^{\top}]$. It follows from \eqref{eq: Q closed form} that
\begin{align}
    \tilde \theta^*_{x_0, \epsilon} &= \begin{bmatrix}
         A(\epsilon)^{-1} & 0 \\
         0 & (\epsilon/\nu)^{2}I_d
    \end{bmatrix}\mathbb{E}\left[\mathbbm{1}_{\{X \in \mathcal{B}(0, \epsilon)\}}\begin{pmatrix}
        1 \\ X/\epsilon
    \end{pmatrix} m(X)\right] \\
    &= \begin{bmatrix}
        A(\epsilon)^{-1} \mathbb{E}[\mathbbm{1}_{\{X \in \mathcal{B}(0, \epsilon)\}}m(X)] \\
        \frac{\epsilon}{\nu^2}\mathbb{E}[\mathbbm{1}_{\{X \in \mathcal{B}(0, \epsilon)\}} Xm(X)]
    \end{bmatrix}\label{theta tilde simplification}
\end{align}
with $\nu^2 = \mathbb{E}_{p^*}[\mathbbm{1}_{\{X \in \mathcal{B}(0, \epsilon)\}}X_i^2]$ and $A(\epsilon) = \mathbb{P}_{p^*}(X \in \mathcal{B}(0, \epsilon)).$ Thus, $c_0 = A(\epsilon)^{-1} \mathbb{E}[\mathbbm{1}_{\{X \in \mathcal{B}(0, \epsilon)\}}m(X)]$. To solve for $c_*$, we will expand $m$ as follows
\begin{align*}
    m(x) &= f\left(g(0) + \nabla g(0)^{\top}x\right) \\
    &= f(g(0)) + f'(g(0))\nabla g(0)^{\top}x + \frac{1}{2}f''(g(0))(\nabla g(0)^{\top}x)^2 + r_2(\nabla g(0)^\top x)
\end{align*}
for some function $r_2$. Multiplying by $\mathbbm{1}_{\{X \in \mathcal{B}(0, \epsilon)\}}X$ and taking the expectation yields
\begin{align*}
    &\mathbb{E}[\mathbbm{1}_{\{X \in \mathcal{B}(0, \epsilon)\}}X]f(g(0)) = 0 \\
    &\mathbb{E}[\mathbbm{1}_{\{X \in \mathcal{B}(0, \epsilon)\}}XX^{\top}]f'(g(0))\nabla g(0) = \nu^2f'(g(0))\nabla g(0) \\
    & \frac{1}{2}\mathbb{E}[\mathbbm{1}_{\{X \in \mathcal{B}(0, \epsilon)\}}(\nabla g(0)^{\top}X)^2X]f''(g(0)) = 0. \tag{by spherical symmetry}
\end{align*}
For the remainder term we have that $r_2(a) = o(a^2), $ and therefore for every $\delta >0$ and every $a$ such that $|a|\leq \delta$ we can write $|r_2(a)| \le \omega(\delta)a^2$. Because on $\mathcal{B}(0, \epsilon)$ we have $|\nabla g(0)^\top x| \leq \|\nabla g(0)\|_2\epsilon$, it follows that
\[r_2(\nabla g(0)^\top x) \leq \omega(\|\nabla g(0)\|_2\epsilon)\|\nabla g(0)\|_2^2\|x\|_2^2.\]
Multiplying by $\mathbbm{1}_{\{X \in \mathcal{B}(0, \epsilon)\}}X$, taking the expectation and applying Jensens inequality yields
\begin{align*}
    \|\mathbb{E}[\mathbbm{1}_{\{X \in \mathcal{B}(0, \epsilon)\}}X r_2(\nabla g(0)^\top X)]\|_2 &\leq \mathbb{E}[\mathbbm{1}_{\{X \in \mathcal{B}(0, \epsilon)\}}\|X\|_2^3]\omega(\|\nabla g(0)\|_2 \epsilon)\|\nabla g(0)\|_2^2  \\
    &\leq \mathbb{E}[\mathbbm{1}_{\{X \in \mathcal{B}(0, \epsilon)\}}\|X\|_2^2]\omega(\|\nabla g(0)\|_2 \epsilon)\|\nabla g(0)\|_2^2\epsilon.
\end{align*}
By spherical symmetry, $\mathbb{E}[\mathbbm{1}_{\{X \in \mathcal{B}(0, \epsilon)\}}\|X\|_2^2] = d\cdot \nu^2$. Thus, we have
\begin{align*}
    c_* \nabla g(0) &= \frac{\epsilon}{\nu^2}\left(\nu^2f'( g(0))\nabla g(0) + o(d\cdot\nu^2\epsilon)\nabla g(0)\right) \\
    &= \epsilon f'( g(0))\nabla g(0)+ o(\epsilon^2)\nabla g(0).
\end{align*}
\end{proof}

\subsection{Supplementary results for \Cref{thm: decomposition for theta}}

\begin{restatable}{lemma}{}\label{conditional residual bound remainder lemma}
    Under the assumptions of \Cref{thm: decomposition for theta}, let
    $u = Y - \boldsymbol X^\top\theta^*_{x_0,\epsilon}$, so that
    $\mathbb E[u \mid X = x] = f(g(x)) - \boldsymbol x^\top\theta^*_{x_0,\epsilon}$.
    Then
    \begin{equation}
        \sup_{x \in \mathcal B(0,\epsilon)} |\mathbb E[u \mid X = x]| = O(\epsilon).
        \label{eq: conditional residual bound remainder}
    \end{equation}
    If moreover $\epsilon \to 0$, the right side can be replaced by $o(\epsilon)$.
\end{restatable}

\begin{proof}
    If $\epsilon$ is fixed, this is immediate, as $f \circ g$ is bounded on
    $\mathcal B(0,\epsilon)$ (because $g(\mathcal B(0,\epsilon))$ is a compact interval
    and $f \in BV_{\mathrm{loc}}(\mathbb R)$ is bounded on compact intervals),
    $\theta^*_{x_0,\epsilon}$ is a fixed vector, and
    $\|\boldsymbol x\|_2 \le \sqrt 2$ on $\mathcal B(0,\epsilon)$. If $\epsilon \to 0$, we use the hypothesis that $f$ is differentiable at $g(0)$.
    Since $g \in C^2(U)$, we have
    \[
        \sup_{x \in \mathcal B(0,\epsilon)}
        \big| g(x) - g(0) \big| \le C_{x_0}\epsilon
    \]
    for some $C_{x_0} > 0$.
    Differentiability of $f$ at $g(0)$ then implies
    \[
        \delta_\epsilon :=
        \sup_{x \in \mathcal B(0,\epsilon)}
        \big| f(g(x)) - \big(f(g(0)) + f'(g(0))\nabla g(0)^\top x\big) \big|
        = o(\epsilon),
    \]
    since the increments $g(x) - g(0)$ at which the differentiability of $f$ is
    invoked are all bounded by $C'_{x_0}\epsilon$. Now write $L(x) = f(g(0)) + f'(g(0))\nabla g(0)^\top x$ and $r(x) = f(g(x)) - L(x)$,
    so that $\sup_{\mathcal B(0,\epsilon)} |r| = \delta_\epsilon$. Since
    $\boldsymbol x = (1, x/\epsilon)^\top$, the function $L$ is linear in
    $\boldsymbol x$, as $L(x) = \boldsymbol x^\top\theta_L$ with
    $\theta_L = \big(f(g(0)),\, \epsilon f'(g(0))\nabla g(0)\big)$. Substituting
    $f(g(X)) = \boldsymbol X^\top\theta_L + r(X)$ into
    \eqref{eq: population estimator closed form} gives
    \[
        \theta^*_{x_0,\epsilon} - \theta_L
        = Q^{-1}\,\mathbb E\big[w\boldsymbol X r(X)\big],
        \qquad Q = \mathbb E\big[w\boldsymbol X\boldsymbol X^\top\big].
    \]
    Since $w$ vanishes outside $\mathcal B(0,\epsilon)$, where
    $\|\boldsymbol X\|_2 \le \sqrt 2$ and $|r(X)| \le \delta_\epsilon$, and since
    $p^*$ is bounded above on $\mathcal B(0,\epsilon)$,
    \[
        \big\| \mathbb E[w\boldsymbol X r(X)] \big\|_2
        \le \sqrt 2\,\delta_\epsilon\, \mathbb P_{p^*}\big(\mathcal B(0,\epsilon)\big)
        \lesssim \delta_\epsilon\,\epsilon^d .
    \]
    By \Cref{lemma: Q op norm}, $\|Q^{-1}\|_{\operatorname{op}} \lesssim \epsilon^{-d}$,
    so $\|\theta^*_{x_0,\epsilon} - \theta_L\|_2 \lesssim \delta_\epsilon$. Therefore,
    for every $x \in \mathcal B(0,\epsilon)$,
    \[
        |\mathbb E[u \mid X = x]|
        = \big| f(g(x)) - \boldsymbol x^\top\theta^*_{x_0,\epsilon} \big|
        \le |r(x)| + \big|\boldsymbol x^\top(\theta_L - \theta^*_{x_0,\epsilon})\big|
        \le \delta_\epsilon + \sqrt 2\,\|\theta^*_{x_0,\epsilon} - \theta_L\|_2
        \lesssim \delta_\epsilon = o(\epsilon),
    \]
    which proves \eqref{eq: conditional residual bound remainder}.
\end{proof}

\begin{restatable}{lemma}{} \label{lemma: remainder term}
    Under the assumptions of \Cref{thm: decomposition for theta}, for $R_1$ defined in \eqref{eq: expansion with remainder}, we have
    \[
        R_1
        =
        \frac{1}{n}\sum_{i=1}^n
        \frac{\mathbb{E}[a(\mathcal Z)\mid X=X_i]}{p(X_i)}
        +
        o_P\left(\sqrt{\frac{\epsilon^d}{n}}\right).
    \]
\end{restatable}
\begin{proof}
\Cref{eq: conditional residual bound remainder}, together with the local upper and lower bounds on $p^*$ and $p$, implies
\begin{equation}
    \left\|
        \frac{\mathbb{E}[a(\mathcal Z)\mid X=x]}{p(x)}
    \right\|_2
    \lesssim
    \epsilon\mathbbm{1}_{\{x\in\mathcal{B}(0,\epsilon)\}}.
    \label{eq: conditional a bound remainder}
\end{equation}
In particular,
\begin{equation}
    \int
    \left\|
        \frac{\mathbb{E}[a(\mathcal Z)\mid X=x]}{p(x)}
    \right\|_2dx
    =
    O(\epsilon^{d+1}),
    \qquad
    \int
    \left\|
        \frac{\mathbb{E}[a(\mathcal Z)\mid X=x]}{p(x)}
    \right\|_2^2dx
    =
    O(\epsilon^{d+2}).
    \label{eq: conditional a integral bounds}
\end{equation}
Moreover, the moment assumptions on the noise, boundedness of $p^*/p$ on $\mathcal{B}(0,\epsilon)$, and $\|\boldsymbol{X}\|_2\lesssim1$ give
\begin{equation}
    \mathbb{E}\left\|
        \frac{a(\mathcal Z)}{p(X)^2}
    \right\|_2^2
    =
    O(\epsilon^d),
    \qquad
    \mathbb{E}\left\|
        \frac{a(\mathcal Z)}{p(X)}
    \right\|_2^2
    =
    O(\epsilon^d).
    \label{eq: a moment bounds remainder}
\end{equation}
We now expand $R_1$ as
\begin{align}
    R_1
    &=
    \frac{1}{n^2}\sum_{i,j}
    \frac{a(\mathcal Z_i)}{p(X_i)^2}
    \frac{1}{h^d}
    K\left(\frac{X_j-X_i}{h}\right)
    -
    \frac{1}{n}\sum_{i=1}^n
    \frac{a(\mathcal Z_i)}{p(X_i)} \nonumber\\
    &=
    R_{11}
    +
    \frac{K(0)}{nh^d}
    \mathbb{P}_n\left[\frac{a(\mathcal Z)}{p(X)^2}\right]
    -
    \mathbb{P}_n\left[\frac{a(\mathcal Z)}{p(X)}\right],
    \label{eq: R1 repaired split}
\end{align}
where, using the symmetry of $K$,
\[
    R_{11}
    =
    \frac{1}{n^2}\sum_{i<j}
    \left(
        \frac{a(\mathcal Z_i)}{p(X_i)^2}
        +
        \frac{a(\mathcal Z_j)}{p(X_j)^2}
    \right)
    \frac{1}{h^d}K\left(\frac{X_j-X_i}{h}\right).
\]
Define
\[
    t(w_i,w_j)
    =
    \left(
        \frac{a(\mathcal Z_i)}{p(X_i)^2}
        +
        \frac{a(\mathcal Z_j)}{p(X_j)^2}
    \right)
    \frac{1}{h^d}K\left(\frac{X_j-X_i}{h}\right),
\]
$g(w_i)=\mathbb{E}[t(w_i,W)]-\mathbb{E}[t(W_1,W_2)]$, and
$m(w_1,w_2)=t(w_1,w_2)-\mathbb{E}[t(W_1,W_2)]-g(w_1)-g(w_2)$.
Then $\mathbb{E}[m(w,W)]=0$ for every fixed $w$, and the Hoeffding decomposition \citep[Lemma 11.1]{vandervaart1998asymptotic} gives
\begin{equation}
    R_{11}
    =
    \frac{n-1}{2n}\mathbb{E}[t(W_1,W_2)]
    +
    \frac{n-1}{n}\mathbb{P}_ng
    +
    \frac{1}{n^2}\sum_{i<j}m(w_i,w_j).
    \label{eq: repaired hoeffding remainder}
\end{equation}
For fixed $w_i$, conditioning on the second observation yields
\begin{align*}
    \mathbb{E}[t(w_i,W)]
    &=
    \frac{a(\mathcal Z_i)}{p(X_i)^2}
    \int
    \frac{1}{h^d}
    K\left(\frac{x-X_i}{h}\right)p(x)\,dx \\
    &\qquad
    +
    \int
    \frac{1}{h^d}
    K\left(\frac{x-X_i}{h}\right)
    \frac{\mathbb{E}[a(\mathcal Z)\mid X=x]}{p(x)}
    \,dx.
\end{align*}
By symmetry and exchangeability,
\begin{align*}
    \mathbb{E}[t(W_1,W_2)]
    &=
    2\mathbb{E}\left[
        \frac{a(\mathcal Z)}{p(X)^2}
        \int
        \frac{1}{h^d}
        K\left(\frac{x-X}{h}\right)p(x)\,dx
    \right].
\end{align*}
The order-$s$ kernel conditions and the smoothness of $p$ imply, uniformly over the localization region,
\[
    \int
    \frac{1}{h^d}K\left(\frac{x-X}{h}\right)p(x)\,dx
    =
    p(X)+O(h^s).
\]
Since $\mathbb{E}[a(\mathcal Z)/p(X)]=\mathbb{E}[w\boldsymbol{X}u]=0$ by the population first-order conditions, it follows that the contribution of the preceding expectation is entirely due to the kernel approximation error. Substituting these identities into \eqref{eq: repaired hoeffding remainder} and then into \eqref{eq: R1 repaired split}, straightforward cancellation gives
\begin{align}
    R_1
    -
    \mathbb{P}_n\left[
        \frac{\mathbb{E}[a(\mathcal Z)\mid X]}{p(X)}
    \right]
    &=
    \frac{n-1}{n}(\mathbb{P}_n-\mathbb{P})
    \left[
        \frac{a(\mathcal Z)}{p(X)^2}
        \left\{
            \int \frac{1}{h^d}
            K\left(\frac{x-X}{h}\right)p(x)\,dx
            -
            p(X)
        \right\}
    \right] \nonumber\\
    &
    +
    \frac{n-1}{n}\mathbb{P}_n
    \left[
        \int
        \frac{1}{h^d}
        K\left(\frac{x-X}{h}\right)
        \frac{\mathbb{E}[a(\mathcal Z)\mid X=x]}{p(x)}
        \,dx
        -
        \frac{\mathbb{E}[a(\mathcal Z)\mid X]}{p(X)}
    \right] \nonumber\\
    &
    -
    \frac{1}{n}\mathbb{P}_n
    \left[
        \frac{\mathbb{E}[a(\mathcal Z)\mid X]}{p(X)}
        +
        \frac{a(\mathcal Z)}{p(X)}
    \right]
    +
    \frac{1}{n^2}\sum_{i<j}m(w_i,w_j) \nonumber\\
    &
    +
    \frac{K(0)}{nh^d}
    \mathbb{P}_n\left[\frac{a(\mathcal Z)}{p(X)^2}\right].
    \label{eq: repaired exact remainder decomposition}
\end{align}
We now bound each term on the right. For the first term, the uniform kernel-bias bound and
\eqref{eq: a moment bounds remainder} imply
\[
    (\mathbb{P}_n-\mathbb{P})
    \left[
        \frac{a(\mathcal Z)}{p(X)^2}
        \left\{
            \int \frac{1}{h^d}
            K\left(\frac{x-X}{h}\right)p(x)\,dx-p(X)
        \right\}
    \right]
    =
    O_P\left(h^s\sqrt{\frac{\epsilon^d}{n}}\right)
    =
    o_P\left(\sqrt{\frac{\epsilon^d}{n}}\right).
\]
For the second term, Fubini's theorem gives
\begin{multline*}
    \mathbb{E}\left[
        \int
        \frac{1}{h^d}
        K\left(\frac{x-X}{h}\right)
        \frac{\mathbb{E}[a(\mathcal Z)\mid X=x]}{p(x)}
        \,dx
        -
        \frac{\mathbb{E}[a(\mathcal Z)\mid X]}{p(X)}
    \right] \\
     =
    \int
    \frac{\mathbb{E}[a(\mathcal Z)\mid X=x]}{p(x)}
    \left\{
        \int
        \frac{1}{h^d}
        K\left(\frac{x-y}{h}\right)p(y)\,dy-p(x)
    \right\}dx.
\end{multline*}
Thus \eqref{eq: conditional a integral bounds} and the kernel-bias bound imply that the norm of this expectation is $O(\epsilon^{d+1}h^s)$, and
\[
    \frac{\epsilon^{d+1}h^s}{\sqrt{\epsilon^d/n}}
    =
    \epsilon\sqrt{n\epsilon^d}\,h^s=o(1)
\]
since $\epsilon\leq\epsilon_0$ and $\sqrt{n\epsilon^d}\,h^s\rightarrow0$ by assumption. For the centered empirical fluctuation,
\begin{align*}
    \frac{n-1}{n}\mathbb{P}_n
    \left[
        \int
        \frac{1}{h^d}
        K\left(\frac{x-X}{h}\right)
        \frac{\mathbb{E}[a(\mathcal Z)\mid X=x]}{p(x)}
        \,dx
        -
        \frac{\mathbb{E}[a(\mathcal Z)\mid X]}{p(X)}
    \right]
\end{align*}
write $\phi(x)=\mathbb E[a(\mathcal Z)\mid X=x]/p(x)$ and $K_h(x):=h^{-d}K(x/h)$. By symmetry of $K$ the summand is $(K_h*\phi-\phi)(X)$, so by independence the fluctuation has second moment at most
\[
    \frac{1}{n}\,\mathbb{E}\big\|(K_h*\phi)(X)-\phi(X)\big\|_2^2
    \leq
    \frac{\|p\|_\infty}{n}\,\big\|K_h*\phi-\phi\big\|_{L^2(\mathbb R^d)}^2,
\]
where $\|p\|_\infty<\infty$ since $p$ is continuous with compact support. If $\epsilon\rightarrow0$, Young's convolution inequality \citep[Theorem 1.2.12]{grafakos2014classical} and \eqref{eq: conditional a integral bounds} give $\|K_h*\phi-\phi\|_{L^2}\leq(\|K\|_{L^1}+1)\|\phi\|_{L^2}=O(\epsilon^{d/2+1})$, so the fluctuation is $O_P(\epsilon^{d/2+1}/\sqrt n)$ and
\[
    \frac{\epsilon^{d/2+1}/\sqrt n}{\sqrt{\epsilon^d/n}}
    =
    \epsilon=o(1).
\]
If $\epsilon$ is fixed, this ratio is a constant and we instead use $h\rightarrow0$. In particular, since $K\in L^1(\mathbb R^d)$ with $\int K=1$, the family $(K_h)_{h>0}$ is an approximate identity (\citep{grafakos2014classical}[Definition 1.2.15]), and therefore $\|K_h*\phi-\phi\|_{L^2}\rightarrow0$ as $h\rightarrow0$ \citep[Theorem 1.2.19]{grafakos2014classical}, so the fluctuation is $o_P(1/\sqrt n)=o_P(\sqrt{\epsilon^d/n})$. Hence the entire second term in \eqref{eq: repaired exact remainder decomposition} is $o_P(\sqrt{\epsilon^d/n})$ in both regimes. For the third term, the population first-order conditions give $\mathbb{E}[\mathbb{E}[a(\mathcal Z)\mid X]/p(X)]=0$. Thus \eqref{eq: conditional a integral bounds} and \eqref{eq: a moment bounds remainder} imply
\[
    \mathbb{P}_n\left[
        \frac{\mathbb{E}[a(\mathcal Z)\mid X]}{p(X)}
    \right]
    =
    O_P\left(\frac{\epsilon^{d/2+1}}{\sqrt n}\right),
    \qquad
    \mathbb{P}_n\left[\frac{a(\mathcal Z)}{p(X)}\right]
    =
    O_P\left(\sqrt{\frac{\epsilon^d}{n}}\right).
\]
Multiplication by $1/n$ therefore makes the third term
$o_P(\sqrt{\epsilon^d/n})$. For the degenerate U-statistic term, degeneracy implies
\[
\mathbb E\left\|
\frac{1}{n^2}\sum_{i<j}m(W_i,W_j)
\right\|_2^2
\lesssim
\frac{\mathbb E\|m(W_1,W_2)\|_2^2}{n^2}
\lesssim
\frac{\mathbb E\|t(W_1,W_2)\|_2^2}{n^2}.
\]
Conditioning on $X_1$ gives
$\mathbb{E}[h^{-2d}K((X_2-X_1)/h)^2\mid X_1]=O(h^{-d})$, and
\eqref{eq: a moment bounds remainder} therefore yields
$\mathbb{E}\|t(W_1,W_2)\|_2^2=O(\epsilon^d/h^d)$. Consequently
\[
    \frac{1}{n^2}\sum_{i<j}m(W_i,W_j)
    =
    O_P\left(\sqrt{\frac{\epsilon^d}{n^2h^d}}\right)
    =
    o_P\left(\sqrt{\frac{\epsilon^d}{n}}\right),
\]
where the final step follows from $nh^d\to\infty$. Finally, by \eqref{eq: conditional a bound remainder},
$\|\mathbb{E}[a(\mathcal Z)/p(X)^2]\|_2=O(\epsilon^{d+1})$, while
\eqref{eq: a moment bounds remainder} gives
\[
    \mathbb{P}_n\left[\frac{a(\mathcal Z)}{p(X)^2}\right]
    -
    \mathbb{E}\left[\frac{a(\mathcal Z)}{p(X)^2}\right]
    =
    O_P\left(\sqrt{\frac{\epsilon^d}{n}}\right).
\]
Hence the diagonal term satisfies
\[
    \frac{K(0)}{nh^d}
    \mathbb{P}_n\left[\frac{a(\mathcal Z)}{p(X)^2}\right]
    =
    O\left(\frac{\epsilon^{d+1}}{nh^d}\right)
    +
    O_P\left(\frac{\epsilon^{d/2}}{n^{3/2}h^d}\right).
\]
Dividing by $\sqrt{\epsilon^d/n}$, these two terms become $\epsilon\sqrt{n\epsilon^d}/(nh^d)$ and $1/(nh^d)$, which tend to zero because $\epsilon\leq\epsilon_0$, $\sqrt{n\epsilon^d}\log n/(nh^d)\rightarrow0$, and $nh^d/\log n\rightarrow\infty$ by assumption. Hence both terms are
$o_P(\sqrt{\epsilon^d/n})$. Therefore every remainder in
\eqref{eq: repaired exact remainder decomposition} is negligible at the desired scale, and
\[
    R_1
    =
    \frac{1}{n}\sum_{i=1}^n
    \frac{\mathbb{E}[a(\mathcal Z)\mid X=X_i]}{p(X_i)}
    +
    o_P\left(\sqrt{\frac{\epsilon^d}{n}}\right).
\]
\end{proof}

\begin{restatable}{lemma}{} \label{lemma: Lambda trace bound}
    Let $w = \mathbbm{1}_{\{X \in \mathcal{B}(0, \epsilon)\}}\frac{p^*(X)}{p(X)}$ where $p$ is the density of $X$ (with $X \sim p$), let $u = Y - \boldsymbol{X}^{\top}\theta_{x_0, \epsilon}^*$, and finally let $\Lambda = \operatorname{Cov}(w\boldsymbol{X}u)$. Suppose there exists $\epsilon_0>0$ such that $p$ is bounded away from zero and $p^*$ is bounded above on $\mathcal{B}(0,\epsilon_0)$, and suppose that $\sup_{x\in\mathcal{B}(0,\epsilon_0)}\mathbb{E}[|\varepsilon|^3\mid X=x]<\infty$. Then, for all sufficiently small $\epsilon$,
    \[\operatorname{tr}\left(\Lambda\right) \lesssim \epsilon^{d}.\]
\end{restatable}
Note that this lemma uses the uncentered covariance $\operatorname{Cov}(w\boldsymbol{X}u)$ for a trace bound, while the CLT sections use the centered covariance $\operatorname{Cov}(w\boldsymbol{X}(u - \mathbb{E}[u\,|\,\boldsymbol{X}]))$.

\begin{proof} Note that population first order conditions for the optimization in \eqref{eq: population estimator} indicate that $\mathbb{E}[w\boldsymbol{X}u] = 0$. Thus, $\Lambda = \mathbb{E}[w^2u^2\boldsymbol{X}\boldsymbol{X}^{\top}]$, and therefore
\begin{align*}
    \operatorname{tr}(\Lambda)
    &= \mathbb{E}\left[w^2u^2\|\boldsymbol{X}\|_2^2\right] \\
    &\leq 2w_{\max,\epsilon}\mathbb{E}\left[wu^2\right],
\end{align*}
where $w_{\max,\epsilon} = \sup_{x \in \mathcal{B}(0, \epsilon)}p^*(x)/p(x)$ and we have used $\|\boldsymbol{X}\|_2^2\leq 2$ on $\mathcal{B}(0,\epsilon)$ together with $w^2\leq w_{\max,\epsilon}w$. By the local upper and lower bounds on $p^*$ and $p$, $w_{\max,\epsilon}$ is uniformly bounded for all sufficiently small $\epsilon$. Since $\theta^*_{x_0,\epsilon}$ minimizes the population weighted least-squares objective, comparison with $\theta=0$ gives $\mathbb{E}[wu^2]\leq \mathbb{E}[wY^2]$. Moreover, by the definition of $w$,
\begin{align*}
    \mathbb{E}[wY^2]
    &= \int_{\mathcal{B}(0,\epsilon)}p^*(x)\mathbb{E}[Y^2\mid X=x]\,dx.
\end{align*}
Under the model $Y=f(g(X))+\varepsilon$ and $\mathbb{E}[\varepsilon\mid X]=0$, we have $\mathbb{E}[Y^2\mid X=x]=f(g(x))^2+\mathbb{E}[\varepsilon^2\mid X=x]$. Since $f\circ g$ is locally bounded and, by Lyapunov's inequality \citep[Theorem 2.5]{gut2013probability}, $\mathbb{E}[\varepsilon^2\mid X=x]\leq \mathbb{E}[|\varepsilon|^3\mid X=x]^{2/3}$, the assumptions imply that $\sup_{x\in\mathcal{B}(0,\epsilon_0)}\mathbb{E}[Y^2\mid X=x]<\infty$. Hence,
\begin{align*}
    \mathbb{E}[wY^2]
    &\lesssim \mathcal{L}^d(\mathcal{B}(0,\epsilon))
    \lesssim \epsilon^d.
\end{align*}
Combining the preceding bounds yields $\operatorname{tr}(\Lambda)\lesssim\epsilon^d$.
\end{proof}

\subsection{Supplementary results for \Cref{corr: central limit for v}}

\begin{restatable}{lemma}{} \label{lemma: asymptotic var bound}
    Let $Q=\mathbb{E}[w\boldsymbol X\boldsymbol X^\top]$ and $\Lambda=\mathbb{E}[w^2\sigma^2(X)\boldsymbol{X}\boldsymbol{X}^{\top}]$, so that $\Sigma_\epsilon=Q^{-1}\Lambda Q^{-1}$ is the matrix appearing in \Cref{corr: central limit for v}. Further suppose that $p$, $p^*(\cdot)$, and $\sigma^2$ are continuous at $0$, with $p(0)>0$ and $p^*(0)>0$, and that there exists $\epsilon_0>0$ such that $p$ is bounded away from $0$ and $p^*$ and $\sigma^2$ are bounded on $\mathcal{B}(0,\epsilon_0)$. Then
    \[
        \lim_{\epsilon \rightarrow 0}\epsilon^d\Sigma_\epsilon
        = \operatorname{diag}\left(\sigma_\mu^2,\,\Sigma_0\right),
        \qquad
        \sigma_{\mu}^2 = \frac{\sigma^2(0)}{p(0)\mathcal{L}^d(\mathcal{B}(0,1))},
        \qquad
        \Sigma_0 = \frac{(d+2)\sigma^2(0)}{p(0)\mathcal{L}^d(\mathcal{B}(0,1))}I_d.
    \]
\end{restatable}
\begin{proof} Observe that we can rewrite $Q$ as follows
\begin{align*}
    Q &= \mathbb{E}_{p^*}\left[\mathbbm{1}_{\{X \in \mathcal{B}(0, \epsilon)\}}\begin{pmatrix}
        1 \\ X/\epsilon
    \end{pmatrix}\begin{pmatrix}
        1 \\ X/\epsilon
    \end{pmatrix}^{\top}\right] \\
    &= \begin{bmatrix}
         \mathbb{P}_{p^*}(X \in \mathcal{B}(0, \epsilon)) & 0 \\
         0 & (1/\epsilon^2)\mathbb{E}\left[\mathbbm{1}_{\{X \in \mathcal{B}(0, \epsilon)\}}XX^{\top}\right]
    \end{bmatrix} \\
    &= \begin{bmatrix}
         \mathbb{P}_{p^*}(X \in \mathcal{B}(0, \epsilon)) & 0 \\
         0 & (\nu/\epsilon)^2I_d
    \end{bmatrix},
\end{align*}
where $\nu^2 = \mathbb{E}_{p^*}[\mathbbm{1}_{\{X \in  \mathcal{B}(0, \epsilon)\}}X_i^2]$ for any $i \in [d]$. The last two steps follow from the spherical symmetry of $p^*$. It follows that
\begin{align} \label{eq: Q closed form}
    Q^{-1} &= \begin{bmatrix}
         \mathbb{P}_{p^*}(X \in \mathcal{B}(0, \epsilon))^{-1} & 0 \\
         0 & (\epsilon/\nu)^{2}I_d
    \end{bmatrix}.
\end{align}
Next, writing $\psi(x) =p^*(x)^2\sigma^2(x)/p(x)$ and using the definition of $w$,
\begin{align*}
    \Lambda
    = \int_{\mathcal{B}(0, \epsilon)}\psi(x)\,\boldsymbol x\boldsymbol x^{\top}\,dx
    = \begin{bmatrix}
        \displaystyle\int_{\mathcal{B}(0, \epsilon)}\psi(x)\,dx
        & \displaystyle\epsilon^{-1}\int_{\mathcal{B}(0, \epsilon)}\psi(x)\,x^{\top}\,dx \\[3ex]
        \displaystyle\epsilon^{-1}\int_{\mathcal{B}(0, \epsilon)}\psi(x)\,x\,dx
        & \displaystyle\epsilon^{-2}\int_{\mathcal{B}(0, \epsilon)}\psi(x)\,xx^{\top}\,dx
    \end{bmatrix}.
\end{align*}
Combining this with \eqref{eq: Q closed form}, the blocks of $\Sigma_\epsilon=Q^{-1}\Lambda Q^{-1}$ are
\begin{align*}
    [\Sigma_\epsilon]_{11}
    &= \mathbb{P}_{p^*}(X \in \mathcal{B}(0, \epsilon))^{-2}\int_{\mathcal{B}(0, \epsilon)}\psi(x)\,dx, \\
    [\Sigma_\epsilon]_{1,2:d+1}
    &= \mathbb{P}_{p^*}(X \in \mathcal{B}(0, \epsilon))^{-1}\,(\epsilon/\nu)^{2}\,\epsilon^{-1}\int_{\mathcal{B}(0, \epsilon)}\psi(x)\,x^{\top}\,dx, \\
    V\Sigma_\epsilon V^{\top}
    &= (\epsilon/\nu)^{4}\,\epsilon^{-2}\int_{\mathcal{B}(0, \epsilon)}\psi(x)\,xx^{\top}\,dx,
\end{align*}
where $V=\begin{bmatrix}0 & I_d\end{bmatrix}$ selects the bottom-right $d\times d$ block. Now apply the change of variables $x=\epsilon z$, so that $dx=\epsilon^d dz$. Then
\[
    \mathbb{P}_{p^*}(X \in \mathcal{B}(0, \epsilon)) = \epsilon^d\int_{\mathcal{B}(0,1)}p^*(\epsilon z)\,dz,
    \qquad
    \nu^2 = \epsilon^{d+2}\int_{\mathcal{B}(0,1)}z_i^2\,p^*(\epsilon z)\,dz,
\]
and $\int_{\mathcal{B}(0,\epsilon)}\psi(x)\,x^{\otimes k}\,dx=\epsilon^{d+k}\int_{\mathcal{B}(0,1)}\psi(\epsilon z)\,z^{\otimes k}\,dz$ for $k\in\{0,1,2\}$. Substituting,
\begin{align}
    \epsilon^d[\Sigma_\epsilon]_{11}
    &= \frac{\int_{\mathcal{B}(0, 1)}\psi(\epsilon z)\,dz}{\left[\int_{\mathcal{B}(0, 1)}p^*(\epsilon z)\,dz\right]^2},
    \qquad
    \epsilon^d[\Sigma_\epsilon]_{1,2:d+1}
    = \frac{\int_{\mathcal{B}(0, 1)}\psi(\epsilon z)\,z^{\top}\,dz}{\int_{\mathcal{B}(0, 1)}p^*(\epsilon z)\,dz\cdot\int_{\mathcal{B}(0, 1)}z_i^2\,p^*(\epsilon z)\,dz},
    \nonumber\\
    \epsilon^dV\Sigma_\epsilon V^{\top}
    &= \frac{\int_{\mathcal{B}(0, 1)}\psi(\epsilon z)\,zz^{\top}\,dz}{\left[\int_{\mathcal{B}(0, 1)}z_i^2\,p^*(\epsilon z)\,dz\right]^2}. \label{eq: asymp var intermediate}
\end{align}
By continuity at $0$, for every $z\in\mathcal{B}(0,1)$ we have $p^*(\epsilon z)\rightarrow p^*(0)$ and $\psi(\epsilon z)\rightarrow\psi(0)=p^*(0)^2\sigma^2(0)/p(0)$. Moreover, the local boundedness assumptions and the lower bound on $p$ give $\sup_{\mathcal{B}(0,\epsilon_0)}\psi<\infty$, so the constant functions provide integrable dominating functions on $\mathcal{B}(0,1)$. Thus, by the dominated convergence theorem and \Cref{prop: spherical cov},
\begin{align*}
    \int_{\mathcal{B}(0,1)}\psi(\epsilon z)\,dz &\rightarrow \psi(0)\,\mathcal{L}^d(\mathcal{B}(0,1)),
    &
    \int_{\mathcal{B}(0,1)}p^*(\epsilon z)\,dz &\rightarrow p^*(0)\,\mathcal{L}^d(\mathcal{B}(0,1)), \\
    \int_{\mathcal{B}(0,1)}\psi(\epsilon z)\,z\,dz &\rightarrow \psi(0)\int_{\mathcal{B}(0,1)}z\,dz = 0,
    &
    \int_{\mathcal{B}(0,1)}z_i^2\,p^*(\epsilon z)\,dz &\rightarrow \frac{p^*(0)\,\mathcal{L}^d(\mathcal{B}(0,1))}{d+2}, \\
    \int_{\mathcal{B}(0,1)}\psi(\epsilon z)\,zz^{\top}\,dz &\rightarrow \frac{\psi(0)\,\mathcal{L}^d(\mathcal{B}(0,1))}{d+2}I_d.
\end{align*}
Plugging these limits into the three expressions above yields
\[
    \lim_{\epsilon \rightarrow 0}\epsilon^d[\Sigma_\epsilon]_{11}
    = \frac{\psi(0)}{p^*(0)^2\,\mathcal{L}^d(\mathcal{B}(0,1))}
    = \frac{\sigma^2(0)}{p(0)\mathcal{L}^d(\mathcal{B}(0,1))},
    \qquad
    \lim_{\epsilon \rightarrow 0}\epsilon^d[\Sigma_\epsilon]_{1,2:d+1} = 0,
\]
and
\[
    \lim_{\epsilon \rightarrow 0}\epsilon^dV\Sigma_\epsilon V^{\top}
    = \frac{(d+2)\,\psi(0)}{p^*(0)^2\,\mathcal{L}^d(\mathcal{B}(0,1))}I_d
    = \frac{(d+2)\sigma^2(0)}{p(0)\mathcal{L}^d(\mathcal{B}(0,1))}I_d,
\]
which together give $\lim_{\epsilon\rightarrow0}\epsilon^d\Sigma_\epsilon=\operatorname{diag}(\sigma_\mu^2,\Sigma_0)$.
\end{proof}

\subsection{Miscellaneous results}

\begin{restatable}[Boundedness of $\gamma^{(k)}$]{rmk}{} \label{remark: deriv bound}
    The curve $\gamma$ has first, second and third derivatives that are bounded in $\ell_2$ norm. Namely, for any $\gamma \in C^3(I)$ and $k \in \{1,2,3\}$ we have 
    \[\sup_{t\in I}\|\gamma^{(k)}(t)\|_2 \leq M_k\]
    for some $|M_k| < \infty$.
\end{restatable}
\begin{proof} Since $\gamma \in C^3(I)$, $\gamma^{(k)}$ is continuous for $k \leq 3$. Since $I$ is a compact set, by the Extreme Value Theorem $\gamma^{(k)}$ must be bounded on $I$.
\end{proof}

\begin{restatable}[Restricted spherical symmetry]{prop}{} \label{prop: conditional spherical symmetry}
    Suppose $X$ has a density $f(x)$ satisfying spherical symmetry (\Cref{def: spherical symmetry}) with mean $0$. Then for any $\epsilon > 0$ such that $\mathbb{P}(X\in\mathcal{B}(0,\epsilon))>0$, the distribution associated with the density $\tilde{f}(x) \propto \mathbbm{1}_{\{x \in \mathcal{B}(0, \epsilon)\}}f(x)$ is also spherically symmetric.
\end{restatable}
\begin{proof} Since $\mathbb{P}(X\in\mathcal{B}(0,\epsilon))>0$, we can write
\[
    \tilde f(x)
    =
    \frac{\mathbbm{1}_{\{x\in\mathcal{B}(0,\epsilon)\}}f(x)}
    {\mathbb{P}(X\in\mathcal{B}(0,\epsilon))}.
\]
For any orthogonal matrix $Q$, we have $Qx\in\mathcal{B}(0,\epsilon)$ if and only if $x\in\mathcal{B}(0,\epsilon)$. Moreover, spherical symmetry of $f$ implies $f(Qx)=f(x)$. Hence
\begin{align*}
    \tilde f(Qx)
    &=
    \frac{\mathbbm{1}_{\{Qx\in\mathcal{B}(0,\epsilon)\}}f(Qx)}
    {\mathbb{P}(X\in\mathcal{B}(0,\epsilon))} \\
    &=
    \frac{\mathbbm{1}_{\{x\in\mathcal{B}(0,\epsilon)\}}f(x)}
    {\mathbb{P}(X\in\mathcal{B}(0,\epsilon))}
    =
    \tilde f(x).
\end{align*}
Thus $\tilde f$ is spherically symmetric.

\end{proof}

\begin{restatable}[Perturbation]{lemma}{} \label{lemma: perturbation}
    Suppose $w\geq 0$, $\mathbb{E}[w\boldsymbol{X}\boldsymbol{X}^{\top}] \succ 0$, and $\mathbb{E}[w\|\boldsymbol{X}\|_2^2]$, $\mathbb{E}[wY^2]$, and $\mathbb{E}[w\xi(X)^2]$ are finite. Define
    \[\theta^* = \arg\min_{\theta}\mathbb{E}\big[w\big(Y - \theta^{\top}\boldsymbol{X}\big)^2\big] \qquad \text{and} \qquad \tilde{\theta}^* = \arg\min_{\theta}\mathbb{E}\big[w\big(Y - \xi(X) - \theta^{\top}\boldsymbol{X}\big)^2\big].\]
    Then
    \[\theta^* - \tilde{\theta}^* = \mathbb{E}[w\boldsymbol{X}\boldsymbol{X}^{\top}]^{-1}\mathbb{E}[w\boldsymbol{X}\xi(X)].\]
\end{restatable}
\begin{proof} The assumed integrability guarantees that the objectives and the relevant cross-moments are finite, while $\mathbb{E}[w\boldsymbol{X}\boldsymbol{X}^\top]\succ0$ makes both objectives strictly convex. Thus, the first order optimality conditions uniquely characterize the minimizers and give
\[
    \theta^* = \mathbb{E}[w\boldsymbol{X}\boldsymbol{X}^{\top}]^{-1}\mathbb{E}[w\boldsymbol{X}Y]
    \qquad \text{and} \qquad
    \tilde{\theta}^* = \mathbb{E}[w\boldsymbol{X}\boldsymbol{X}^{\top}]^{-1}\mathbb{E}[w\boldsymbol{X}(Y-\xi(X))].
\]
Taking the difference,
\begin{align*}
    \theta^* - \tilde{\theta}^*
    &= \mathbb{E}[w\boldsymbol{X}\boldsymbol{X}^{\top}]^{-1}\mathbb{E}[w\boldsymbol{X}(Y-Y+\xi(X))] \\
    &= \mathbb{E}[w\boldsymbol{X}\boldsymbol{X}^{\top}]^{-1}\mathbb{E}[w\boldsymbol{X}\xi(X)].
\end{align*}
\end{proof}

\begin{restatable}[Projection norm bound]{lemma}{} \label{lemma: pnorm bound} Suppose that $\boldsymbol{X} = \begin{bmatrix}
    1 & X/\epsilon
\end{bmatrix}^{\top}$ where $X \sim p$, and assume that we are operating under the hypotheses of \Cref{thm: population prop}. Then we have
\begin{equation}
   \left\|\mathbb{E}[w\boldsymbol{X}\boldsymbol{X}^{\top}]^{-1}\mathbb{E}[w\boldsymbol{X}]\right\|_2 =  1.
\end{equation}
\end{restatable}
\begin{proof} Let $e = \begin{bmatrix}
    1 & 0 & \dots & 0
\end{bmatrix}^{\top} \in \mathbb{R}^{d+1}$. Then for every $x \in \mathbb{R}^d$ we have
\begin{align*}
    \boldsymbol{x}\boldsymbol{x}^{\top}e &= \boldsymbol{x}(\boldsymbol{x}^{\top}e) \\
    &= \boldsymbol{x}.
\end{align*}
Multiplying by $w$ and taking the expectation yields
\[\mathbb{E}[w\boldsymbol{X}\boldsymbol{X}^{\top}]e = \mathbb{E}[w\boldsymbol{X}]\]
which, if the matrix on the LHS is invertible, implies
\[\mathbb{E}[w\boldsymbol{X}\boldsymbol{X}^{\top}]^{-1}\mathbb{E}[w\boldsymbol{X}] = e.\]
Taking the norm completes the proof.

\end{proof}

\begin{restatable}{lemma}{}\label{lemma: operator norm and var proxy}
    Suppose there exists $\epsilon_0>0$ such that $p$ is bounded away from zero and $p,p^*$ are bounded above on $\mathcal{B}(0,\epsilon_0)$. Let $w_{\max} = \sup_{x \in \mathcal{B}(0, \epsilon)}p^*(x)/p(x)$. Then the matrix $Y_i = w_i\boldsymbol{X}_i\boldsymbol{X}_i^{\top} - \mathbb{E}[w\boldsymbol{X}\boldsymbol{X}^{\top}]$ has operator norm bounded by $4w_{\max}$ and variance proxy $\sigma^2 = \|\sum_{i = 1}^n \mathbb{E}[Y_i^2]\|_{\operatorname{op}} \lesssim n\epsilon^d$.
\end{restatable} 
\begin{proof} Since $\|\boldsymbol{X}_i\|_2^2\leq2$ whenever $w_i\neq0$, we have
\begin{align*}
    \|Y_i\|_{\operatorname{op}}
    &\leq w_i\|\boldsymbol{X}_i\boldsymbol{X}_i^{\top}\|_{\operatorname{op}}
    +\|\mathbb{E}[w\boldsymbol{X}\boldsymbol{X}^{\top}]\|_{\operatorname{op}} \\
    &\leq 2w_{\max}+2w_{\max}
    =4w_{\max}.
\end{align*}
Moreover, since the $X_i$'s are i.i.d.,
\[
    \sigma^2=n\|\mathbb{E}[Y^2]\|_{\operatorname{op}}.
\]
By expanding the square and using $\mathbb{E}[Y]=0$,
\[
    \mathbb{E}[Y^2]
    =
    \mathbb{E}[(w\boldsymbol{X}\boldsymbol{X}^{\top})^2]
    -
    \left(\mathbb{E}[w\boldsymbol{X}\boldsymbol{X}^{\top}]\right)^2
    \preceq
    \mathbb{E}[(w\boldsymbol{X}\boldsymbol{X}^{\top})^2].
\]
Since $(w\boldsymbol{X}\boldsymbol{X}^{\top})^2=w^2\|\boldsymbol{X}\|_2^2\boldsymbol{X}\boldsymbol{X}^{\top}$ and $\|\boldsymbol{X}\|_2^2\leq2$ on $\mathcal{B}(0,\epsilon)$,
\begin{align*}
    \|\mathbb{E}[Y^2]\|_{\operatorname{op}}
    &\leq \mathbb{E}\|(w\boldsymbol{X}\boldsymbol{X}^{\top})^2\|_{\operatorname{op}} \\
    &= \mathbb{E}[w^2\|\boldsymbol{X}\|_2^4] \\
    &\leq 4w_{\max}^2\mathbb{P}(X\in\mathcal{B}(0,\epsilon))
    \lesssim \epsilon^d,
\end{align*}
where the final bound follows from the local upper bound on $p$. Therefore, $\sigma^2\lesssim n\epsilon^d$.
\end{proof}

\begin{restatable}{lemma}{}
    Let $Q = \mathbb{E}\left[w\boldsymbol{X}\boldsymbol{X}^{\top}\right]$
    and assume that $p$ and $p^*$ are lower bounded by some $m > 0$ and upper bounded by some finite value on $\mathcal{B}(0, \epsilon)$. We then have
    $\lambda_{\min}(Q) \gtrsim \epsilon^d$ where $\lambda_{\min}(A)$ denotes the minimum eigenvalue of the matrix $A$. 
    \label{lemma: Q op norm}
\end{restatable}
\begin{proof} We have
\begin{align*}
    \lambda_{\min}(Q) &= \min_{\|v\|_2 = 1}v^{\top}Qv \\
    &= \min_{\|v\|_2 = 1} \mathbb{E}[wv^{\top}\boldsymbol{X}\boldsymbol{X}^{\top}v] \\
    &= \min_{\|v\|_2 = 1} \mathbb{E}[w(\boldsymbol{X}^{\top}v)^2]
\end{align*}
Now let $w_{\min} = \inf_{x \in \mathcal{B}(0, \epsilon)}p^*(x)/p(x)$ -- it follows that
\begin{align*}
    \lambda_{\min}(Q) &\gtrsim \min_{\|v\|_2 = 1}\mathbb{E}[\mathbbm{1}_{\{X \in \mathcal{B}(0, \epsilon)\}}(\boldsymbol{X}^{\top}v)^2] \\
    &= \min_{\|v\|_2 = 1}\int_{\mathcal{B}(0, \epsilon)}(\boldsymbol{x}^{\top}v)^2 dp(x) \\
    &\geq m \mathcal{L}^d(\mathcal{B}(0, \epsilon))\min_{\|v\|_2 = 1}\int_{\mathcal{B}(0, \epsilon)}(\boldsymbol{x}^{\top}v)^2 du(x)
\end{align*}
where $u$ is the uniform measure over $\mathcal{B}(0, \epsilon)$. We then have
\begin{align*}
    \lambda_{\min}(Q)
    &\gtrsim \mathcal{L}^d(\mathcal{B}(0,\epsilon))
    \min_{\|v\|_2=1}
    v^\top
    \begin{bmatrix}
        1 & 0 \\
        0 & \mathbb{E}[(X/\epsilon)(X/\epsilon)^\top]
    \end{bmatrix}
    v.
\end{align*}
By \Cref{prop: spherical cov},
\[
    \mathbb{E}[(X/\epsilon)(X/\epsilon)^\top]
    =
    \frac{1}{d+2}I_d,
\]
where the expectation is with respect to the uniform measure on $\mathcal{B}(0,\epsilon)$. Therefore,
\begin{align*}
    \lambda_{\min}(Q)
    &\gtrsim
    \mathcal{L}^d(\mathcal{B}(0,\epsilon))
    \min\left\{1,\frac{1}{d+2}\right\} \\
    &=
    \frac{\mathcal{L}^d(\mathcal{B}(0,\epsilon))}{d+2}
    \gtrsim \epsilon^d.
\end{align*}

\end{proof}

\begin{restatable}{lemma}{}
    Let
    \[
        Q = \mathbb{E}\left[w\boldsymbol{X}\boldsymbol{X}^{\top}\right]
        \qquad\text{and}\qquad
        \hat{Q} = \frac{1}{n}\sum_{i = 1}^n\hat{w}_i\boldsymbol{X}_i\boldsymbol{X}_i^{\top},
    \]
    and suppose the assumptions of \Cref{lemma: Q op norm} and \cref{assumption: lce and covariate density} hold. Then we have 
    \[
        \left\|\hat{Q}^{-1} - Q^{-1}\right\|_{\operatorname{op}}
        =
        O_P\left(\frac{1}{\epsilon^{3d/2}\sqrt{n}} + \frac{1}{n\epsilon^{2d}}\right)
        + o_P\left(\frac{1}{\epsilon^d}\right)
    \]
    when $h\rightarrow0$, $nh^d/\log n\rightarrow\infty$, and $n\epsilon^d\rightarrow\infty$.
    \label{lemma: Q convergence}
\end{restatable}
\begin{proof} Before establishing the convergence of $\hat Q$, we will bound $\hat w_i-w_i$. Let $\Delta(x)=\hat p(x)-p(x)$. A second-order Taylor expansion of $t\mapsto1/t$ gives
\begin{align*}
    |\hat{w}_i - w_i|
    &\leq \mathbbm{1}_{\{X_i \in \mathcal{B}(0, \epsilon)\}}p^*(X_i)
    \left[\frac{\sup_{x \in \mathcal{B}(0, \epsilon)}|\Delta(x)|}{m^2}
    + \xi_i^{-3}\sup_{x \in \mathcal{B}(0, \epsilon)}|\Delta(x)|^2\right],
\end{align*}
where $\xi_i$ lies between $\hat p(X_i)$ and $p(X_i)$. By \Cref{lemma: uniform convergence of kde},
\[
    \sup_{x \in \mathcal{B}(0, \epsilon)}|\Delta(x)|
    =
    O_P\left(\sqrt{\frac{\log n}{nh^d}}+h^s\right)
    =
    o_P(1),
\]
where the final equality follows from $h\rightarrow0$ and $nh^d/\log n\rightarrow\infty$. Thus, by the same argument used in the proof of \Cref{thm: decomposition for theta}, $\xi_i^{-3}$ is uniformly bounded with probability approaching $1$, and therefore
\[
    |\hat w_i-w_i|
    =
    \mathbbm{1}_{\{X_i\in\mathcal B(0,\epsilon)\}}p^*(X_i)o_P(1)
\]
uniformly over $i$. Now define $\tilde Q=n^{-1}\sum_{i=1}^nw_i\boldsymbol{X}_i\boldsymbol{X}_i^\top$. Since $p^*$ is locally bounded and $\|\boldsymbol{X}_i\|_2^2\leq2$ on $\mathcal B(0,\epsilon)$, the preceding uniform bound implies
\begin{align*}
    \|\hat Q-\tilde Q\|_{\operatorname{op}}
    \leq \frac{1}{n}\sum_{i=1}^n
    |\hat w_i-w_i|\,\|\boldsymbol{X}_i\boldsymbol{X}_i^\top\|_{\operatorname{op}} \leq
    o_P(1)\frac{1}{n}\sum_{i=1}^n
    \mathbbm{1}_{\{X_i\in\mathcal B(0,\epsilon)\}} =o_P(\epsilon^d),
\end{align*}
where the final equality follows from
\[
\frac{1}{n}\sum_{i=1}^n
\mathbbm{1}_{\{X_i\in\mathcal B(0,\epsilon)\}}
=O_P(\epsilon^d),
\]
since $\mathbb P(X\in\mathcal B(0,\epsilon))=O(\epsilon^d)$ by the local upper bound on $p$. Moreover, \Cref{lemma: operator norm and var proxy} and the matrix Bernstein inequality (\Cref{thm: matrix bernstein}) imply
\begin{align*}
    \|\tilde Q-Q\|_{\operatorname{op}}
    &= O_P\left(\frac{\sqrt{n\epsilon^d}}{n}+\frac{1}{n}\right) = O_P\left(\sqrt{\frac{\epsilon^d}{n}}+\frac{1}{n}\right).
\end{align*}
Hence
\[
    \|\hat Q-Q\|_{\operatorname{op}}
    =
    O_P\left(\sqrt{\frac{\epsilon^d}{n}}+\frac{1}{n}\right)
    +o_P(\epsilon^d).
\]
By \Cref{lemma: Q op norm}, $\lambda_{\min}(Q)\gtrsim\epsilon^d$. Since $n\epsilon^d\rightarrow\infty$,
\[
    \|\hat Q-Q\|_{\operatorname{op}}=o_P(\epsilon^d).
\]
Weyl's inequality therefore implies $\lambda_{\min}(\hat Q)\gtrsim\epsilon^d$ with probability approaching $1$, and consequently $\|Q^{-1}\|_{\operatorname{op}},\|\hat Q^{-1}\|_{\operatorname{op}}\lesssim\epsilon^{-d}$ with probability approaching $1$. Finally,
\begin{align*}
    \|\hat Q^{-1}-Q^{-1}\|_{\operatorname{op}}
    &\leq \|Q^{-1}\|_{\operatorname{op}}
    \|Q-\hat Q\|_{\operatorname{op}}
    \|\hat Q^{-1}\|_{\operatorname{op}} \\
    &=
    O_P\left(\frac{1}{\epsilon^{3d/2}\sqrt n}
    +\frac{1}{n\epsilon^{2d}}\right)
    +o_P\left(\frac{1}{\epsilon^d}\right).
\end{align*}

\end{proof}

\begin{restatable}{lemma}{} \label{lemma: pd}
Suppose $p$ and $p^*$ are continuous at $0$,
with $p(0)>0$ and $p^*(0)>0$.
Then, for all sufficiently small $\epsilon>0$,
the matrix
$Q=\mathbb E_p[w\boldsymbol X\boldsymbol X^\top]$
is positive definite.
\end{restatable}
\noindent By continuity and positivity at $0$, both $p$ and $p^*$ are strictly positive throughout
$\mathcal B(0,\epsilon)$ for all sufficiently
small $\epsilon>0$. Thus, expanding $w$ yields
\[
Q=\mathbb E_p[w\boldsymbol X\boldsymbol X^\top]
=\mathbb E_{p^*}
[\mathbbm{1}_{\{X\in\mathcal B(0,\epsilon)\}}
\boldsymbol X\boldsymbol X^\top].
\]
Thus, for any $v=(v_0,v_1^\top)^\top\in\mathbb{R}^{d+1}$ with $v\neq0$,
\[
    v^\top\mathbb{E}_{p}[w\boldsymbol{X}\boldsymbol{X}^{\top}]v
    =
    \int_{\mathcal{B}(0,\epsilon)}
    \left(v_0+\frac{v_1^\top x}{\epsilon}\right)^2
    p^*(x)\,dx.
\]
The affine function $x\mapsto v_0+v_1^\top x/\epsilon$ is not identically zero when $v\neq0$, and its zero set has Lebesgue measure zero unless $v_1=0$, in which case it is the nonzero constant $v_0$. Since $p^*(x)>0$ almost everywhere on $\mathcal{B}(0,\epsilon)$, the integral is strictly positive. Hence
\[
    \mathbb{E}_{p}[w\boldsymbol{X}\boldsymbol{X}^{\top}]\succ0.
\]

\qed{}

\begin{restatable}[Covariance of $\text{Unif}(\mathcal{B}(0,\epsilon))$]{prop}{} \label{prop: spherical cov}
    Suppose we have $X$ distributed according to a uniform probability measure over $\mathcal{B}(0,\epsilon)$ in $\mathbb{R}^d$. Then the covariance matrix of $X$ is $\frac{\epsilon^2}{d+2}I_d$.
\end{restatable}
\begin{proof} By symmetry, the covariance matrix must be a multiple of $I_d$. To find the constant, we need to understand the distribution of $\|X\|_2$ -- thus we will compute the radial pdf. Note that the cdf of $\|X\|_2$ can be written as
\begin{align*}
    \mathbb{P}(\|X\|_2 \leq r) &= \frac{\mathcal{L}^d(\mathcal{B}(0,r))}{\mathcal{L}^d(\mathcal{B}(0,\epsilon))} \\
    &= \left(\frac{r}{\epsilon }\right)^d
\end{align*}
where $\mathcal{L}$ denotes the $d$-dimensional Lebesgue measure on $\mathbb{R}^d$. The radial pdf can be obtained by differentiating the cdf with respect to $r$,
\begin{align*}
    f_r(r) &= \frac{d}{dr}\mathbb{P}(\|X\|_2\leq r) \\
    &= \frac{dr^{d-1}}{\epsilon^d}.
\end{align*}
Now we can compute the covariance scaling factor as 
\begin{align}
    \mathbb{E}[X_i^2] &= \frac{1}{d}\mathbb{E}\|X\|_2^2 \nonumber  =\frac{1}{\epsilon^d}\int_{0}^{\epsilon}r^2r^{d-1}dr \nonumber  = \frac{1}{\epsilon^d}\frac{1}{d+2}r^{d+2} \Bigg|^{r=\epsilon}_{r=0} \nonumber = \frac{\epsilon^2}{d+2}. \label{eq: uniform eps-ball coordinate variance}
\end{align}
\end{proof}

\subsection{External results}

\begin{restatable}[Matrix Bernstein, \citet{vershynin2018high} Theorem 5.4.1]{thm}{}
    Let $A_1, \dots, A_n$ be independent, mean-zero, $d\times d$ symmetric random matrices such that $\|A_i\|_{\operatorname{op}} \leq K$ almost surely for all $i$. Then for every $t \geq 0$ we have
    \[\mathbb{P}\left(\left\|\sum_{i = 1}^nA_i\right\|_{\operatorname{op}} \geq t\right) \leq 2d\exp\left(-\frac{t^2/2}{\sigma^2 + Kt/3}\right)\]
    where $\sigma^2 = \left\|\sum_{i = 1}^n\mathbb{E}A_i^2\right\|_{\operatorname{op}}$. \label{thm: matrix bernstein}
\end{restatable}

\begin{restatable}[Weyl's Inequality, \citet{bhatia2013matrix} Theorem \Romannum{3}.2.1]{thm}{}  \label{thm: weyl}
    Let $A, B$ be $d\times d$ Hermitian matrices. Then
    \[\lambda_j(A+B) \leq \lambda_i(A) + \lambda_{j - i+1}(B) \qquad \text{for $i\leq j$ }\]
    where $\lambda_i(A)$ denotes the $i^{\text{th}}$ largest eigenvalue of the matrix $A$. 
\end{restatable}

\begin{restatable}[Multivariate Lyapunov Central Limit Theorem]{thm}{} \label{thm: lyapunov clt}
    Let $\{U_{n, i}: 1 \leq i \leq n < \infty\}$ be a triangular array of independent random variables in $\mathbb{R}^d$ with mean zero and finite second moment. Define $V_n = \sum_{i= 1}^n \operatorname{Cov}(U_{n,i})$, and let $v_n^2 = \lambda_{\min}(V_n)$ for all $n$. If, for some $\delta > 0$, Lyapunov's condition
    \[\frac{1}{v_n^{2+\delta}}\sum_{i = 1}^n \mathbb{E}\left[\|U_{n, i}\|_2^{2+\delta}\right] = o(1)\]
    holds, then $V_n^{-1/2}\sum_{i = 1}^nU_{n, i} \overset{d}{\rightarrow} N(0, I_d)$.
\end{restatable}

\section{Further experimental details}\label{experimental details}

\paragraph{Curves and data.} The synthetic experiments use the model in \eqref{eq: the model}, with $f$ the identity and $\varepsilon_i\sim\mathcal{N}(0,0.01^2)$. The first three curves in \Cref{fig:simulation-curves} are
\[
\gamma(t)=
\begin{cases}
    (t,t), & \text{Identity},\\
    (0.5+0.4\cos(2\pi t),\;0.5+0.4\sin(2\pi t)), & \text{Circle},\\
    (t,\;0.5+0.4\sin(2\pi t)), & \text{Sinusoid},
\end{cases}
\qquad t\in[0,1].
\]
The polygon joins $(0.15,0.20)$, $(0.85,0.25)$, $(0.75,0.80)$, $(0.35,0.90)$, and $(0.10,0.55)$ in order, with arc length rescaled to $[0,1]$. For $d>2$, the remaining curve coordinates are zero. Projection models take $g=\Pi$, and covariates are uniform on $[0,1]^d$, except that the lifted experiment replaces the third coordinate by independent $\mathcal{N}(0,10^{-6})$ noise.

\paragraph{Fixed-point comparisons.} For \Cref{fig:fixed-point-planar,fig:fixed-point-lifted}, we use $B=100$ independent training trials at $n\in\{500,1000,2500,5000,10000\}$, with $x_0$ fixed across trials and sample sizes. Training samples are nested, so increasing $n$ adds observations while retaining previous $(X_i,Y_i)$. In the lifted experiment, the curves and inference points remain at $x_3=0$. Each method uses one configuration per curve and dimension, shared by velocity and regression. LLSIR's localization radius $\epsilon$ and any finite Gaussian reference-density standard deviation scale as $n^{-1/(d+4)}$, while other settings are fixed. NSVM's slice count is capped at $\lfloor n/8\rfloor$. Configurations were selected using 50 seeded trials at $n\in\{5000,7500,10000\}$, which are included in the plotted averages. Nodes report mean squared error with one Monte Carlo standard error.

\paragraph{Density reweighting experiment.} \Cref{fig:reweighting} repeats the planar fixed-point protocol with covariates drawn from a Gaussian mixture truncated to $[0,1]^2$ and constructed around the fixed inference points. Unlike the main experiments, where the localization radius shrinks as $n^{-1/(d+4)}$, here the radius and the reference standard deviation are held fixed across sample sizes at the values the shared schedule assigns at $n=10^4$.

\paragraph{Bounded-variation link experiment.} \Cref{fig:reweighting-bv} repeats the planar fixed-point experiment unchanged except that the link function is replaced by the bounded-variation link $f(t)=t+0.25\sum_j\mathbbm{1}\{t \geq s_j\}$ applied to the projection index. The bounded variation link has one jump per distinct inference point at $s_j = \Pi(x_0)+0.02$. Every localization window in the schedule contains its jump, while $f'=1$ almost everywhere leaves the velocity target unchanged. 

\paragraph{Dimension dependence.} The tables use $d\in\{2,3,4,5,6\}$ and $B=200$ independent trials, each with nested training samples at
\[
    n\in\{5000,7155,10238,14651,20965,30000\}.
\]
Each trial evaluates five fixed points $x_0$. Both the velocity and the conditional mean come from the same density-weighted local linear fit, with Gaussian reference density $p^*$ of standard deviation $\epsilon/2$, and the conditional mean estimate is the intercept of this fit. No tuning or NSVM fitting is performed. For each trial, squared errors are averaged over the five query points before fitting a line against $\log n$ on the log-log scale. Tables report the mean and sample standard deviation of all 200 complete trial-specific slopes.

\paragraph{Cellular reprogramming.} The UMAP visualization in \Cref{fig:reprogramming-umap} uses the saved PCA coordinates, 60 neighbors, minimum distance $0.4$, and seed 42. Directions are mapped to UMAP by local ridge regression over 60 expression-space neighbors, with penalty $0.01$ times the mean local covariance eigenvalue. UMAP does not affect the predictions.

\end{appendices}
\bibliography{main}

@article{li1991sliced,
  title={Sliced inverse regression for dimension reduction},
  author={Li, Ker-Chau},
  journal={Journal of the American Statistical Association},
  volume={86},
  number={414},
  pages={316--327},
  year={1991},
  publisher={Taylor \& Francis}
}

@article{gaiffas2007optimal,
  author  = {Ga{\"i}ffas, St{\'e}phane and Lecu{\'e}, Guillaume},
  title   = {Optimal Rates and Adaptation in the Single-Index Model Using Aggregation},
  journal = {Electronic Journal of Statistics},
  volume  = {1},
  pages   = {538--573},
  year    = {2007},
  doi     = {10.1214/07-EJS077}
}

@article{li1989regression,
  title={Regression analysis under link violation},
  author={Li, Ker-Chau and Duan, Naihua},
  journal={The Annals of Statistics},
  volume={17},
  number={3},
  pages={1009--1052},
  year={1989},
  publisher={Institute of Mathematical Statistics}
}

@article{chen2013estimation,
  title={Estimation in single-index panel data models with heterogeneous link functions},
  author={Chen, Jia and Gao, Jiti and Li, Degui},
  journal={Econometric Reviews},
  volume={32},
  number={8},
  pages={928--955},
  year={2013},
  publisher={Taylor \& Francis}
}

@incollection{lee2003smooth,
  title={Smooth manifolds},
  author={Lee, John M},
  booktitle={Introduction to smooth manifolds},
  pages={1--29},
  year={2003},
  publisher={Springer}
}

@article{federer1959curvature,
  title={Curvature measures},
  author={Federer, Herbert},
  journal={Transactions of the American Mathematical Society},
  volume={93},
  number={3},
  pages={418--491},
  year={1959},
  publisher={JSTOR}
}

@inbook{Newey_Ruud_2005, place={Cambridge}, title={Density Weighted Linear Least Squares}, booktitle={Identification and Inference for Econometric Models: Essays in Honor of Thomas Rothenberg}, publisher={Cambridge University Press}, author={Newey, Whitney K. and Ruud, Paul A.}, editor={Andrews, Donald W. K. and Stock, James H.Editors}, year={2005}, pages={554–573}}

@article{hansen2008uniform,
  title={Uniform convergence rates for kernel estimation with dependent data},
  author={Hansen, Bruce E},
  journal={Econometric Theory},
  volume={24},
  number={3},
  pages={726--748},
  year={2008},
  publisher={Cambridge University Press}
}

@article{li2005contour,
  title={Contour regression: a general approach to dimension reduction},
  author={Li, Bing and Zha, Hongyuan and Chiaromonte, Francesca},
  year={2005}
}

@article{li2007directional,
  title={On directional regression for dimension reduction},
  author={Li, Bing and Wang, Shaoli},
  journal={Journal of the American Statistical Association},
  volume={102},
  number={479},
  pages={997--1008},
  year={2007},
  publisher={Taylor \& Francis}
}

@article{dennis2000save,
  title={SAVE: a method for dimension reduction and graphics in regression},
  author={Dennis Cook, R},
  journal={Communications in statistics-Theory and methods},
  volume={29},
  number={9-10},
  pages={2109--2121},
  year={2000},
  publisher={Taylor \& Francis}
}

@article{genovese2012minimax,
  title={Minimax manifold estimation},
  author={Genovese, Christopher R and Perone-Pacifico, Marco and Verdinelli, Isabella and Wasserman, Larry},
  journal={The Journal of Machine Learning Research},
  volume={13},
  number={1},
  pages={1263--1291},
  year={2012},
  publisher={JMLR. org}
}

@inproceedings{balakrishnan2012minimax,
  title={Minimax rates for homology inference},
  author={Balakrishnan, Sivaraman and Rinaldo, Alesandro and Sheehy, Don and Singh, Aarti and Wasserman, Larry},
  booktitle={Artificial Intelligence and Statistics},
  pages={64--72},
  year={2012},
  organization={PMLR}
}

@inproceedings{kokotnoisy,
  title={The Noisy Laplacian: a Threshold Phenomenon for Non-Linear Dimension Reduction},
  author={Kokot, Alex and Murad, Octavian-Vlad and Meila, Marina},
  booktitle={Forty-second International Conference on Machine Learning}
}

@inproceedings{
saidi2025recovering,
title={Recovering Manifold Structure Using Ollivier Ricci Curvature},
author={Tristan Luca Saidi and Abigail Hickok and Andrew J. Blumberg},
booktitle={The Thirteenth International Conference on Learning Representations},
year={2025},
url={https://openreview.net/forum?id=aX7X9z3vQS}
}

@article{kremer2015nearest,
  title={Nearest Neighbor Density Ratio Estimation for Large-Scale Applications in Astronomy},
  author={Kremer, Jan and Gieseke, Fabian and Pedersen, Kim Steenstrup and Igel, Christian},
  journal={Astronomy and Computing},
  volume={12},
  pages={67--72},
  year={2015},
  doi={10.1016/j.ascom.2015.06.005}
}

@book{ambrosio2000functions,
  author    = {Ambrosio, Luigi and Fusco, Nicola
               and Pallara, Diego},
  title     = {Functions of Bounded Variation
               and Free Discontinuity Problems},
  series    = {Oxford Mathematical Monographs},
  publisher = {Oxford University Press},
  year      = {2000},
  isbn      = {9780198502456}
}

@book{hirsch1976differential,
  author    = {Hirsch, Morris W.},
  title     = {Differential Topology},
  series    = {Graduate Texts in Mathematics},
  volume    = {33},
  publisher = {Springer},
  year      = {1976},
  doi       = {10.1007/978-1-4684-9449-5}
}

@book{federer1969geometric,
  author    = {Federer, Herbert},
  title     = {Geometric Measure Theory},
  series    = {Grundlehren der mathematischen Wissenschaften},
  volume    = {153},
  publisher = {Springer},
  year      = {1969},
  doi       = {10.1007/978-3-642-62010-2}
}

@inproceedings{menon2016linking,
  title={Linking Losses for Density Ratio and Class-Probability Estimation},
  author={Menon, Aditya Krishna and Ong, Cheng Soon},
  booktitle={Proceedings of the 33rd International Conference on Machine Learning},
  volume={48},
  pages={304--313},
  year={2016},
  publisher={PMLR}
}

@book{vershynin2018high,
  title={High-dimensional probability: An introduction with applications in data science},
  author={Vershynin, Roman},
  volume={47},
  year={2018},
  publisher={Cambridge university press}
}

@book{bhatia2013matrix,
  title={Matrix analysis},
  author={Bhatia, Rajendra},
  volume={169},
  year={2013},
  publisher={Springer Science \& Business Media}
}

@article{Schmidt_Hieber_2020,
   title={Nonparametric regression using deep neural networks with ReLU activation function},
   volume={48},
   ISSN={0090-5364},
   url={http://dx.doi.org/10.1214/19-AOS1875},
   DOI={10.1214/19-aos1875},
   number={4},
   journal={The Annals of Statistics},
   publisher={Institute of Mathematical Statistics},
   author={Schmidt-Hieber, Johannes},
   year={2020},
   month=aug }

@book{gut2013probability,
  title={Probability: A Graduate Course},
  author={Gut, Allan},
  edition={2},
  year={2013},
  publisher={Springer},
  address={New York}
}

@book{grafakos2014classical,
  title={Classical Fourier Analysis},
  author={Grafakos, Loukas},
  edition={3},
  series={Graduate Texts in Mathematics},
  volume={249},
  year={2014},
  publisher={Springer},
  address={New York}
}

@article{juditsky2009nonparametric,
  author  = {Juditsky, Anatoli B. and Lepski, Oleg V. and Tsybakov, Alexandre B.},
  title   = {Nonparametric Estimation of Composite Functions},
  journal = {The Annals of Statistics},
  year    = {2009},
  volume  = {37},
  number  = {3},
  pages   = {1360--1404},
  doi     = {10.1214/08-AOS611}
}

@article{hardle1989average,
  author  = {H{\"a}rdle, Wolfgang and Stoker, Thomas M.},
  title   = {Investigating Smooth Multiple Regression by the Method of Average Derivatives},
  journal = {Journal of the American Statistical Association},
  year    = {1989},
  volume  = {84},
  number  = {408},
  pages   = {986--995},
  doi     = {10.1080/01621459.1989.10478863}
}

@article{powell1989semiparametric,
  author  = {Powell, James L. and Stock, James H. and Stoker, Thomas M.},
  title   = {Semiparametric Estimation of Index Coefficients},
  journal = {Econometrica},
  year    = {1989},
  volume  = {57},
  number  = {6},
  pages   = {1403--1430},
  doi     = {10.2307/1913713}
}

@article{ichimura1993semiparametric,
  author  = {Ichimura, Hidehiko},
  title   = {Semiparametric Least Squares ({SLS}) and Weighted {SLS} Estimation of Single-Index Models},
  journal = {Journal of Econometrics},
  year    = {1993},
  volume  = {58},
  number  = {1--2},
  pages   = {71--120},
  doi     = {10.1016/0304-4076(93)90114-K}
}

@article{hardle1993optimal,
  author  = {H{\"a}rdle, Wolfgang and Hall, Peter and Ichimura, Hidehiko},
  title   = {Optimal Smoothing in Single-Index Models},
  journal = {The Annals of Statistics},
  year    = {1993},
  volume  = {21},
  number  = {1},
  pages   = {157--178},
  doi     = {10.1214/aos/1176349020}
}

@article{hristache2001direct,
  author  = {Hristache, Marian and Juditsky, Anatoli and Spokoiny, Vladimir},
  title   = {Direct Estimation of the Index Coefficient in a Single-Index Model},
  journal = {The Annals of Statistics},
  year    = {2001},
  volume  = {29},
  number  = {3},
  pages   = {593--623},
  doi     = {10.1214/aos/1009210682}
}

@article{hristache2001structure,
  author  = {Hristache, Marian and Juditsky, Anatoli and Polzehl, J{\"o}rg and Spokoiny, Vladimir},
  title   = {Structure Adaptive Approach for Dimension Reduction},
  journal = {The Annals of Statistics},
  year    = {2001},
  volume  = {29},
  number  = {6},
  pages   = {1537--1566},
  doi     = {10.1214/aos/1015345954}
}

@article{cook2002dimension,
  author  = {Cook, R. Dennis and Li, Bing},
  title   = {Dimension Reduction for Conditional Mean in Regression},
  journal = {The Annals of Statistics},
  year    = {2002},
  volume  = {30},
  number  = {2},
  pages   = {455--474},
  doi     = {10.1214/aos/1021379861}
}

@article{xia2002adaptive,
  author  = {Xia, Yingcun and Tong, Howell and Li, W. K. and Zhu, Li-Xing},
  title   = {An Adaptive Estimation of Dimension Reduction Space},
  journal = {Journal of the Royal Statistical Society: Series B},
  year    = {2002},
  volume  = {64},
  number  = {3},
  pages   = {363--410},
  doi     = {10.1111/1467-9868.03411}
}

@article{xia2007constructive,
  author  = {Xia, Yingcun},
  title   = {A Constructive Approach to the Estimation of Dimension Reduction Directions},
  journal = {The Annals of Statistics},
  year    = {2007},
  volume  = {35},
  number  = {6},
  doi     = {10.1214/009053607000000352}
}

@article{fukumizu2009kernel,
  author  = {Fukumizu, Kenji and Bach, Francis R. and Jordan, Michael I.},
  title   = {Kernel Dimension Reduction in Regression},
  journal = {The Annals of Statistics},
  year    = {2009},
  volume  = {37},
  number  = {4},
  pages   = {1871--1905},
  doi     = {10.1214/08-AOS637}
}

@article{lee2013nonlinear,
  author  = {Lee, Kuang-Yao and Li, Bing and Chiaromonte, Francesca},
  title   = {A General Theory for Nonlinear Sufficient Dimension Reduction: Formulation and Estimation},
  journal = {The Annals of Statistics},
  year    = {2013},
  volume  = {41},
  number  = {1},
  pages   = {221--249},
  doi     = {10.1214/12-AOS1071}
}

@article{chen2024deep,
  author  = {Chen, YinFeng and Jiao, YuLing and Qiu, Rui and Yu, Zhou},
  title   = {Deep Nonlinear Sufficient Dimension Reduction},
  journal = {The Annals of Statistics},
  year    = {2024},
  volume  = {52},
  number  = {3},
  pages   = {1201--1226},
  doi     = {10.1214/24-AOS2390}
}

@article{tang2026belted,
  author  = {Tang, Yin and Li, Bing},
  title   = {Belted and Ensembled Neural Network for Linear and Nonlinear Sufficient Dimension Reduction},
  journal = {Journal of the American Statistical Association},
  year    = {2026},
  volume  = {121},
  number  = {554},
  pages   = {1244--1255},
  doi     = {10.1080/01621459.2025.2590775}
}

@article{ruppert1994multivariate,
  author  = {Ruppert, David and Wand, M. P.},
  title   = {Multivariate Locally Weighted Least Squares Regression},
  journal = {The Annals of Statistics},
  year    = {1994},
  volume  = {22},
  number  = {3},
  pages   = {1346--1370},
  doi     = {10.1214/aos/1176325632}
}

@article{lu1996partial,
  author  = {Lu, Zhan-Qian},
  title   = {Multivariate Locally Weighted Polynomial Fitting and Partial Derivative Estimation},
  journal = {Journal of Multivariate Analysis},
  year    = {1996},
  volume  = {59},
  number  = {2},
  pages   = {187--205},
  doi     = {10.1006/jmva.1996.0060}
}

@article{bickel2007local,
  author  = {Bickel, Peter J. and Li, Bo},
  title   = {Local Polynomial Regression on Unknown Manifolds},
  journal = {IMS Lecture Notes--Monograph Series},
  year    = {2007},
  volume  = {54},
  pages   = {177--186},
  doi     = {10.1214/074921707000000148}
}

@article{cheng2013local,
  author  = {Cheng, Ming-Yen and Wu, Hau-Tieng},
  title   = {Local Linear Regression on Manifolds and Its Geometric Interpretation},
  journal = {Journal of the American Statistical Association},
  year    = {2013},
  volume  = {108},
  number  = {504},
  pages   = {1421--1434},
  doi     = {10.1080/01621459.2013.827984}
}

@article{wu2026conditional,
  author  = {Wu, Yantao and Maggioni, Mauro},
  title   = {Conditional Regression for the Nonlinear Single-Variable Model},
  journal = {Journal of Machine Learning Research},
  year    = {2026},
  volume  = {27},
  number  = {155},
  pages   = {1--75}
}

@article{ruud1986consistent,
  title={Consistent estimation of limited dependent variable models despite misspecification of distribution},
  author={Ruud, Paul A},
  journal={Journal of Econometrics},
  volume={32},
  number={1},
  pages={157--187},
  year={1986},
  publisher={Elsevier}
}

@book{tsybakov2009introduction,
  title={Introduction to nonparametric estimation},
  author={Tsybakov, Alexandre B and Zaiats, Vladimir},
  volume={11},
  year={2009},
  publisher={Springer}
}

@article{mityagin2020zero,
  author  = {Mityagin, Boris S.},
  title   = {The zero set of a real analytic function},
  journal = {Mathematical Notes},
  volume  = {107},
  number  = {3},
  pages   = {529--530},
  year    = {2020},
  doi     = {10.1134/S0001434620030189}
}

@phdthesis{zeng2026rank,
  author  = {Zeng, Lang},
  title   = {Rank-Based Deep Learning for Survival Analysis},
  school  = {University of Pittsburgh},
  year    = {2026},
  note    = {ProQuest Dissertations \& Theses, 32674948}
}

@book{vandervaart1998asymptotic,
  author    = {van der Vaart, A. W.},
  title     = {Asymptotic Statistics},
  series    = {Cambridge Series in Statistical and Probabilistic Mathematics},
  publisher = {Cambridge University Press},
  address   = {Cambridge},
  year      = {1998}
}

@book{rudin1976principles,
  author    = {Rudin, Walter},
  title     = {Principles of Mathematical Analysis},
  edition   = {3},
  publisher = {McGraw-Hill},
  year      = {1976}
}

@book{rudin1987real,
  author    = {Rudin, Walter},
  title     = {Real and Complex Analysis},
  edition   = {3},
  publisher = {McGraw-Hill},
  address   = {New York},
  year      = {1987}
}

@article{schiebinger2019reprogramming,
  title={Optimal-Transport Analysis of Single-Cell Gene Expression Identifies Developmental Trajectories in Reprogramming},
  author={Schiebinger, Geoffrey and Shu, Jian and Tabaka, Marcin and Cleary, Brian and Subramanian, Vidya and Solomon, Aryeh and Gould, Joshua and Liu, Siyan and Lin, Stacie and Berube, Peter and Lee, Lia and Chen, Jenny and Brumbaugh, Justin and Rigollet, Philippe and Hochedlinger, Konrad and Jaenisch, Rudolf and Regev, Aviv and Lander, Eric S.},
  journal={Cell},
  volume={176},
  number={4},
  pages={928--943.e22},
  year={2019},
  doi={10.1016/j.cell.2019.01.006}
}

@article{mcinnes2018umap,
  title={{UMAP}: Uniform Manifold Approximation and Projection for Dimension Reduction},
  author={McInnes, Leland and Healy, John and Melville, James},
  journal={arXiv preprint arXiv:1802.03426},
  year={2018},
  doi={10.48550/arXiv.1802.03426}
}

@article{horowitz2007rate,
  title={Rate-optimal estimation for a general class of nonparametric regression models with unknown link functions},
  author={Horowitz, Joel L and Mammen, Enno},
  year={2007}
}

@article{kereta2021nonlinear,
  title={Nonlinear generalization of the monotone single index model},
  author={Kereta, {\v{Z}}eljko and Klock, Timo and Naumova, Valeriya},
  journal={Information and Inference: A Journal of the IMA},
  volume={10},
  number={3},
  pages={987--1029},
  year={2021},
  publisher={Oxford University Press}
}

@article{horowitz2001nonparametric,
  title={Nonparametric estimation of a generalized additive model with an unknown link function},
  author={Horowitz, Joel L},
  journal={Econometrica},
  volume={69},
  number={2},
  pages={499--513},
  year={2001},
  publisher={JSTOR}
}

@article{lu1996multivariate,
  title={Multivariate locally weighted polynomial fitting and partial derivative estimation},
  author={Lu, Zhan-Qian},
  journal={journal of multivariate analysis},
  volume={59},
  number={2},
  pages={187--205},
  year={1996},
  publisher={Elsevier}
}

@book{fan2018local,
  title={Local polynomial modelling and its applications: monographs on statistics and applied probability 66},
  author={Fan, Jianqing},
  year={2018},
  publisher={Routledge}
}

\end{document}